\documentclass{elsarticle}
\pdfoutput=1

\usepackage[utf8]{inputenc}
\usepackage{xspace}
\usepackage{tikz}
\usepackage{graphicx}
\usepackage{subfigure}
\usepackage{comment}
\usepackage{placeins}
\usepackage[autostyle]{csquotes}
\usepackage{amsmath}
\usepackage{todonotes}
\usepackage{anslistings}
\usepackage{enumitem}

\usepackage{hyperref}
\hypersetup{
    colorlinks,
    linkcolor={red!50!black},
    citecolor={blue!50!black},
    urlcolor={blue!80!black}
}

\newcommand{\enablelinenumbers}{}
\newread\datafile
\openin\datafile=linenumbers.txt
\ifeof\datafile
\else
\closein\datafile
\input{linenumbers.txt}
\fi

\usetikzlibrary{positioning}

\journal{Computers \& Mathematics with Applications}

\setlist[description]{style=multiline,leftmargin=2cm,labelwidth=3.5em,itemsep=0pt}

\newcommand{\dune}{\textsc{Dune}\xspace}
\newcommand{\dunecommon}{dune-common\xspace}

\newcommand{\dunegrid}{dune-grid\xspace}
\newcommand{\duneistl}{dune-istl\xspace}

\newcommand{\opm}{OPM\xspace}

\newcommand{\opmdata}{opm-data\xspace}

\newcommand{\opmmaterial}{opm-material\xspace}

\newcommand{\opmtests}{opm-tests\xspace}

\newcommand{\opmflow}{OPM Flow\xspace}

\newcommand{\opmversion}{2019.10\xspace}

\newcommand{\petsc}{PETSc\xspace}

\newcommand{\COto}{CO\ensuremath{_\mathsf{2}}\xspace}

\newcommand{\code}[1]{\texttt{#1}}
\newcommand{\Code}[1]{\texttt{#1}}

\newcommand{\Kb}{{\bf K}}
\newcommand{\vb}{{\bf v}}
\newcommand{\gb}{{\bf g}}
\newcommand{\nb}{{\bf n}}

\newcommand{\ub}{{\bf u}}
\newcommand{\cb}{{\bf c}}

\graphicspath{{./figures/spe1/}{./figures/spe3/}{./figures/spe9/}}

\begin{document}

\begin{frontmatter}

\title{The Open Porous Media Flow Reservoir Simulator}

\author[1]{Atgeirr Fl{\o} Rasmussen}
\author[2]{Tor Harald Sandve}
\author[1]{Kai Bao}
\author[3]{Andreas Lauser}
\author[4]{Joakim Hove}
\author[1]{B{\aa}rd Skaflestad}
\author[2]{Robert Kl\"ofkorn}
\author[5]{Markus Blatt}
\author[6]{Alf Birger Rustad}
\author[2]{Ove S{\ae}vareid}
\author[1]{Knut-Andreas Lie}
\author[7]{Andreas Thune}
\address[1]{SINTEF Digital}
\address[2]{NORCE Norwegian Research Centre AS}
\address[3]{Poware Software Solutions}
\address[4]{DataGR}
\address[5]{Dr.~Blatt HPC-Simulation-Software \& Services}
\address[6]{Equinor ASA}
\address[7]{Simula Research Laboratory}

\begin{abstract}
  The Open Porous Media (\opm) initiative is a community effort
  that encourages open innovation and reproducible research for
  simulation of porous media processes. \opm coordinates
  collaborative software development, maintains and distributes
  open-source software and open data sets, and seeks to ensure that
  these are available under a free license in a long-term perspective.

  In this paper, we present \opmflow, which is a reservoir simulator
  developed for industrial use, as well as some of the individual
  components used to make \opmflow. The descriptions apply to the \opmversion 
  release of \opm.
\end{abstract}

\end{frontmatter}

\enablelinenumbers


\section{Introduction}


The Open Porous Media (\opm) initiative was started in 2009 to
encourage open innovation and reproducible research on modelling and
simulation of porous media processes. \opm was initially founded as a
collaboration between groups at Equinor (formerly Statoil), SINTEF, the University of
Stuttgart, and the University of Bergen, but over time, several other groups
and individuals have joined and contributed. What today forms the \opm suite of
software, has mainly been developed by SINTEF, NORCE (formerly IRIS),
Equinor, Ceetron Solutions,
Poware Software Solutions, and Dr.~Blatt HPC-Simulation-Software \& Services.

The initial vision was to create long-lasting, efficient, and
well-maintained, open-source software for simulating flow and transport in porous
media. The scope has later been extended to also provide open data sets, thus
making it easier to benchmark, compare, and test different
mathematical models, computational methods, and software implementations.


\begin{figure}
  \center
  \scalebox{.85}{\tikzstyle{module}=[black, rectangle, rounded corners, fill=blue!20, minimum height=6mm, draw]
\tikzstyle{depend}=[black, thick, ->, >=stealth]
\tikzstyle{optional}=[black, thick, dashed, ->, >=stealth]
\begin{tikzpicture}[scale=1.0]

  \node [module] (opm-common) at (5,4) {\Code{opm-common}};
  \node [module] (opm-material) at (6,3) {\Code{opm-material}};
  \node [module] (opm-grid) at (3,3) {\Code{opm-grid}};
  \node [module] (opm-upscaling) at (2,2) {\Code{opm-upscaling}};
  \node [module] (opm-models) at (5, 2) {\Code{opm-models}};
  \node [module] (opm-simulators) at (5,1) {\Code{opm-simulators}};

  \draw [depend] (opm-material) to (opm-common);
  \draw [depend] (opm-grid) to (opm-common);
  \draw [depend] (opm-models) to (opm-material);
  \draw [depend] (opm-models) to (opm-grid);
  \draw [depend] (opm-upscaling) to (opm-grid);
  \draw [depend] (opm-simulators) to (opm-models);

\end{tikzpicture}}
  \caption{Current module structure as of the \opmversion release.
    Arrows indicate inter-module dependencies: a module depends on
    another module if it has an arrow pointing towards it.
    Since the \opm software suite is in active development, the module
    organization is likely to keep changing in the future.}
  \label{fig:modules}
\end{figure}
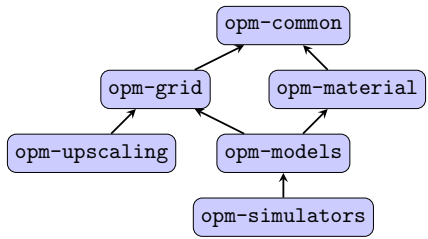

\opm has so far primarily focussed on tools for reservoir simulation, and consists today
of the \opmflow\ reservoir simulator, upscaling tools, and a selection of supporting and
experimental software pieces. The corresponding source code is divided into six modules,
as shown in Figure~\ref{fig:modules}, and is organized as a number of git repositories
hosted on \url{github.com/OPM}. The data repositories \opmdata and \opmtests are also
hosted there, and contain example cases and data used for integration testing.
Herein, we only give an in-depth description of \opmflow,
which in turn uses or builds on existing frameworks and libraries such as
\href{http://www.dune-project.org}{\dune} \citep{dunepaperI:08}, DuMuX \citep{DuMuX},
Zoltan \citep{ZoltanIsorropiaOverview2012}, and \href{http://www.boost.org}{Boost} to
reduce implementation and maintenance cost and improve software quality.  The Matlab
Reservoir Simulation Toolbox \citep{MRST, MRST-AD, mrst-book} has also been a significant
source of ideas and concepts. In addition to the simulator tools, \opm contains the
\href{http://resinsight.org}{ResInsight} postprocessor, which has been used for all the 3D
visualization herein. Apart from computational routines underlying flow diagnostics
\cite{MKL14:diagnostics}, ResInsight is independent of the other \opm modules.

All software in \opm is distributed under the GNU General Public License (GPL),
\href{http://www.gnu.org/licenses/gpl-3.0.en.html}{version 3}, with the option of using
newer versions of the license. Datasets are distributed under the Open Database License,
\href{http://opendatacommons.org/licenses/odbl/1.0/}{version 1.0}, with individual content
under the Database Contents License,
\href{http://opendatacommons.org/licenses/dbcl/1.0/}{version 1.0}.

Code development in \opm follows an open model: All source code
contributions are made using the GitHub pull request
mechanism. Anyone can create such a request, which asks the
maintainers to merge some change (bugfix, improvement, or new feature)
into the master branch of one of the module
repositories. Each request is reviewed and typically merged by
the maintainer after some discussion and possible revisions of the
proposed changes. \opm generally
accepts contributions of many kinds, not just software improvements
and additions, but also use-case reports, documentation, and bug
reports.


\section{\opmflow} \label{sec:flow}

\opmflow is developed to serve two main
purposes: 1) as an open-source alternative and
drop-in replacement for contemporary commercial simulators to assist
oil companies in planning and managing (conventional) hydrocarbon
assets, and 2) as a research and prototyping platform
that enables its users to implement, test, and validate new models and
computational methods in a realistic and industry-relevant
setting.  To this end, \opmflow aims to represent
reservoir geology, fluid behaviour, and description of wells and
production facilities  as in commercial simulators and
hence offers standard
fully-implicit discretizations of black-oil type models and supports
industry-standard input and output formats. The simulator is implemented using
automatic differentiation (AD) \cite{Neidinger2010} to avoid error-prone derivation and
coding of analytical Jacobians for the residual equations, which also
makes it easier to extend the simulator with new model equations, e.g.,
for CO$_2$ injection or enhanced oil recovery (EOR).
Equally important, using AD makes it realtively simple to extend the simulator to compute
sensitivities and gradients with respect to various types of model parameters, using e.g.,
an adjoint method \cite{Jansen:adjoints}.

This paper does not aim to explain how to install or run \opmflow. For
that we refer to the web site, \href{http://opm-project.org}{opm-project.org}, or the \opmflow manual
\cite{flowmanual}.

\begin{figure}
\centering
\includegraphics[width=0.95\textwidth]{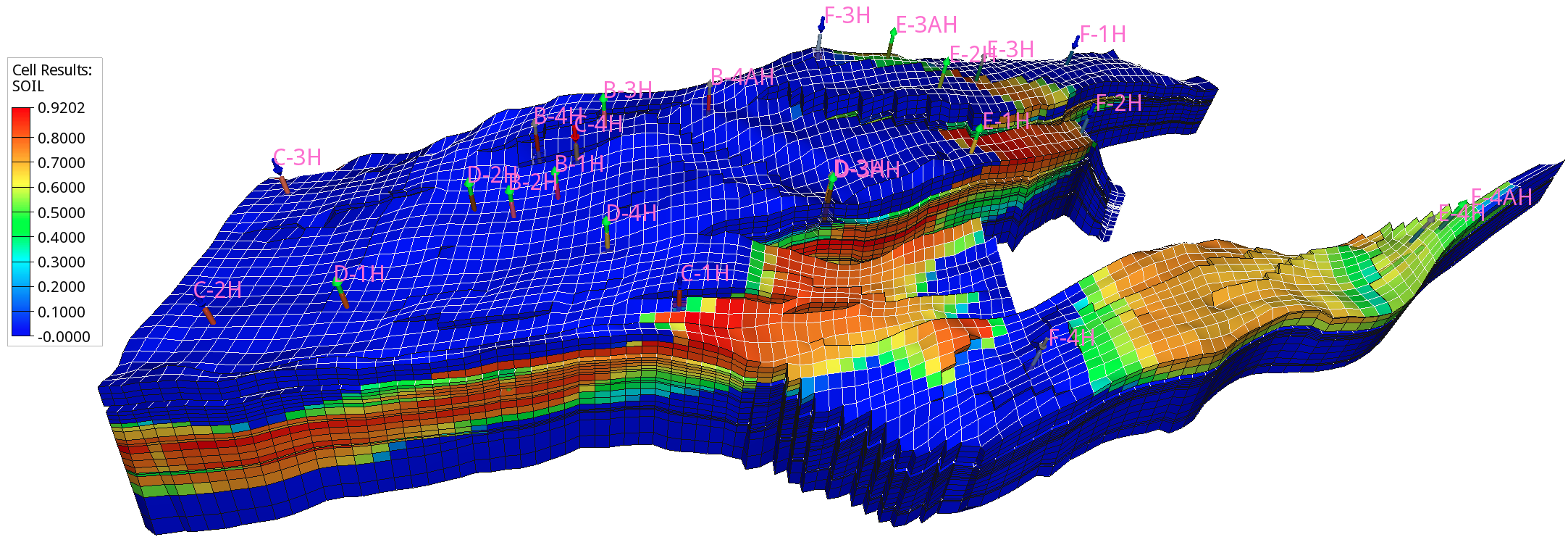}
\caption{The Norne reservoir showing oil saturation and wells after
  approximately 3.5 years of production. The z-direction has been
  exaggerated 5 times.}
\label{fig:norne-soil}
\end{figure}

\subsection{The black-oil model}
\label{sec:black-oil}

The black-oil equations constitute the most widely used flow model in
simulation of hydrocarbon reservoirs. The model is based on the
premise that we have three different fluid phases (aqueous, oleic, and
gaseous) and three (pseudo)components: water, oil, and gas.  Oil and gas
are not single hydrocarbon species, but represent all hydrocarbon species that
exist in liquid and vapor form at suface conditions.  Mixing is allowed, in
the sense that both oil and gas can be found in the
oleic phase, in the gaseous phase, or in both. The amounts of dissolved
gas in the oleic phase and vaporized oil in the gaseous phase must
be kept track of. Below, we will use subscripts $w, o, g$ to indicate
quantities related to each of the phases or components.

\subsubsection{Model equations}

The black-oil equations can be deduced from conservation of mass for each
component with suitable closure relationships such as Darcy's law and initial
and boundary conditions.  The equations are discretized in space with
an upwind finite-volume scheme using a two-point flux approximation
and in time using an implicit (backward) Euler scheme. The resulting
equations are solved simultaneously in a fully implicit formulation by
a Newton-type linearization with a properly preconditioned, iterative
linear solver. We give both the continuous and discrete equations.

\paragraph{Continuous equations}

The conservation laws form a system of partial differential
equations, one for each pseudocomponent $\alpha$:
\begin{align}\label{eq:materialbalance-cont}
\frac{\partial}{\partial t} \left(\phi_{\rm ref} A_\alpha \right)
 + \nabla \cdot \ub_\alpha + q_\alpha & = 0,
\end{align}
where the accumulation terms and fluxes are given by
\begin{subequations}
\label{eq:acc}
\begin{align}
  A_w &= m_{\phi} b_{w} s_{w},                          & \ub_w &= b_{w} \vb_{w}, \\
  A_o &= m_{\phi} (b_{o} s_{o} + r_{og}b_{g}s_{g}), & \ub_o &= b_{o} \vb_{o} +
r_{og}b_{g}\vb_{g}, \\
  A_g &= m_{\phi} (b_{g} s_{g} + r_{go}b_{o}s_{o}), & \ub_g &= b_{g} \vb_{g} +
r_{go}b_{o}\vb_{o}.
\end{align}
\end{subequations}
and the phase fluxes are given by Darcy's law:
\begin{align}
  \label{eq:Darcy-mph}
  \vb_\alpha = -\lambda_\alpha \Kb (\nabla p_\alpha - \rho_\alpha \gb).
\end{align}
In addition, the following closure relations should hold:
\begin{align}
s_w + s_o + s_g &= 1, \label{eq:fluids-fill-porevolume} \\
p_{c,ow} & = p_{o} - p_{w}, \label{eq:oil-water-capillary} \\
p_{c,og} & = p_{o} - p_{g}; \label{eq:oil-gas-capillary}
\end{align}
see \ref{appendix:nomenclature:cont} for definitions of the
symbols used.

\paragraph{Discrete equations}

In the following, let subscript $i$ denote a discrete quantity defined
in cell $i$ and subscript $ij$ denote a discrete quantity defined at
the connection between two cells $i$ and $j$. Two cells share a
connection if they are adjacent geometrically in the computational
grid or share an explicit non-neighbor connection (NNC; provided by
the user). For example
$v_{o, ij}$ is the oil flux from cell $i$ to cell $j$. For
oriented quantities such as fluxes, the orientation is taken to be
from cell $i$ to cell $j$, and the quantity is skew-symmetric
($v_{o,ij} = -v_{o,ji}$), whereas non-oriented quantities such as the
transmissibility $T_{ij}$ are symmetric ($T_{ij} = T_{ji}$).
Quantities with superscript 0 are taken at the start of the discrete
time step, other quantities are at the end of the time
step. Superscripts or subscripts applied to an expression in
parenthesis apply to each element in the expression.

The discretized equations and residuals are, for each
pseudo-component $\alpha$ and cell $i$:
\begin{align}\label{eq:materialbalance}
  R_{\alpha,i} = \frac{\phi_{{\rm ref},i} V_i}{\Delta t} \left( A_{\alpha,i} -
A_{\alpha,i}^0\right)
  + \sum_{j \in C(i)} u_{\alpha,ij} + q_{\alpha,i} & = 0,
\end{align}
where $A_\alpha$ are as in \eqref{eq:acc} and 
$u_\alpha$ are fluxes defined similar to the velocities $\ub_\alpha$:
\begin{subequations}
\begin{align}
 u_w &= b_{w} v_{w}, \\
 u_o &= b_{o} v_{o} + r_{og}b_{g}v_{g}, \\
 u_g &= b_{g} v_{g} + r_{go}b_{o}v_{o}.
\end{align}
\end{subequations}
The relations \eqref{eq:fluids-fill-porevolume},
\eqref{eq:oil-water-capillary} and \eqref{eq:oil-gas-capillary} hold
for each cell $i$.
The fluxes are given for each connection $ij$ by:
\begin{align}
  (b_{\alpha} v_{\alpha})_{ij}
  & = (b_\alpha \lambda_{\alpha} m_T)_{U(\alpha, ij)}
    T_{ij} \Delta \Phi_{\alpha, ij} \label{eq:flux1} \\
  (r_{\beta\alpha} b_{\alpha} v_{\alpha})_{ij}
  & = (r_{\beta\alpha} b_\alpha \lambda_{\alpha} m_T)_{U(\alpha, ij)}
    T_{ij} \Delta \Phi_{\alpha, ij} \label{eq:flux2} \\
  \Delta \Phi_{\alpha,ij} & = p_{\alpha,i} - p_{\alpha,j}
                         - g \rho_{\alpha,ij} (z_i - z_j) \label{eq:head} \\
  \rho_{\alpha,ij} & = (\rho_{\alpha,i} + \rho_{\alpha,j})/2 \\
  U(\alpha, ij) & = \begin{cases}
       i, & \Delta \Phi_{\alpha,ij} \geq 0, \\
       j, & \Delta \Phi_{\alpha,ij} < 0. \\
     \end{cases}
\end{align}
See \ref{appendix:nomenclature:disc} for definitions of the
symbols used above.

\subsubsection{Choice of primary variables}

For a grid with $n$ cells, the material balances \eqref{eq:materialbalance} give us $3n$
equations. (When only two phases are defined, either oil/water or oil/gas, the system is
simplified and we solve only two mass balance equations, giving 2n equations.)
In addition, we
have equations and unknowns from the well model, which will be
described in Section~\ref{sec:wellmodels}. An important question
is which quantities we should use as primary variables, i.e., the
unknowns we solve for. There is some flexibility, as we have
relations connecting many of the quantities.

Our basic choice for non-miscible flow is to use the pressure of one
of the phases (which we take to be oil pressure $p_o$), $s_w$, and
$s_g$ as our primary variables. The pressure $p_o$ will then behave
mostly in an elliptic fashion, whereas the saturations $s_w$ and $s_g$
are more hyperbolic.

For miscible flow, it is a little more complicated: Since the gaseous
phase may disappear if all the gas dissolves into the oleic phase, and
similarly the oleic phase may disappear if all the oil vaporizes into
the gaseous phase, we cannot always use $s_g$ for our third variable.
Instead we use the third variable to track the composition of the phase that
has {\em not} disappeared. We therefore use either $s_g$, $r_{go}$, or
$r_{og}$ as our third primary variable, depending on the fluid state,
and we call that variable $x$:
\begin{equation}
  x = \begin{cases}
    s_g, &  \text{all three phases present}, \\
    r_{go}, &  \text{no gaseous phase}, \\
    r_{og}, &  \text{no oleic phase}. \\
  \end{cases}
  \label{eq:primaryVariableMeaning}
\end{equation}
This choice is made separately for each cell $i$ depending on its
state.

\subsubsection{Initial and boundary conditions}

Initial conditions are defined by the initial values of
pressure $p$, saturations $s_\alpha$, and, if applicable, the mixing
ratios $r_{go}$ and $r_{og}$. These values can be specified by the
user directly in the input data, or computed from an equilibration
procedure that ensures the initial state is in vertical equilibrium.
This procedure takes as input the fluid pressure at a
reference depth, the depths of the water-oil and oil-gas contacts,
and the capillary pressures at those contact depths. It then solves
the following ordinary differential equations:
\begin{equation}
  \frac{\mathrm{d}p_{\alpha}}{\mathrm{d}z} =\rho_{\alpha}(z,p_{\alpha})\, g.
\end{equation}
The fluid density typically depends on the pressure, and possibly
also on its composition; see Section~\ref{sec:pvt} for details. The
equations are solved numerically using a 4th-order Runge-Kutta method, with
the order of the phases decided by the location of the datum depth: First, we
solve for the phase in whose zone the datum is located, using the
datum pressure to fixate the pressure solution. We then solve for phases
corresponding to the zones above and below. To fixate the
phase pressures, we use the already computed phase pressure(s)
evaluated at the zone contact depth and the input capillary pressures.

Good estimates of the initial water saturation of the reservoir are often known 
from seismic and well logs. This data can be used to improve the initial state. 
The equilibration procedure is the same as described above, 
but the capillary pressure is scaled to match the initial water saturation. 
Each cell will thus have a its own scaling of the capillary pressure that 
needs to be taken into account during the simulations. 

\opmflow has so far mainly been developed to simulate reservoirs having
no fluid communication with the surrounding rock; the default boundary
conditions are therefore no-flow Neumann conditions. In that case, the
only way the reservoir communicates with the outside is through
the coupled well models, see Section~\ref{sec:wellmodels}. In recent
versions, support for more general boundary conditions has been
added, as well as support for some aquifer models that allow the
reservoir domain to be in communication with surrounding water-carrying rock.

\subsubsection{Rock properties}
\label{sec:rockprops}

Rock {\em porosity}, usually denoted $\phi$, is the void fraction of the
bulk volume that is able to store and transmit fluids.
It usually depends on pressure, which we model
as $\phi = m_\phi(p)\phi_{\rm ref}$, where the multiplier $m_\phi$ is a
function of pressure and $\phi_{\rm ref}$ is a reference porosity. This model
does not account for the possibility of irreversible changes to the
rock, such as fracturing. We model porosity as cell-by-cell piecewise
constant.

The rock {\em permeability} (or absolute permeability), usually
denoted $\Kb$, measures the ability of a porous medium to
transmit a single fluid phase at certain pressure conditions;
see Darcy's law \eqref{eq:Darcy-mph}.
For an isotropic porous medium the permeability is a scalar, but in general
$\Kb$ is a tensor, indicating that the medium's resistance to flow
depends on the flow direction. A typical example includes layered
rocks in subsurface reservoirs, which are typically more permeable in
the directions tangential to the layer than in the normal
direction. We model permeability as a cell-by-cell piecewise constant
tensor. The permeability is not used directly in the discrete
formulation above, but enters the equation through the
transmissibility factor; see \ref{appendix:trans} for details.

\subsubsection{Relative permeability and capillary pressure}
\label{sec:relperm-cappress}
The rock's ability to transmit fluids is reduced when more than one fluid phase are
present. This is modelled by introducing a saturation-dependent factor $k_{r,\alpha}$
called \emph{relative permeability}. (In \eqref{eq:Darcy-mph}, we instead use the
\emph{mobility} $\lambda_{\alpha}=k_{r,\alpha}/\mu_{\alpha}$, where $\mu_\alpha$ denotes
the viscosity.)  For a three-phase system, one could imagine that relative permeabilities
were functions of two independent arguments. A more common and natural approach is to
assemble the three-phase properties stepwise from "two-phase" relations. The water
relative permeability is then assumed to be a function of water saturation only,
$k_{r,w}(s_{w})$, and its domain is confined by the connate water saturation $s_{wco}$ and
maximum saturation $s_{w,max}$.  We complement the water/oil sub-system by a relative
permeability for oil $k_{r,ow}(s_{o})$, with oil saturations bounded by
$s_{o}\in [1-s_{w,max} ,1-s_{wco}]$. Similarly for the gas phase, we have a relative
permeability $k_{r,g}(s_{g})$, and endpoints $s_{g,min}$ and $s_{g,max}$ respectively.
Whereas the connate water $s_{wco}$ is a significant parameter that signals presence of
water also in the gas cap, it is usual to assume $s_{g,min}=0$.  Thus, the oil relative
permeability for the gas/oil/connate-water subsystem, $k_{r,og}(s_{o})$, has a domain
bounded by $s_{o}\in [1-s_{g,max},1-s_{wco}]$.

Both sub-systems are also equipped with a capillary pressure relation, 
and assuming normal wettability conditions (water wetting relative oil) $p_{c,ow}(s_{w})$ will be
a decreasing function of $s_{w}$ while (oil wetting relative gas) $p_{c,og}(s_{g})$ is an increasing 
function of $s_{g}$.

A simple and robust approach for combining $k_{r,ow}$ and $k_{r,og}$ into a true
three-phase relative permeability $k_{r,o}$ is to assume local segregation in each cell
and use a volume-weighted average between the water and gas zones
\begin{align}\label{eq:gwseg}
  k_{r,o}(s_{o}) = \frac{ (s_{w} - s_{wco})k_{r,ow}(s_{o}) + s_{g}k_{r,og}(s_{o})}
                      { s_{w} - s_{wco} + s_{g} }.
\end{align}
For each property, an arbitrary number of alternative tables can be defined and assigned
cellwise throughout the grid.  The functions may also be modified by the introduction of
cellwise transformations, known as endpoint scaling, that provide local adaptivity.

\opmflow provides some support for hysteresis effects \citep{killough1976hysteresis} that
typically will arise from reversal of saturation changes.  For relative permeabilities,
the possibility is limited to the non-wetting phases, i.e., the oil phase in the water/oil
sub-system and the gas in the gas/oil system.  In addition to the usual (drainage)
relative permeability $k_{r,n}(s_{n})$, a second (imbibition) curve $k_{r,n}^{imb}(s_{n})$
must be supplied.  Consider a drainage process that is reversed after reaching some
saturation $s_{n}^{hst}$.  Following Carlson's method \citep{carlson1981hysteresis}, a
so-called scanning curve is obtained by finding an appropriate saturation translation
$\Delta_{n}^{hst}$ from the relation
$k_{r,n}^{imb}(s_{n}^{hst}+\Delta_{n}^{hst})=k_{r,n}(s_{n}^{hst})$.  The imbibition
process will then proceed according to the relative permeability curve
$k_{r,n}^{imb}(s_{n}+\Delta_{n}^{hst})$.

\subsubsection{Pressure-volume-temperature (PVT) relationships}\label{sec:pvt}

Pseudocomponents, or components for short, essentially correspond to the identifiable
phases at standard surface conditions, and each component is assigned a surface density,
$\rho_{S,\alpha}$.  At reservoir conditions, the fluid volume will consist of certain mix
of components, referred to as the composition.  In the standard black-oil formulation,
pressure is essentially the single intensive quantity, thus implicitly assuming
steady-state heat distribution.  (This can to some extent be generalized by partitioning
the reservoir into multiple PVT regions, each with a separate set of properties.)

Water has a simple PVT behavior, with a one-to-one correspondence between the
water component and the aqueous phase in terms of the water shrinkage factor
$b_w=\rho_w/\rho_{S,w}$ that relates the density $\rho_w$ at reservoir conditions to the density $\rho_{S,w}$ at surface conditions. The factor is a function of water pressure, and it is parameterized in terms of 
a constant compressibility, a reference pressure, and a corresponding
reference factor. So $b_w = b_w(p_w; c_w,p_{w,ref},b_{w,ref})$.
Water viscosity, $\mu_w$, is considered pressure dependent, and the fraction $\mu_w/b_w$ is 
also specified as a constant "compressibility" quantity.

The oleic phase behavior can be specified at various levels of complexity, and the
simplest formulation available is similar to that of water.  There is also a slightly
more flexible variant available, in which $b_o$ and $\mu_o$ are given as tabulated
functions of the oil pressure $p_o$, thus relaxing the constant compressibility form.
These versions do not permit any dissolved gas and are consequently referred to as "dead
oil" formulations.

For a system with gas dissolved into the oleic phase, we must specify the dissolution capacity
of the oleic phase, represented as the maximum gas-oil ratio $r_s$, as a function of pressure. 
For each value of $r_s$, the corresponding pressure is referred to as the bubble point pressure $p_{bpp}$.
Thus, the actual ratio of dissolved gas at given reservoir conditions
obeys $r_{go}\leq r_s(p_o)$ with equality if $p_o\leq p_{bpp}$.
Shrinkage factors $b_o$ and viscosities $\mu_o$ must be supplied accordingly, and in particular, 
properties must be defined also for undersaturated conditions,
i.e., pressures above $p_{bpp}$. A simulator must be able to handle
the change from having no free gas at pressures above $p_{bpp}$ to
having a gas phase present at lower pressures. An example is given in
Section~\ref{sec:spe9}; see Figure~\ref{fig:spe9_sgas_swat}.
Oil containing dissolved gas is known as "live oil".

The properties of the gaseous phase mirror those of the oleic, and the gas equivalent to dead oil 
is called "dry gas", referring to the exclusion of vaporized oil from the gaseous phase.
This opposed to "wet gas", which in analogy to the live-oil model, requires specification 
of a maximum vaporized oil-gas ratio $r_v$ that controls the amount of oil in the gaseous phase.
\opmflow can also have a fourth phase to support solvent modelling, and these PVT properties
are defined similar to those of dry gas.

\opmflow provides a mechanism to contain or tune the behavior when combining live-oil and wet-gas properties.
The values of both $r_s$ and $r_v$ may be scaled back by some positive powers of the fraction $(s_o/s_{o,max})$, 
where the denominator refers to the cellwise historical maximum of the oil saturation $s_o$.

\subsection{Well models}\label{sec:wellmodels}

\begin{figure}
  \centering
  \includegraphics[width=0.95\textwidth]{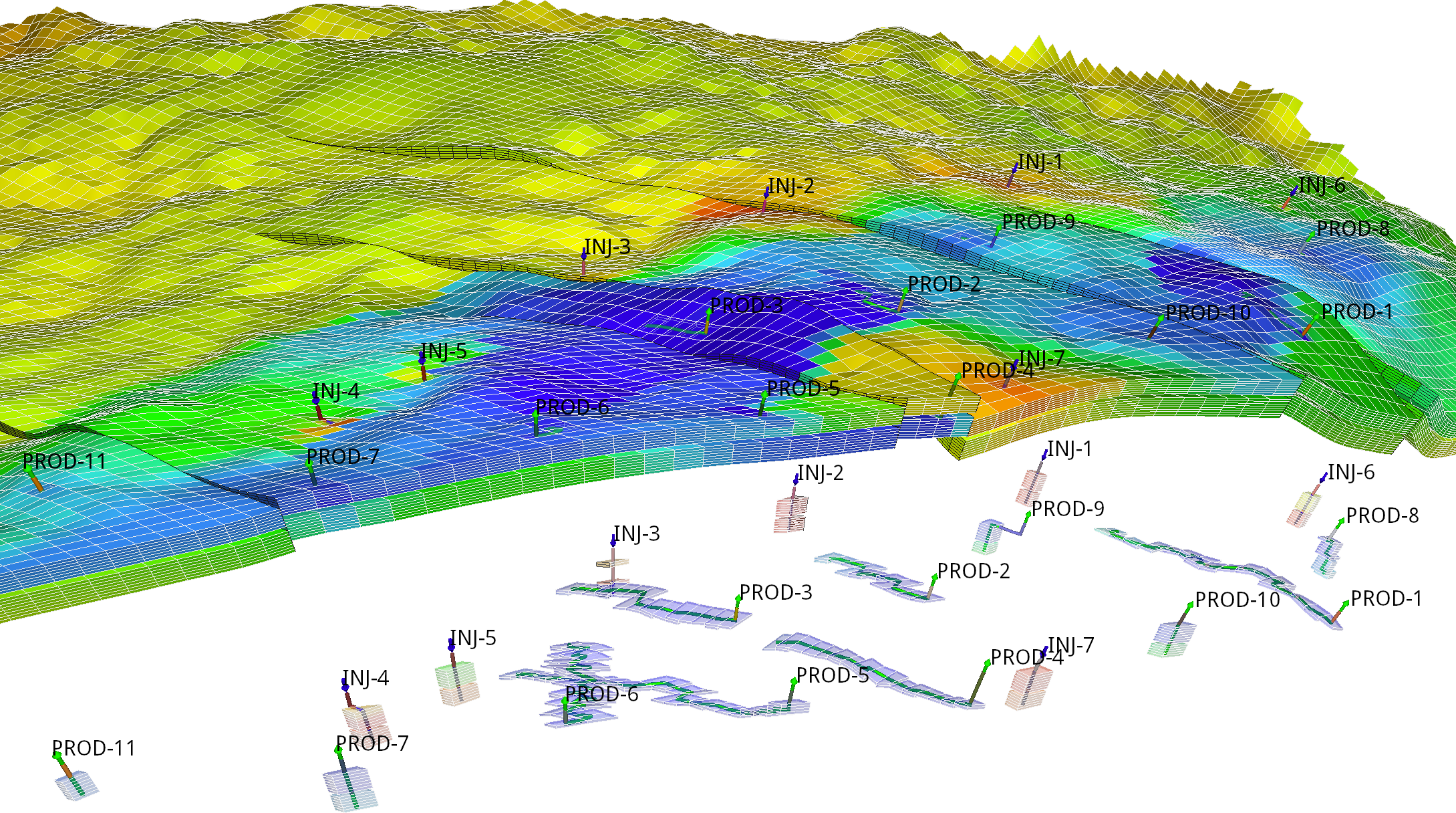}
  \caption{Case from the \href{http://ww.isapp2.com/optimization-challenge.html}{OLYMPUS
      Optimization Challenge} \citep{fonseca2018overview}. Top: pressure at the end of the
  simulation time. Bottom: wells and the cells they are connected to.}
\label{fig:olympus}
\end{figure}

\opm provides two different well models. The 'standard' well model describes the flow
conditions within each well with a single set of primary variables
\citep{holmes1983enhancements}. This works adequately in describing the majority of wells
in reservoir simulation, as a result, it has been used as the default well model. More
advanced multi-segment well models \citep{holmes1998application} discretize the wellbore
with 'segments' to provide more detailed, flexible, and accurate wellbore simulation,
especially for multi-lateral wells and long horizontal wells. In either case, wells
interact with reservoir formation through connections that represent the flow paths
between the wellbore and single grid blocks. The calculation of inflow rates through
connections plays an important role within well modeling. Figure~\ref{fig:olympus} shows
both vertical and mostly horizontal wells in a realistic reservoir model.

\subsubsection{Standard well model}

For the standard well model, a single set of primary variables is introduced
for each well. For a three-phase black oil system, there are four primary
variables: $Q_t$ is the weighted total flow rate, the weighted
fractions of water $F_w$ and gas $F_g$ describe the fluid composition
within the wellbore \citep{holmes1983enhancements}, and the fourth is the
bottom-hole pressure $p_{bhp}$, i.e., the pressure in the wellbore
at the datum depth. The following equations relate
the primary variables to the flow rates:
\begin{align}
    Q_t &= \sum_{\alpha \in \{o, g, w\}} g_{\alpha} Q_{\alpha
          } \label{eq:weighted_total_rate}, \quad
    F_w = \frac{g_w Q_w}{Q_t}, \quad F_g = \frac{g_g Q_g}{Q_t}
\end{align}
Here, $Q_{\alpha}$ is the flow rate of component $\alpha$ under surface
conditions and $g_\alpha$ is a weighting factor. For gas, this factor
is typically chosen to be a small value like
$0.01$ to avoid gas fractions close to unity
\citep{holmes1983enhancements}.

Volumetric inflow rates at reservoir conditions are calculated as
\begin{equation}
q_{\alpha,j}^r = T_{w,j} M_{\alpha,j} [p_j - (p_{bhp,w} + h_{w,j})],
\label{eq:wellinflow}
\end{equation} 
where $q_{\alpha,j}^r$ is the flow rate of phase $\alpha$ through connection $j$. The
other symbols are described in \ref{appendix:nomenclature:well}.  The rate
is negative for flow from wellbore to reservoir, and positive for flow
into the opposite direction. The pressure differences $h_{w,j}$ between connection $j$ and
the datum point are computed explicitly based on the fluid composition in the wellbore at
the start of the timestep.

To keep the system closed, we introduce conservation equations for each component,
\begin{equation}
  \label{eq:conseq-swell}
R_{\alpha, w} = \frac{A_{\alpha, w} - A_{\alpha, w}^0}{\Delta t}  + Q_\alpha - \sum_{j\in C(w)} q_{\alpha, j} = 0,
\end{equation}
that are solved in a fully implicit and coupled way with the black-oil equations
\eqref{eq:materialbalance}.  Here, $C(w)$ is the set of connections of
the well $w$, $q_{\alpha,j}$ is the flow rate of phase $\alpha$
through connection $j$ under surface condition, which can be calculated from
$q_{\alpha,j}^r$ defined in \eqref{eq:wellinflow}; the relation is similar to the one
between component and mass fluxes \eqref{eq:acc}. The storage term $A_{\alpha, w}$
describes the amount of component $\alpha$ in the wellbore (small volume), and it is
introduced for better stability of the well solution.

In addition, we need equations that model how the wells are controlled.  The well control
for a well controlled by a prescribed bhp target reads
\begin{equation}\label{eq:well-control-bhp}
R_{c, w} = p_{bhp, w} - p_{bhp, w}^{target} = 0,
\end{equation}
whereas for a rate-controlled well we have
\begin{equation}\label{eq:well-control-rate}
R_{c, w} = Q_\alpha - Q_\alpha^{target} = 0.
\end{equation}
where $Q_\alpha^{target}$ is the desired surface-volume rate of the
controlled component $\alpha$, typically oil rate for a production well.

Under some conditions, fluids from the reservoir formation can enter
the wellbore through some connections and get reinjected into the
formation through other connections of the same well. This phenomenon
is called cross-flow, and is modeled by the standard well model. Since
the model assumes that the fluid composition is uniform throughout the
wellbore (given by $F_g$ and $F_w$), the same composition will be
injected through all injecting connections. When this is not accurate
enough, a multi-segment well model can be used.

\subsubsection{Multi-segment well model}

\begin{figure}
\subfigure[]
{\includegraphics[width=0.6\textwidth]{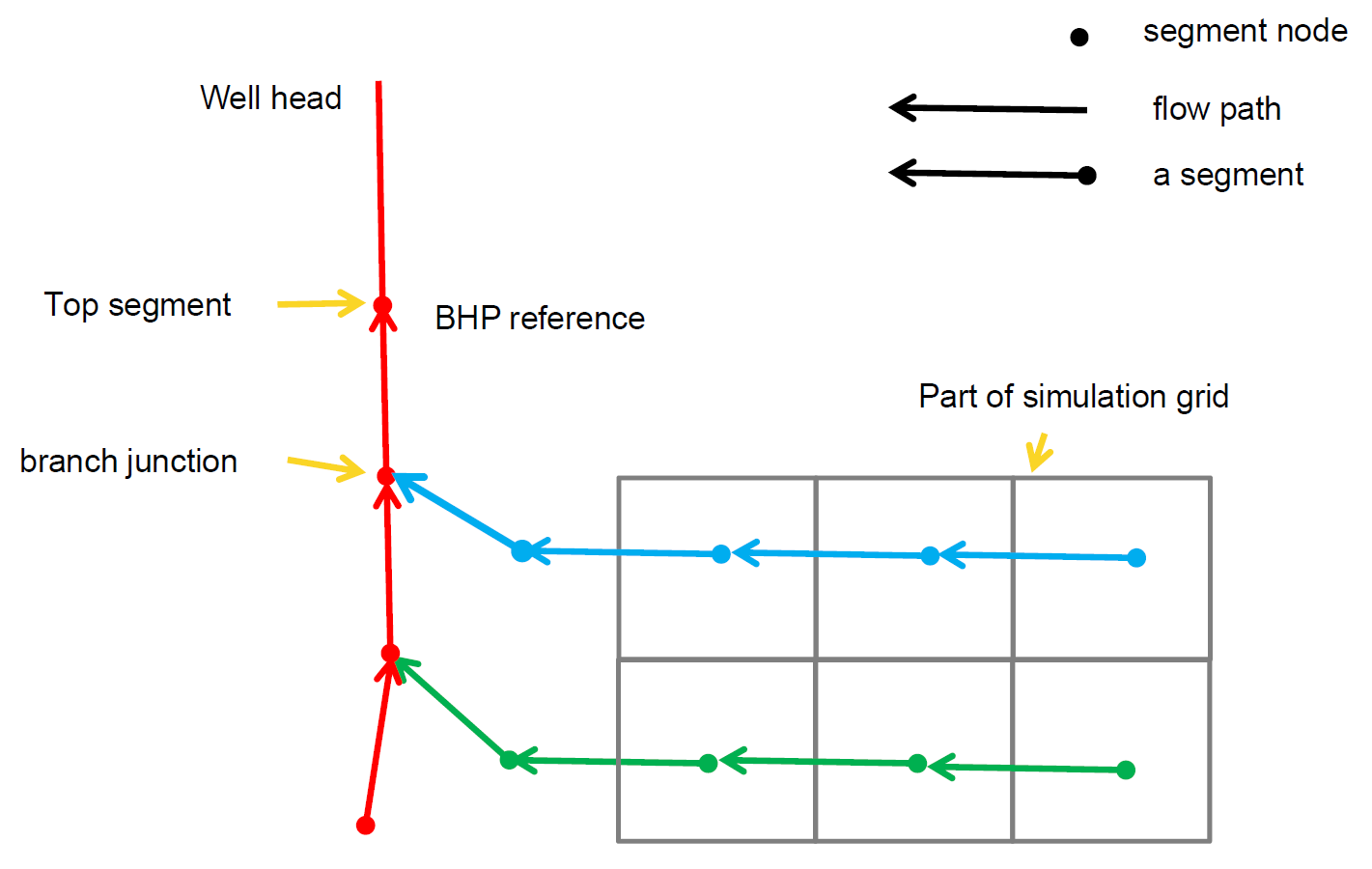}\label{fig:msw}}
\subfigure[]
{\includegraphics[width=0.3\textwidth]{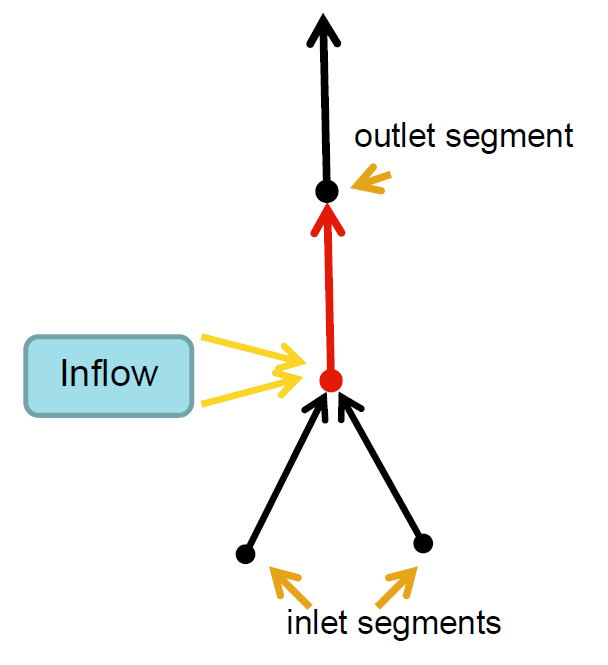} \label{fig:segment}}
\caption{A illustration of the multi-segment well and the segment structure.}
\label{fig:msw_illustration}
\end{figure}

So-called multi-segment well models \cite{holmes1998application} were introduced 
to simulate more advanced wells, such as multilateral wells, horizontal wells, and inflow
control devices. With a multi-segment model, the wellbore is divided
into an arbitrary number of segments (Figure~\ref{fig:msw_illustration}). Including more
segments can give a more accurate simulation, at the cost of higher computational
cost. Each segment consists of two parts: a segment node and a flow path to the
neighboring segment in the direction towards the well head, defined as the outlet
segment (Figure~\ref{fig:segment}). Most segments also have inlet segment neighbors in the
direction away from the well head. In the well shown in Figure~\ref{fig:msw}, there are
three branches, including one main branch (in red color) and two lateral
branches (in green and blue colors). For multilateral wells, a segment node
must be placed at the branch junction. Segments at branch junction points have additional
inlet segments and segments at the end of branches do not have inlet segments. The top
segment of each well does not have an outlet segment. The pressure at this segment
is the bottom-hole pressure of the well, whereas the component rates equal the component rates
for the well as a whole.

\begin{figure}
\centering
{\includegraphics[width=0.5\textwidth]{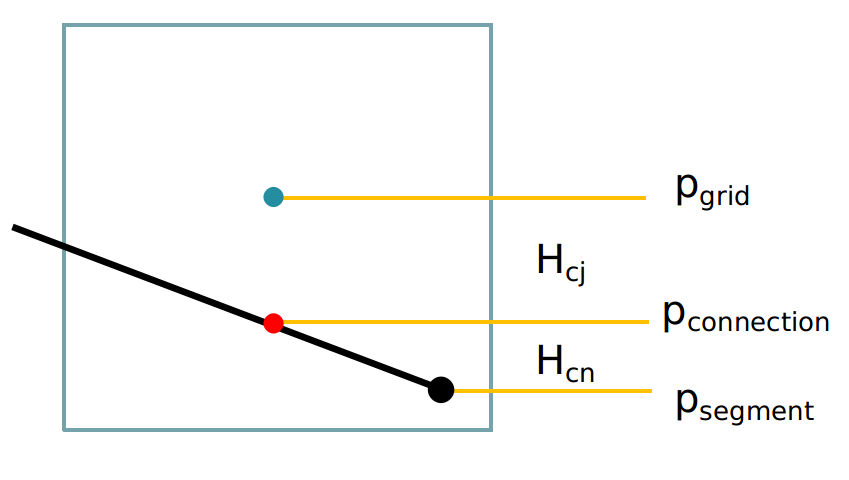}}
\caption{Hydrostatic pressure drops used in the multi-segment well
model.}
\label{fig:msw_inflow}
\end{figure}

Each segment has the node pressure $p$ as a primary variable in addition to the primary
variables $Q_t$, $F_w$, and $F_g$ introduced for the standard well model (see
\eqref{eq:weighted_total_rate}). Likewise, the component equations \eqref{eq:conseq-swell}
are extended to include additional terms $Q_{\alpha, i}$ that
represent the flow rate from the inlet segments $I(n)$ of segment $n$ as
well as the inflow from the reservoir through the segment's
connections $C(n)$.
\begin{equation}
R_{\alpha,n} = \frac{A_{\alpha, n} - A_{\alpha, n}^0}{\Delta t} - \sum_{i\in I(n)} Q_{\alpha, i} - \sum_{j\in C(n)}
q_{\alpha, j} + Q_{\alpha, n} = 0.
\end{equation}
Each segment can thus receive inflow from more than one connection or no
connection through the segment node, whereas each connection can only contribute to one
segment. In the standard well model, the connection is located at the centroid of the
grid block. In the multi-segment well model, the connection and
segment node can be located any place in the grid block. The inflow calculation of
the connection in the multi-segment
well model is as follows:
\begin{equation}
q_{\alpha, j}^r = T_{w,j} M_{\alpha, j} (p_{j} + H_{cj} - p_n - H_{nc}).
\label{eq:msw_inflow}
\end{equation}
Here, $p_{j}$ is the pressure at the cell center of the grid block that contains
connection $j$; $H_{cj}$ is the hydrostatic pressure difference between the cell
center and the connection; $p_n$ is the pressure of segment $n$; and $H_{nc}$ is the
hydrostatic pressure difference between the connection and the segment
(Figure~\ref{fig:msw_inflow}).

The last equation for each segment describes the pressure relationships between the
segment $n$ and its outlet segment $m$:
\begin{equation}
  R_{p, n} = p_{n} - p_{m} - H_h - H_f - H_a = 0.
\end{equation}
Here, the $H$ terms represent the hydrostatic, frictional, and acceleration pressure drops
between the segments.

The top segment does not have a outlet segment, and hence the pressure equation is
replaced by a well control equation, which is the same as for the standard well model,
\eqref{eq:well-control-bhp} or \eqref{eq:well-control-rate}.

\begin{figure}
\centering
{\includegraphics[width=0.5\textwidth]{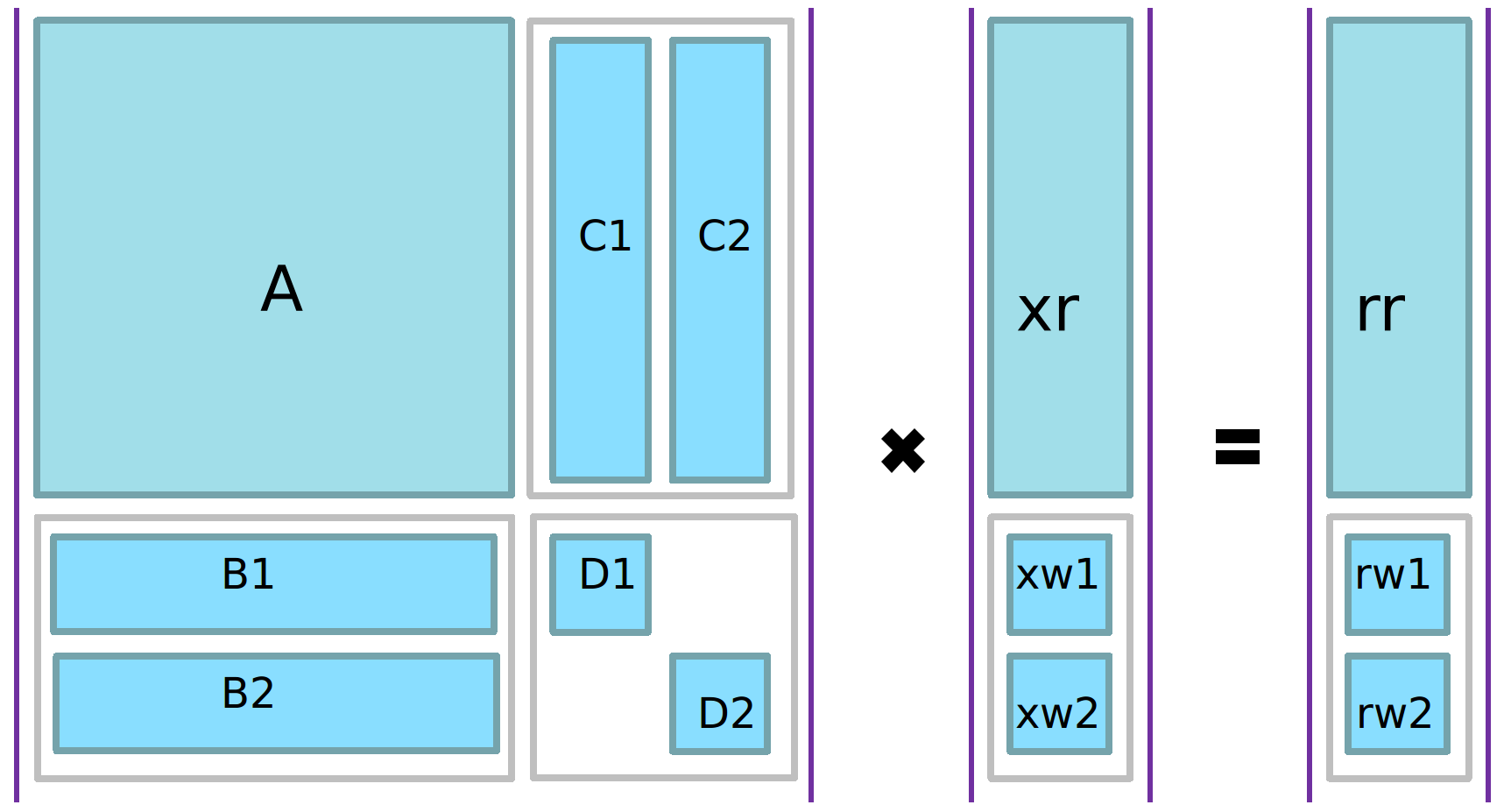}}
\caption{Partition of the matrix system for a case with two
  wells. Here, $A$ contains the reservoir flow equations, $D_1$ and
  $D_2$ contain the linearized well equations for each of the two
  wells, whereas the $B$ and $C$ matrices are coupling entries between
  the reservoir and well models.}
\label{fig:matrix_system}
\end{figure}

As with the standard well model, the well
equations are solved with the reservoir mass conservation equations in a coupled, fully implicit
way. The well equations are added to the global system as shown in Figure~\ref{fig:matrix_system}
using a Schur complement approach:

\begin{align}
  \label{eq:wellequation}
(A - \sum_{w}C_wD_w^{-1}B_w)\, x_r &= R_r - \sum_{w}C_wD_w^{-1}\, R_w, \\
x_w &= D_w^{-1}(R_w - B_w \, x_r).
\end{align}

\subsection{Solution strategy}

Altogether, the reservoir and well equations described thus far define a large system of
(fully implicit) nonlinear equations, which we can write on compact residual form as
$R(y)=0$, where $y$ is the vector of primary variables.  This system is solved using a
Newton--Raphson type method. Let $y_n = (p_o, s_w, x) $ be the primary variables after $n$
Newton iterations. Given an initial state $y_0$, the solution of $R(y)=0$
can be found by iteratively solving
\begin{equation}
\label{eq:linsystem}
J(y_n) (y_{n+1} - y_n) = -R(y_n) 
\end{equation}
until $R(y_n)$ is less then some prescribed tolerance. Creating the
Jacobian matrix $J(y_n)$ and solving this linearized problem is the
core computational task of the simulator, and its performance depends largely on how this
part of the code is programmed. The linearization and the solution
procedure for \eqref{eq:linsystem} is explained in detail in the following
subsections, but first some words about convergence of the nonlinear
solver.

Because the solution at the end of the previous time step is used
as initial guess, the Newton--Raphson is expected to converge
rapidly if a sufficiently small time step is used. For large time steps and/or large
changes in the solution, typically caused by changing well controls, the
Newton--Raphson method may converge poorly or not at all. One simple
way of improving the convergence is to restrict the change allowed
per iteration and prevent the solution from jumping far across the boundary
between saturated and undersaturated states in a single iteration.
This so-called Appleyard chop technique
\citep{appleyard1983nestedfactorization}
is implemented in \opmflow, and is often sufficient for the method to
converge. Still, time steps may need to be cut to make the solver converge
for some difficult cases.

Note also that since the third primary variable $x$ may change
interpretation from one iteration to the next, we are not strictly
using the Newton--Raphson method as such, but a close
relative. Changing primary variable may disturb the convergence, and we
therefore add a small threshold to the variable-switching logic in
\eqref{eq:primaryVariableMeaning} to prevent oscillation between
states.
Lack of smoothness of the residuals can also come from other sources.
For example, the functions for relative permeability and capillary
pressure can be almost discontinuous in some cases, they can also have
hysteretic behaviour.

The third point that affects the convergence of the nonlinear solver
is the accuracy of the linear solvers. Solving the linear system
to a loose accuracy may lead to poor convergence of the
nonlinear solver. The tuning of the time-stepping procedure, and the
tolerance used in the linear and nonlinear solvers are highly related,
and optimal choices are case- and problem-dependent.

From the user side, there are several command-line
options that can be used to control the solver. In particular,
\code{--tolerance-mb}
controls the mass balance error as a reservoir-average saturation
error, and
\code{--tolerance-cnv}
controls the maximal local residual error.
For more information about runtime options and other aspects of
running \opmflow, see Section 2.2 of the \opmflow manual
\citep{flowmanual}. Running \opmflow with the option \code{--help} will
print a list of available parameters.

\subsubsection{Solving the system of linear equations}

Depending on fluid properties and other reservoir-specific parameters, the 
system of equations \eqref{eq:materialbalance} 
could exhibit elliptic or degenerate parabolic behavior or
could even be hyperbolic for specific setups. Therefore, solving the linearized
system \eqref{eq:linsystem} stemming from these equations is challenging, 
since it is likely non-symmetric and ill-conditioned. 
Instead of solving equation \eqref{eq:linsystem} directly, the solution is approximated 
by an iterative linear solver. Two variants are available, a 
stabilized Bi-conjugate Gradient method (BiCG-stab) 
and a restarted Generalized Minimal Residual (GMRes)  
solver. By default, BiCG-stab is used.
In addition, several options are available for preconditioning of the linear system, 
the default choice being the incomplete lower triangular--upper triangular factorization with level of fill in $0$ (ILU0).
More sophisticated choices are based on an algebraic multigrid (AMG) method for
preconditioning, such as a recently added 
Constraint Pressure Residual (CPR) preconditioner (originally introduced in
\cite{wallis:83}) using the approach of \citep{Scheichl2003Decoupling}.

Implementation of the AMG and the CPR preconditioner 
is further discussed in Section~\ref{sec:parallelization} and the solver itself
in \cite{blattamg}. The \opmflow manual \citep{flowmanual} gives an overview of
parameters for choosing solvers and preconditioners, as well as other related parameters.
The solvers and preconditioners are implemented in the \dune module \duneistl 
and described in \cite{ISTL,ISTLParallel}.
In the current version, \opmflow assembles the Jacobian matrix into the 
block compressed row storage (\code{BCRSMatrix}) 
data structures provided by \duneistl. Each matrix entry is of the type 
\code{FieldMatrix}, which is a dense matrix with fixed row and column length provided by \dunecommon.
Blocked matrix storage enables better performance on modern
CPUs compared to the standard approach that stores each matrix entry separately. 
Likewise, the unknown $\Delta y$ and the right hand side $-R(y_n)$ are stored in a
\code{BlockVector} from the \duneistl package.

The interface between \opmflow and linear solver packages has been designed in a
flexible way so that besides \duneistl other linear algebra packages could be used. 
For example, an experimental implementation using the \petsc library \cite{petsc-user-ref} 
has been successfully tested. This is, however, not yet part of the main branch of the source code. 

\subsection{Automatic differentiation in \opmflow}

The coefficients of the linearized systems required by the nonlinear Newton solver (e.g.,
$J(y_n)$ in equation \eqref{eq:linsystem}) have traditionally been computed by evaluating
closed-form expressions obtained by differentiating the discretized flow equations analytically.
Differentiating these flow equations manually and programming the resulting formulas 
is both time-consuming and error-prone. The problem is particularly 
pronounced when extending an existing simulator with new functional relationships. 
If a quantity is set to depend on a different or new (primary) variable, 
the changes in derivatives will propagate through the chain rule to 
the Jacobians of all quantities that depend on the modified quantity. 
Thus, even very simple extensions in functional dependencies may 
cause large modifications throughout the simulator code.

Automatic differentiation (AD) is a way to mitigate this problem; see for example
\citep{griewank:2008} or \citep{Neidinger2010} for an introduction.  There are many
examples of AD tools used for scientific computing, such as
\href{http://trilinos.org/packages/sacado}{Sacado}, ADOL-C \citep{adol-c} or ADETL; the
latter used for the \href{http://supri-b.stanford.edu/research-areas/ad-gprs}{AD-GPRS}
reservoir simulator.

The AD mechanism implemented in \opm works by treating
each quantity in the program not as just a single value, but as a
value paired with values for the \enquote{relevant} partial derivatives. 
Whenever the value of a new quantity is computed, a combination of the chain rule
and standard differentiation rules for elementary unary and binary operations 
is used to compute the correct value of all the corresponding partial derivatives. 
This is called the forward approach to AD, in contrast to
backward AD, which stores the adjoints of each quantity.

The drawback is that an AD library
must be learned, or developed from scratch if no existing library
matches the needs of the application at hand. In \opm's case, we
decided to develop a library from scratch, since no existing
packages provided the desired flexibility and performance, and we also
could not find a suitable freely licensed package. Further, experience from 
developing a similar library in MRST \citep{MRST-AD} made the
implementation effort of \opm's AD library relatively small.

The cost of AD for the user can be lower performance compared
to a well-tuned hand-written implementation. We have made numerous
improvements to our automatic differentiation implementation
to reduce this overhead, but it is clear that by investing
sufficient effort into a hand-written code, one can possibly get
better performance. Writing such optimal Jacobian code for arbitrary
extensions is not a trivial task, and due to the increased chance for
errors is not guaranteed to give any speed-up at all.

\subsubsection{General approach}

The \opm implementation of AD introduces a class that mimics the behaviour of the built-in
floating point types of C++ (\code{double} and \code{float}) as closely as possible. The
AD object contains a value and a fixed number of derivatives and defines all basic
arithmetic operators as well as common mathematical functions. Also, to allow easy
comparisons with code that uses standard floating-point objects, all functions that work
on AD-objects can also be used with objects of built-in floating-point types. In \opm an
implementation class called \code{Evaluation} is provided for this purpose:

\begin{lstlisting}[style=cppstyle]
  // ValueT is typically double or float
  template <class ValueT, int numDerivs>
  class Evaluation;
\end{lstlisting}

This class overloads all arithmetic operators so that objects of type \code{Evaluation}
can be used in the same way as plain \code{double} or \code{float} types. An example code
snippet for the implementation of the multiplication operator looks as follows: 

\begin{c++}
// Multiplication operator 
Evaluation& operator*=(const Evaluation& other)
{
    // The values are multiplied, whereas the derivatives
    // follow the product rule, i.e., (u*v)' = (v'u + u'v).
    const ValueType u = this->value();
    const ValueType v = other.value();
    // Value, usually position 0
    data_[valuepos_] *= v ;
    //  Derivatives usually start at position 1
    for (int i = dstart_; i < dend_; ++i) {
        data_[i] = data_[i] * v + other.data_[i] * u;
    }
    return *this;
}
\end{c++}
For improved performance the implementation of the \code{Evaluation} class is
specialized for values with up to $12$ partial derivatives, whereas a default implementation based on \code{for} loops is used for higher numbers of
derivatives. The \opmmaterial module
provides a python script implementing a simple code generation algorithm for 
the specialized \code{Evaluation classes} for any number, if
necessary. Typical three-phase black-oil equations use three
partial derivatives, corresponding to the number of
primary variables per cell.

\subsubsection{Application-specific localization}

\begin{figure}
\center
\includegraphics[width=.45\linewidth]{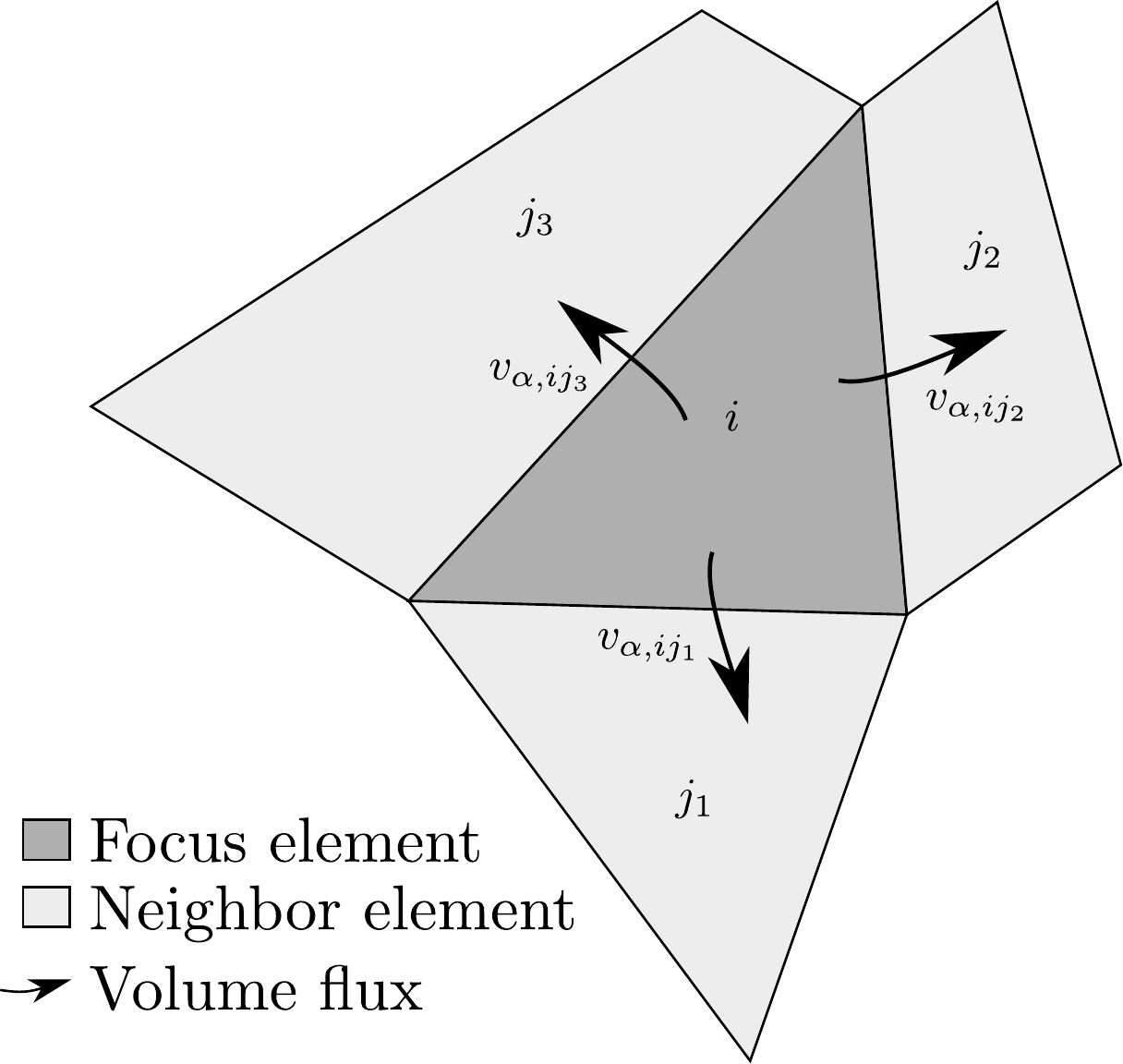}
\caption{Localized linearization of element $i$ and its neighbors $j_{\left<\cdot\right>}$.
  Equations~\eqref{eq:materialbalance} are evaluated locally with
  derivatives of the primary variables of element~$i$.}
\label{fig:localized_linearization}
\end{figure}

For performance reasons, the implementation of automatic
differentiation in \opm only supports carrying a fixed number of
derivatives, which needs to be specified at compile-time. Yet, the
number of unknowns for the discretized conservation
equations~\eqref{eq:materialbalance} is only known at run-time because
it depends on the number of grid cells.

This problem is solved using a
localized linearization scheme as illustrated by
Figure~\ref{fig:localized_linearization}: The residuals of the mass
conservation equations are computed for a single element, which is
called the \textit{focus} element. In addition to the values of the
residuals of the focus element, the partial derivatives of the
residuals of all elements in the local neighbourhood are calculated with
regard to the fixed number of primary variables of the focus element.
For more details on this approach, confer \cite{lauser2018local}.

The two-point flux-approximation (TPFA) scheme used ensures that the residuals of any
cell only depend on the variables of the cell itself and its immediate
neighbours, which lets the localized linearization scheme be applied.

Once the local residual vector and its dense Jacobian matrix have been
computed, they are transferred to a variable-sized vector for the
residual and a variable-sized sparse matrix for its Jacobian matrix
representing the linearized system of equations for the whole
spatial domain. This vector and the matrix are subsequently passed to
the linear solver.

\subsection{Parallelization}
\label{sec:parallelization}

Parallelization of \opmflow follows a hybrid approach with a combination of 
multi-threading and MPI. The assembly of the Jacobian matrix
is multi-threaded through OpenMP, and controllable through the OpenMP 
environment variables. The assembly part is less dependent on 
memory bandwidth, and this enables us to exploit more CPUs on a multicore 
machine than the scaling possible for the linear solve part. In particular, 
it allows for improved performance on CPUs with hyper-threading. Writing binary
results to file is also threaded to insulate the reservoir 
simulation from slow disk systems, as further commented on in Section~\ref{sec:fileio}. 
The \code{CpGrid} class implementing stratigraphic grids is based on the \dune
grid interface and \opmflow therefore naturally inherits
the MPI parallel interface of the \dune modules \dunegrid and \duneistl (see
\cite{dunepaperI:08,dunepaperII:08,ISTLParallel}). This keeps the implementation effort
as small as possible. Our main goal is to reuse
sequential algorithms whenever possible and only introduce parallelism specific code
where absolutely necessary. In particular, the discretization, and algorithms for 
adaptive time stepping and the Newton--Raphson nonlinear solver are the
same for sequential and parallel runs. For the latter, only the
convergence criterion will be switched to a parallel one as it needs
some global communication.

Standardized parallel file formats for oil reservoir data are
currently lacking and the ECLIPSE file format is inherently sequential in
nature (see Section~\ref{sec:fileio}). 
At the beginning of a simulation, a process therefore needs
to read the complete input file and set up a grid for the entire domain. 
This is a potential bottleneck for scalability to large number of cores, which
has not yet been reached. 
Using the entire grid read initially, the transmissibilities representing cell--cell connections are
computed. A high value means that a change of pressure or
saturation in a cell will influence the properties of the neighboring
cell strongly. Therefore, the current domain decomposition in \opmflow 
is based on keeping cells connected through high transmissibilities on the same process 
when partitioning. 
This is done to ease the coarsening phase of the algebraic 
multigrid solver and improve convergence of other linear solvers. 
The implementation currently does not allow one well to be distributed across multiple 
subdomains.

\begin{figure}
\centering
\includegraphics[width=.5\linewidth]{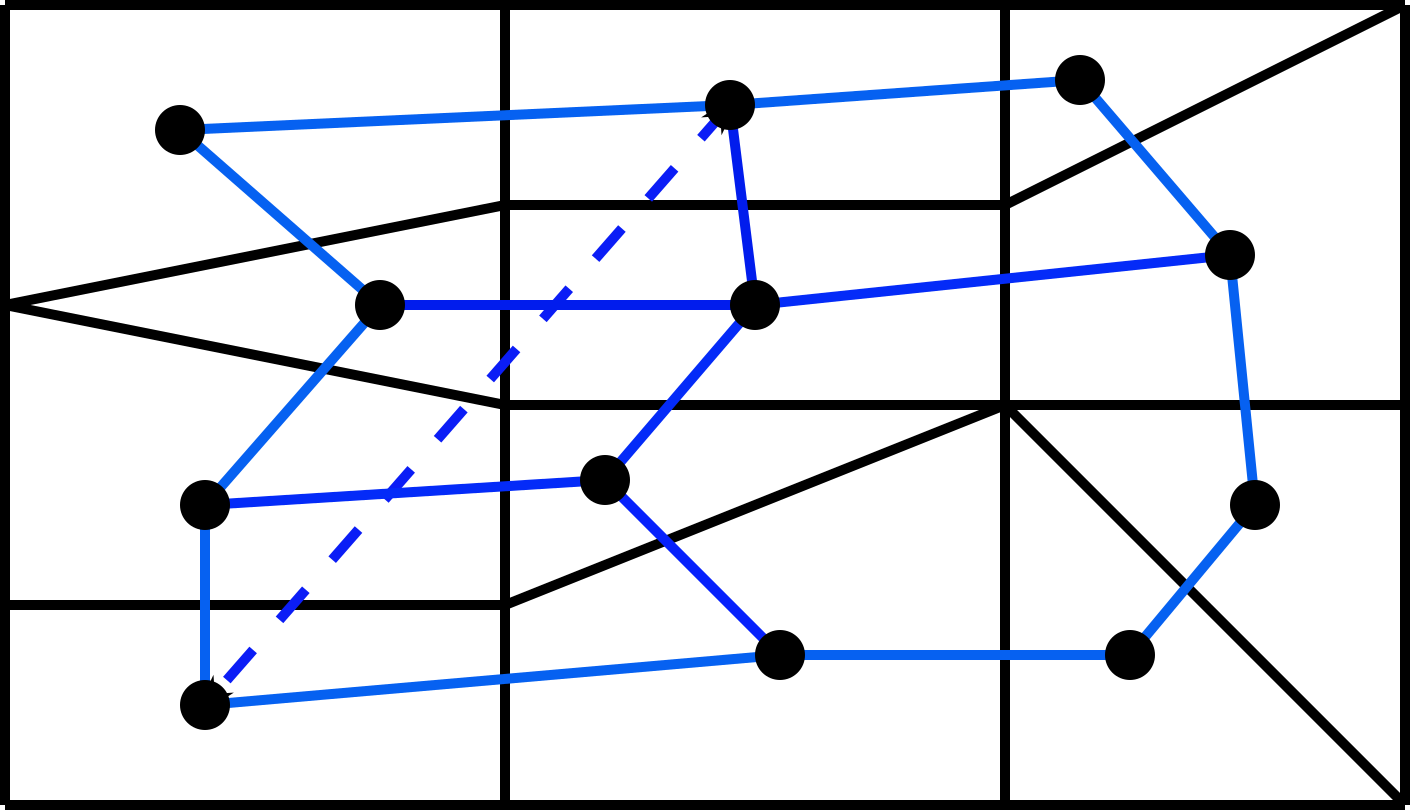}
\caption{\label{fig:org3ce8b2a}
Graph of a grid with one well, where 
  the black dots are the vertices associated with cells in the grid 
  and the blue lines the edges associated with the
cell faces. The dashed blue line is an edge introduced by a well
connecting two non-neighboring cells. 
  }
\end{figure}

At the start of the simulation, the global grid is partitioned 
using the graph partitioning software \href{http://www.cs.sandia.gov/Zoltan}{Zoltan} 
\citep{zoltan,ZoltanHypergraphIPDPS06}
on the grid graph defined with cells as vertices and edges graph
either representing a face between two cells or a connection
between two cells via a well. Figure~\ref{fig:org3ce8b2a} shows a conceptual illustration.
An edge associated with a cell face is assigned the value of the transmissibility, scaled by
\(10^{18}\), whereas any edge associated with a
well connection gets the maximum weight available
to ensure that all cells connected by a well end up in the same subdomain.
This will give partitions in which domain boundaries are at faces having low transmissibility.
In a post-processing step after the initial load balancing, we ensure 
that all cells connected by a well are really contained in one
subdomain since our edge weights are merely a strong suggestion to the
load balancer.
As a result, the calculation of 
the well model can be done sequentially on the process that stores
the well without additional communication. 

\begin{figure}
\centering
  \subfigure[Interior and ghost layer]{
    \label{fig:org937a0ca}
    \includegraphics[width=.45\textwidth]{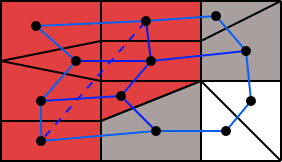}}
  \subfigure[Communication]{
    \label{fig:org5209e15}
    \includegraphics[width=.45\textwidth]{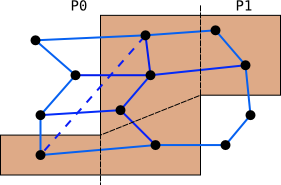}}
  \caption{(a) Interior (red) and ghost (grey) cells stored on a process, and
  (b) grid cells involved in communication.
  }
\end{figure}

The flux calculation, which is a central part of any 
cell-centered finite-volume scheme, needs to access information from 
neighboring cells. 
\opmflow therefore not only stores the grid cells assigned to one process, but
also a halo layer of so-called ghost cells at the process boundary.
During the simulation, the values attached to these ghost cell are updated
using point-to-point communication. Figure~\ref{fig:org937a0ca} illustrates the
local grid stored on one process. 

This decomposition is used to set up a distributed view of the 
\code{CpGrid} class. It implements the parallel \dune grid interface 
(\cite{dunepaperI:08,dunepaperII:08}) allowing values attached to interior cells
to be easily communicated to other
processes, where these cells belong to the ghost layer. 
For the small example in Figure~\ref{fig:org5209e15}, only the brown cells will be
involved in the communication.

As a result, the sequential discretization algorithms can be reused almost unchanged,
since all necessary data (neighboring and wells) is available on one process and only one
communication step during each linearization step of the nonlinear solver is needed to
update the ghost layer.

The iterative solvers in \duneistl are parallelization
agnostic and can be turned into parallel ones by
simply using parallel versions of the preconditioner, scalar product,
and linear operator. All preconditioners, except
AMG, apply the sequential operator on the degrees of freedom
attached to the interior cells only and update the rest using one
step of interior-ghost communication. When used in AMG as smoothers, these
parallel preconditioners are called hybrid smoothers
(see \cite{hybrid}). This approach works very well in the AMG solver provided by
\duneistl \cite{blattamg}, as it uses data agglomeration onto fewer
processes whenever the number of unknowns per process dops below a
prescribed threshold. Thus on coarser levels fewer processes are
involved in the computation until on the coarsest level only one process computes
and a direct coarse solver is used. This remedies the missing
smoothing across the processor partition boundaries on
the finer levels and keeps the ratio between time used for computation between
communication steps and time needed for communication high. 
In addition, a block ILU0 (one block for each process) or a CPR preconditioner is provided. 
%
Both the existing parallel scalar product
and linear operator from \duneistl are reused without modifications.

\subsection{I/O in \opmflow}
\label{sec:fileio}

To be industrially relevant, \opmflow must read and write
files in standard formats well known to the industry. \opmflow
implements support for the input format used by the
ECLIPSE simulator from Schlumberger and also writes output files compatible
with the output from said simulator, because this is the dominant file
formats supported by most pre- and post-processing tools.

\subsubsection{Reading input files}
The input for a simulation case consists of many separate data
items. The computational grid, petrophysical properties, fluid
properties, and well scheduling and behavior must all be specified,
along with other options controlling for example output.

It is common for commercial simulators to do this in the form of an
input file set, often called a {\em deck} (named for the decks of
punched cards used in the early age of simulation). An input deck
consists of many keywords that set properties or trigger functionality
in the simulator, and may be distributed across several files (using
an INCLUDE keyword). The deck is organized into sections dealing with
separate parts, and is similar to a
programming language that manipulates the reservoir simulator like a
state machine. The sections are:
 \\[-1ex]
\begin{small}
\begin{tabular}{l l}
\\
\hline \\[-2ex]
\texttt{RUNSPEC} & Dimensions, phases present. \\
\texttt{GRID} & Grid topology and geometry, faults, petrophysical properties. \\
\texttt{EDIT} & Modifying inputs specified in the GRID section, multipliers. \\
\texttt{PROPS} & Fluid properties. \\
\texttt{REGIONS} & Define regions for properties and output. \\
\texttt{SOLUTION} & Initial solution. \\
\texttt{SUMMARY} & Choose variables for time-series output. \\
\texttt{SCHEDULE} & Parameters and controls for wells and well groups, time stepping. \\
\hline \\
\end{tabular}
\end{small}

The input format is well documented in the \opmflow
manual \citep{flowmanual}. The sections are described in Chapter 3.4
of the manual, and each supported keyword is described in the following
chapters. The list of supported keywords is extensive, and the
parsing performed in \opmflow has been tested on a selection of 
different input decks from industry. For obvious reasons, \opmflow does not yet
support all keywords and features available
in the ECLIPSE simulator. Still, \opmflow is fast reaching a point where
the functionality available covers the most common use cases, and is
wide enough that it can be feasible to modify unsupported cases to
adapt their reservoir description to what is currently supported.  We therefore
believe the current parsing framework is sufficient to make \opmflow an
attractive alternative in the industry.

Parsing in \opmflow is a two-stage process. In the first stage, basic
string parsing is performed, comments are stripped, included files are read in,
default values and multipliers are resolved, and numerical items are converted to
the correct type. This produces an instance of
class \texttt{Opm::Deck}, which is essentially a container of
\texttt{Opm::DeckKeyword} objects. The second stage in the parsing process
consists of creating an instance of type \texttt{Opm::EclipseState}, which is 
an object at a much higher semantic level, in which the
interaction between keywords has been taken into account, e.g., all the
\texttt{EDIT} section modifiers have been applied to produce consistent
geological input fields and the various well-related keywords from the
\texttt{SCHEDULE} section have been assembled into well objects.

The keywords recognized by \opmflow are specified with a simple JSON schema.
A new keyword can be supported by adding a JSON specification
and registering it with the build system, then \opmflow will be able to
parse the new keyword and create \texttt{Opm::DeckKeyword} instances to represent it.
Taking a new keyword into account in the \texttt{Opm::EclipseState} class
requires changes to the code depending on the semantics of
the new keyword. To complete the addition of a new feature, the
numerical code can then be extended to use the new information
available from the \texttt{Opm::EclipseState} object.

\subsubsection{Writing output files}

\opmflow can output a range of different file types. Summary files contain
time-series for well or reservoir data. Restart files, along with INIT
and EGRID files, can be used for restarting a simulation from a report
step or checkpoint, as well as for visualization of the reservoir and
the solution variables. RFT files are used by engineers to get more
information about reservoir behavior.

The files are ECLIPSE-compatible in the sense that they are correctly
rendered in independent third-party viewers like Petrel, S3Graf, and
RMS, in addition to \opm ResInsight. These binary output
files are also compatible to the point that Schlumberger's
simulator ECLIPSE can restart a simulation from these files.

The input keywords \texttt{RPTRST}, \texttt{RPTSOL}, and \texttt{RPTSCHED} are
used to configure which properties should be written to the restart files, and
also for which report steps output should be enabled, most of this output
configuration is supported in \opmflow.

The \texttt{SUMMARY} section is used to configure
which result vectors should be added to the summary output. \opmflow will use the
configuration inferred from this section when creating summary
output. The list of available summary vectors is extensive and \opmflow does not
support them all, but a wide range of field, group, well, region, and block
vectors are supported.
In addition, \opmflow has its own unique PRT file, which has a
report formatted just like ECLIPSE for compatibility.
However, everything else in the PRT file is unique for
\opmflow. The same goes for the terminal output, which is not the
same as ECLIPSE. The ambition is to make output more useful for the
end user.

VTK-format output files for result fields can also be written for visualization with,
e.g., Paraview. This feature can be activated with a command-line option, see the manual
\citep{flowmanual} for this and other output-related options.

\subsubsection{Parallel file I/O}

The ECLIPSE file format is inherently sequential, but since this file format is a de-facto standard in the
industry, the parallel simulation program also needs to handle the format. 
Whereas initial reading of the deck is less performance critical, since it is
only done once, the reoccurring writing of simulation data during the simulation
poses a potential performance bottleneck. To absorb the penalties imposed by
slow file systems, such as network file systems, the actual output is queued in a 
separate output thread that no longer needs to by synchronized with the simulation.
To cater to the sequential file format at hand, a dedicated I/O rank collects all
data from all processes and then the sequential output routines can be used as
before. Again, this is a potential show stopper for scaling to large number of
cores; this has not been an issue yet, but may need to be addressed in future
releases of the software. 
To avoid performances penalties, the communication between the I/O rank
and all other ranks is done asynchronously, such that each simulation core only
places an \code{MPI\_Isend} and continues with simulation, whereas the I/O rank 
only places \code{MPI\_Irecv} in combination with \code{MPI\_Iprobe} and
\code{MPI\_Waitall}. 

\subsection{Extended models}

Whereas the family of black-oil models presented thus far is sufficient for many
real-world cases, it lacks essential features for describing chemically enhanced oil
recovery. This section briefly outlines two extended models implemented in \opmflow, a solvent
model and a polymer model, that add one or a few extra components and model their effect
on the fluid flow.

\subsubsection{Solvent}\label{sec:solvent}

Miscible gas injection is a common EOR technique. The injected gas mixes with the hydrocarbons and enhances the sweep efficiency by mobilizing the residual oil in the reservoir. An important application for the solvent module in \opm is to explore the combined goal of \COto storage with EOR. The \COto is miscible with oil at pressures above a certain threshold, and injecting \COto is thus a commonly used approach to enhanced recovery. 

The solvent module extends the black-oil model with a fourth component representing the injected gas. In addition to the extra conservation equation required for the solvent, the solvent modifies the relative permeability, the capillary pressure, the residual saturations, the viscosity, and the density of the hydrocarbons. The modeling strategy we use is to compute the properties for the fully miscible and the immiscible case and interpolate between these limits using a miscibility function $M$ depending on pressure and solvent saturation. 

Large contrasts in viscosity and density between the injected gas and the oil lead to fingering effects. Resolving individual fingers requires very fine grid resolution and is not feasible for field-scale simulations. The Todd--Longstaff model \citep{todd1972development} introduces an empirical parameter $\omega$ to formulate effective parameters that take into account the effect of fingers on the fine-scale dispersion between the miscible components to overcome this obstacle. The solvent module in \opm follows the implementation suggested in \cite{todd1972development, chase1984numerical}, the most important details are included here. 

The conservation equation for the solvent component, $s$, is 
\begin{align}\label{eq:materialbalance-cont-solvent}
\frac{\partial}{\partial t} \left(\phi_{\rm ref} A_s \right)
 + \nabla \cdot \ub_s + q_s & = 0,
\end{align}
where the accumulation term and the flux are given by
\begin{gather}  
  A_s = m_{\phi} b_{s} s_{s}, \qquad       
  \ub_s = - b_{s} \lambda_s \Kb (\nabla p_g - \rho_s \gb). 
\end{gather}
In addition, the solvent saturation is added to Equation \eqref{eq:fluids-fill-porevolume}, giving: 
\begin{equation}
s_w + s_o + s_g + s_s = 1.
\label{eq:fluids-fill-porevolume-solvent} 
\end{equation}
The relative permeability of the solvent and the gas component is given as a fraction of the total (gas + solvent) relative permeability:
\begin{gather}
 k_{r,s} = \frac{s_s}{s_g + s_s} k_{r,gt}, \qquad
 k_{r,g} = \frac{s_g}{s_g + s_s} k_{r,gt}.
\end{gather}
One can change the gas and solvent fractions multiplying $k_{r,gt}$ with simple functions of the same fractions to allow a more flexible formulation. The total relative permeability is given as an interpolation between the fully miscible relative permeability and the immiscible one: 
\begin{gather}
 k_{r,gt} = M \frac{s_g + s_s}{s_g + s_s + s_o} k_{r,n}(s_n) +(1 - M) k_{r,g}(s_g + s_s), \\
 k_{r,o} = M \frac{s_o}{s_g + s_s + s_o} k_{r,n}(s_n) +(1 - M) k_{r,o}(s_w, s_g),
\end{gather}
where $M$ is a user-defined parameter that varies from zero to one and may depend on both the solvent fraction and the oil pressure. Water-blocking effects are incorporated in the model by further scaling the normalized saturations by effective mobile saturations that increase with increasing water saturation;  see \cite{chase1984numerical} for details. 

The Todd--Longstaff model formulates the effective viscosities for solvent, gas, and oil as a product of the pure viscosities and the fully mixed viscosity
\begin{gather}
 \mu_{{\rm eff},o} = \mu_{o}^{1 - \omega} \mu_{m,os}^{\omega}, \quad
 \mu_{{\rm eff},g} = \mu_{g}^{1 - \omega} \mu_{m,gs}^{\omega}, \quad
 \mu_{{\rm eff},s} = \mu_{s}^{1 - \omega} \mu_{m,gos}^{\omega},
\end{gather}
where the fully mixed viscosities are defined using standard $\tfrac{1}{4}$-power fluid mixing rule \citep{todd1972development}. Pressure effects on the mixing are modeled by replacing the single parameter $\omega$ with an effective mixing parameter 
\begin{equation}
 \omega_{\rm eff} = \omega \omega_p.
\end{equation}
where $\omega_p$ is a function of pressure \citep{chase1984numerical, jakupsstovu2001upscaling}. 
For effective density calculations confer \cite{chase1984numerical}.  The solvent model is used to simulate the SPE 5 benchmark in Section~\ref{sec:spe5},

\subsubsection{Polymer}

Polymer flooding is another widely used EOR technique. The major
mechanism of polymer flooding is to improve the sweep efficiency by increasing
the viscosity of the drive fluid to improve the mobility ratio. The polymer
model in \opm assumes that the polymer component is only transported
within the water phase and only affects the properties of this phase.
The model also includes the effects of dead pore
space, polymer adsorption in the rock, and permeability reduction effect.
Shear-thinning/thickening non-Newtonian fluid rheology is modelled by a
logarithm look-up calculation method based on the water velocity or shear rate.

The polymer model is developed by adding the following polymer transport equation to the
black-oil model
\begin{equation} \label{eq:polymer-transport}
R_{p,i} = \frac{V_i}{\Delta t}(A_{p,i} - A_{p,i}^0) + \sum_{j\in C(i)} F_{p, ij} + q_{w,i} c_{w,i} = 0,
\end{equation}
where
\begin{gather}
A_{p,i} = \phi_{{\rm ref},i} m_\phi b_r b_w s_w (1-s_{dpv}) c  + \rho_r C^a (1-\phi_{{\rm ref},i}) \\
F_{p} = b_w v_p c \\
(b_w{v}_p)_{ij} = \Bigl(\frac{m_T k_{r,w}}{\mu_{{\rm eff},p} R_k}\Bigr)_{U(w, ij)}  T_{ij} \Delta H_{w, ij}, \\
(b_w{v}_w)_{ij} = \Bigl(\frac{m_T k_{r,w}}{\mu_{{\rm eff},w} R_k}\Bigr)_{U(w, ij)}  T_{ij} \Delta H_{w, ij}.
\end{gather}
Here, $c$ is polymer concentration, $b_r$ is rock expansion factor, $s_{dpv}$ is the dead
pore space, $C^a$ is the polymer adsorption concentration, ${v}_p$ is polymer flux rate,
and $\mu_{{\rm eff},p}$ is the effective polymer viscosity, which is a function of the
polymer concentration $c$ and pressure of the water phase. The viscosity of water is
changed as a result of dissolution of polymer, and we calculate the water flux rate based
on effective water viscosity $\mu_{{\rm eff},w}$, which is also a function of polymer
concentration.  The polymer concentration $c_{w,i}$ used to calculate the polymer rate
through well connections in grid block $i$ is set as the injection concentration for
injectors, whereas $c_{w,i} = c_i \mu_{{\rm eff},w}/\mu_{{\rm eff},p}$ for producers,
where $c_i$ is the polymer concentration in grid block $i$. The effective viscosities are
calculated using the Todd--Longstaff mixing model \citep{todd1972development} to handle
sub-grid fingering effects, similar to the solvent model of Section~\ref{sec:solvent}.

\section{Numerical examples and results}\label{sec:numericalresults}

This section demonstrates \opmflow through a series of numerical examples.
Sections~\ref{sec:spe1} to \ref{sec:spe5} report small synthetic cases exercising
particular parts of the simulator's fluid behavior. Section~\ref{sec:spe9} reports a more
realistic case to highlight the treatment of wells. The Norne case in
Section~\ref{sec:norne} is a real field case and demonstrates that \opmflow can be applied
to complex industrial field cases. The polymer example in Section~\ref{sec:polymer} shows
some of the EOR capabilities of \opmflow.

We conclude the section with some notes on performance. Complete input decks used for the
examples are available from \href{http://github.com/OPM}{github.com/OPM}, and can be run
with default options.

\subsection{SPE 1 benchmark}
\label{sec:spe1}

\begin{figure}
\centering
{\includegraphics[width=0.9\textwidth]{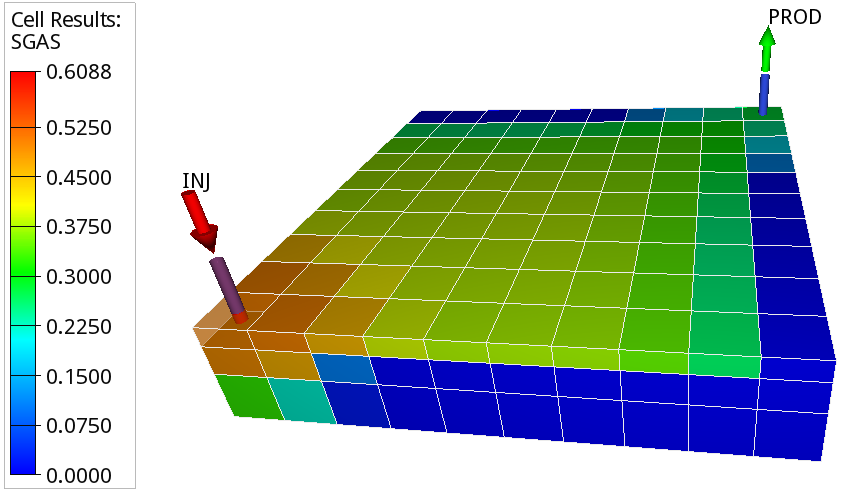}}
\caption{Gas saturation of the SPE 1 benchmark case after a period of gas injection. The z-axis
  has been exaggerated 20 times.}
\label{fig:spe1_sgas}
\end{figure}

\begin{figure}
  \subfigure [Bottom hole pressure of the injector]{%
    \includegraphics[width=0.5\textwidth]{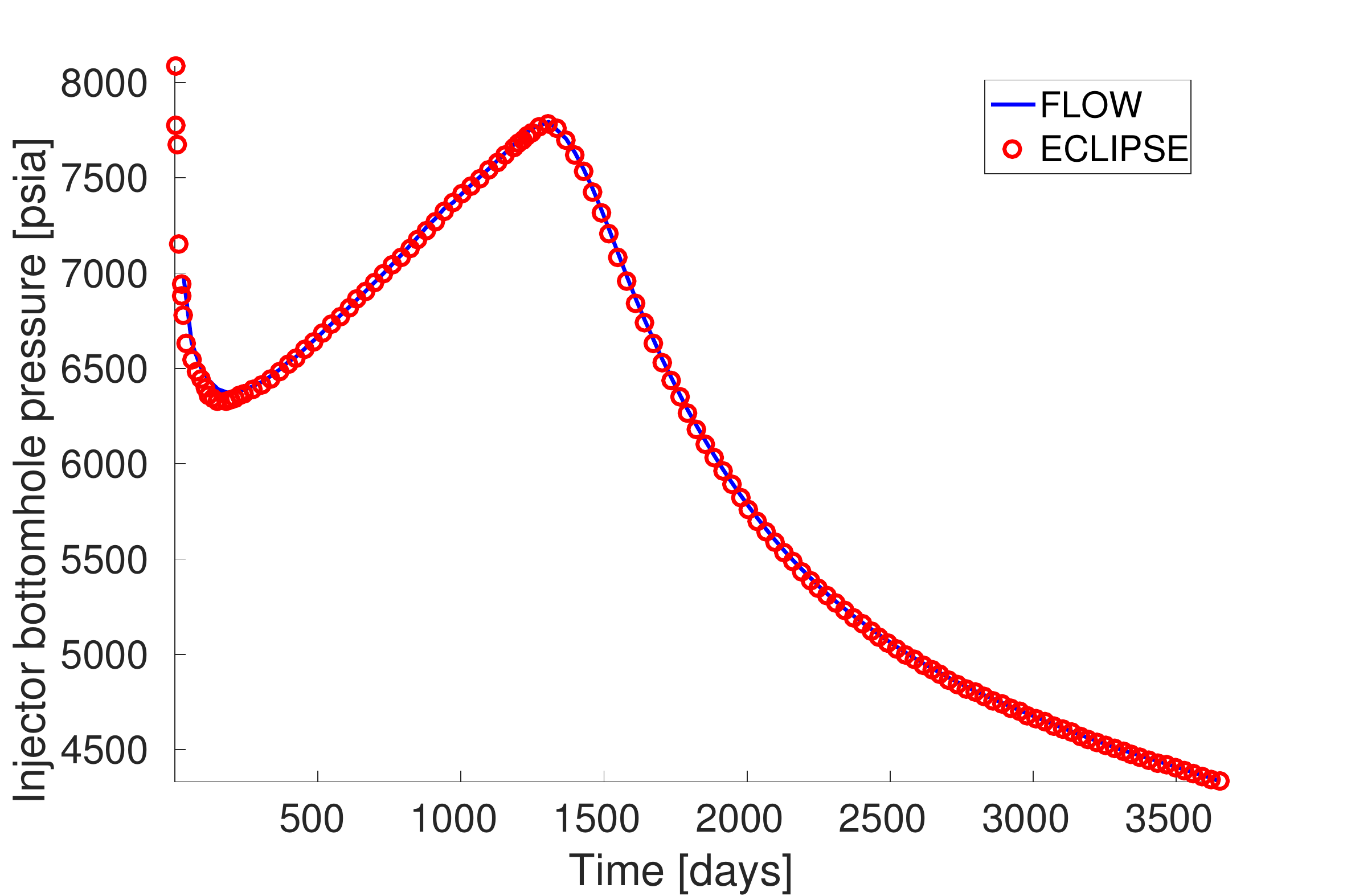}} \
  \subfigure[Bottom hole pressure of the producer]{%
    \includegraphics[width=0.5\textwidth]{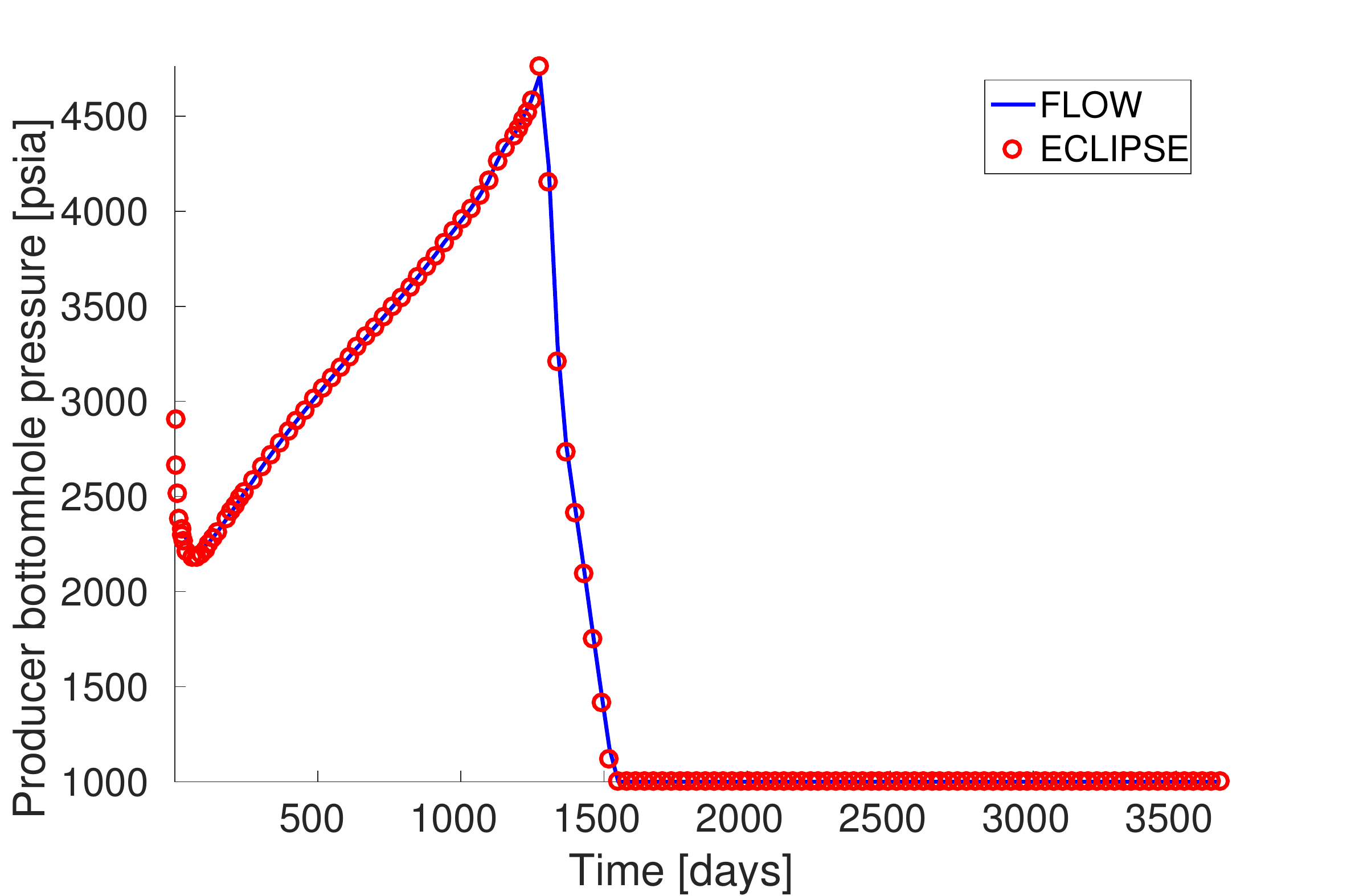}}\\
  \subfigure[Oil production rate of the producer]{%
    \includegraphics[width=0.5\textwidth]{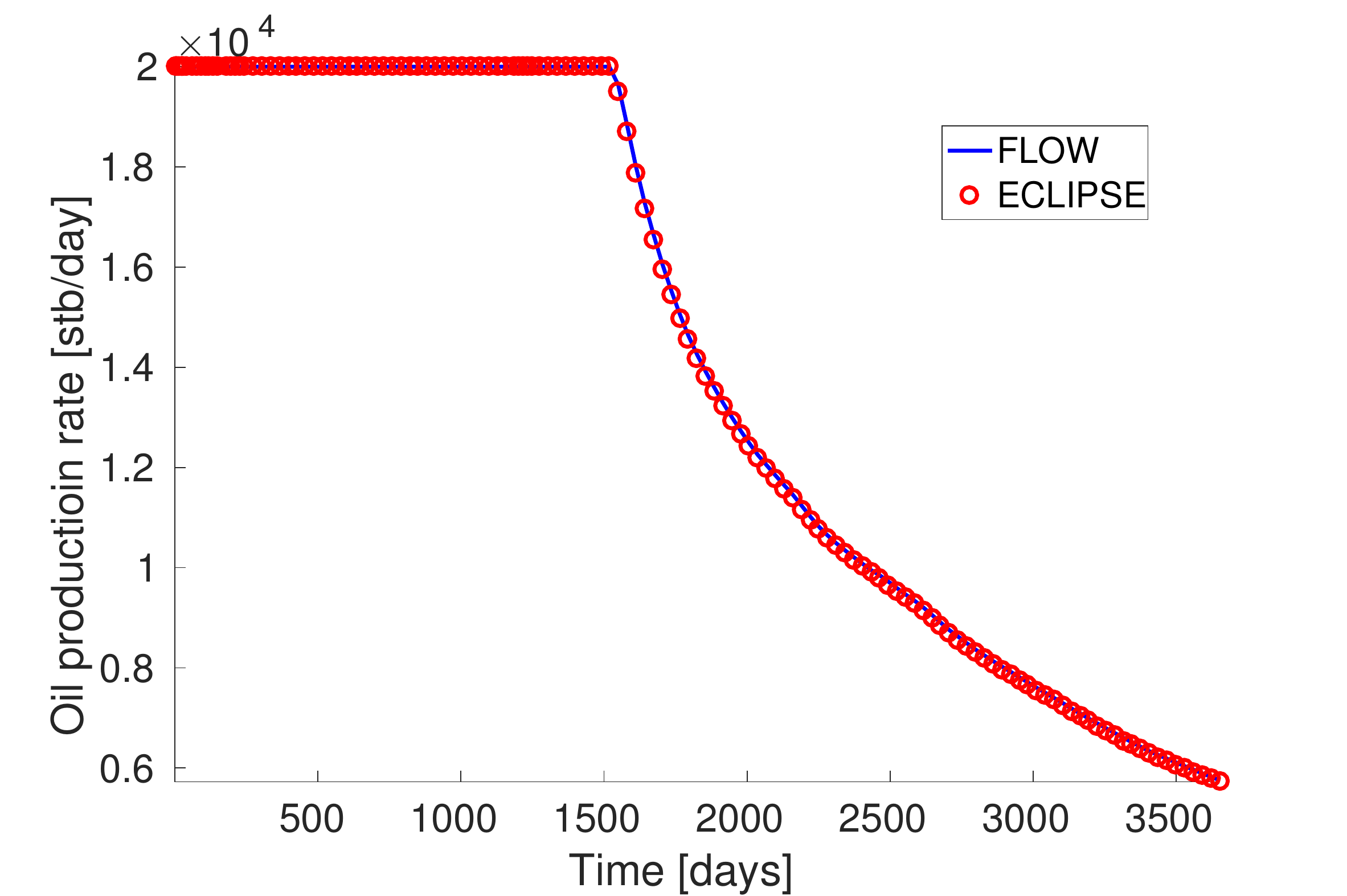}} \
  \subfigure[Gas-oil ratio of the producer]{%
    \includegraphics[width=0.5\textwidth]{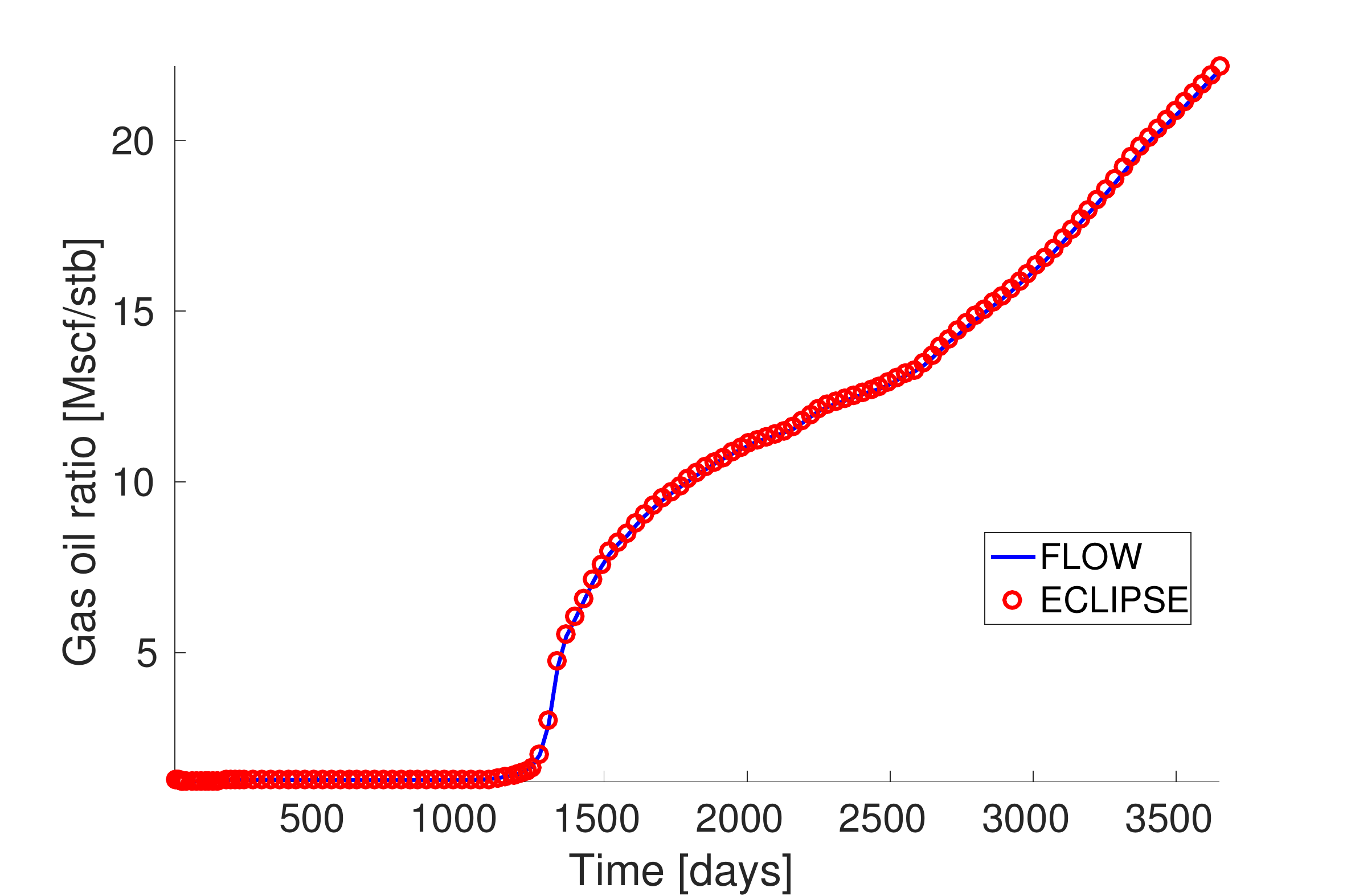}}
  \caption{Well responses computed for the SPE 1 benchmark, Case 2.}
  \label{fig:spe1_results}
\end{figure}

The Society of Petroleum Engineers (SPE) has run a series of
comparative solutions projects with the aim of comparing and
benchmarking different simulators or algorithms. The first such
project (nicknamed SPE 1), was introduced for benchmarking
three-dimensional black oil simulation
\citep{odeh1981comparison}. Two examples were suggested, and here we
use the second case to demonstrate \opmflow. The
grid consists of 10$\times$10$\times$3 cells covering a computational
domain of 10\,000 ft$\times$10\,000 ft$\times$100 ft. The thickness of each layer is
20 ft, 30 ft, and 50 ft respectively. Two wells are located in two opposite corners.
The first well injects gas from
the top layer at a rate of 100 MMscf/day with a BHP limit of 9014
psia. The other well is set to produce oil from the bottom layer with a
production target of 20\,000 stb/day and BHP limit of 1\,000 psia. The reservoir is
initially undersaturated. The participants were asked to report the oil production
over time and the gas-oil ratio (GOR) over time. Figure~\ref{fig:spe1_sgas} shows a view of the
case after a few years of injection, whereas Figure~\ref{fig:spe1_results} verifies
that there is excellent agreement in the well responses
computed by \opmflow and the commercial simulator ECLIPSE 100.




\subsection{SPE 3 benchmark}
\label{sec:spe3}

The third comparative solution project \citep{kenyon1987third} was introduced to study
gas cycling of retrograde condensate reservoirs using compositional modeling.  Based on
the data initially provided as the second case in
\cite{kenyon1987third}, we designed a benchmark test case for the \opmflow
simulator.

Black-oil properties are generated from compositional data using
\href{http://www.pvtsimnova}{PVTsim Nova}. The dimension of the grid is
9$\times$9$\times$4 with 293.3 ft $\times$ 293.3 ft cell sizes in the horizontal direction
and the thickness of each layer being 30 ft, 30 ft, 50 ft and 50 ft, respectively. The
horizontal permeability for each layer is 130 md, 40 md, 20 md, and 150 md
respectively, whereas the vertical permeabilities are 13 md, 4 md, 2 md, and 15 md.

\begin{figure}
  \includegraphics[width=0.5\textwidth]{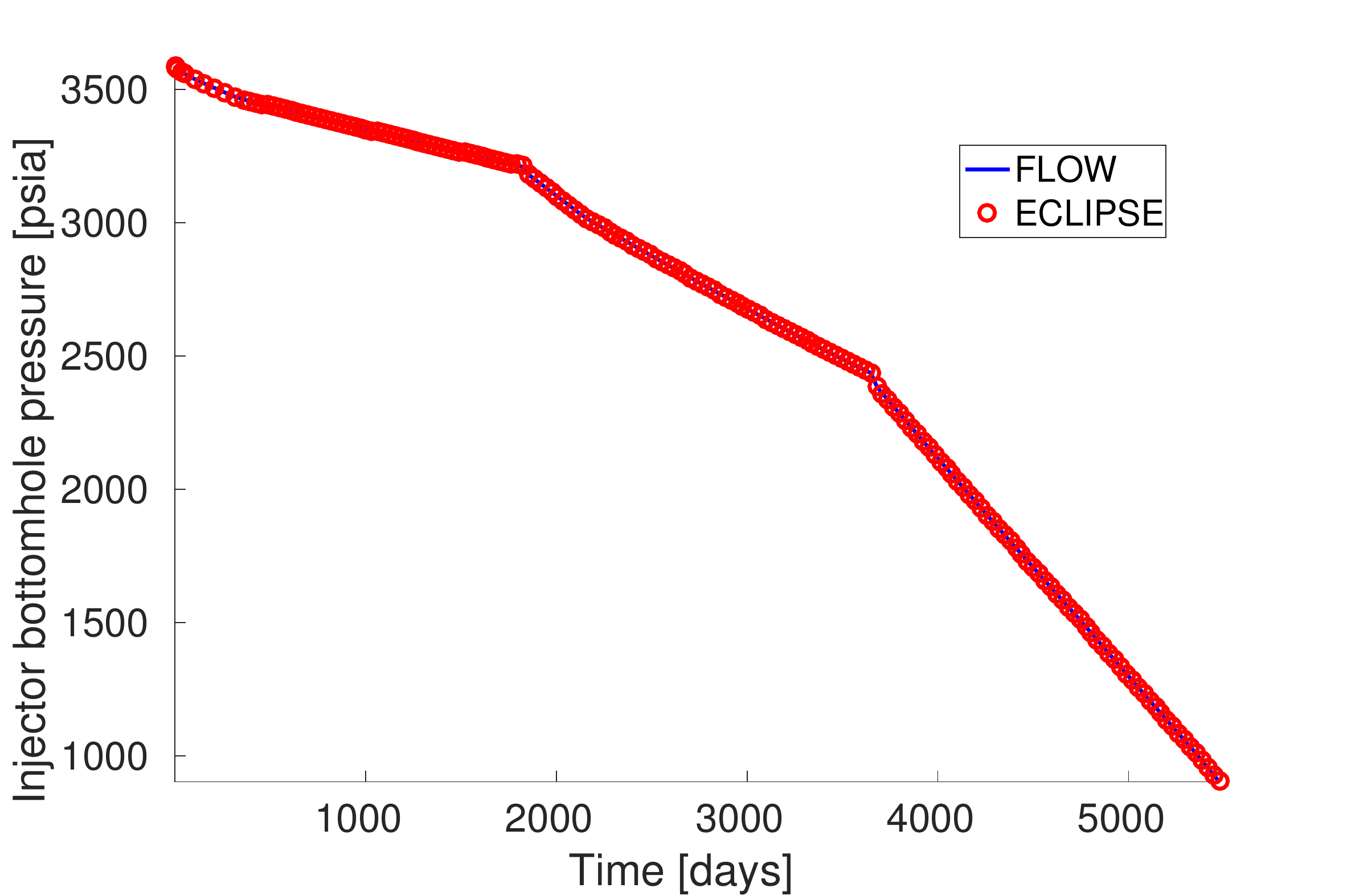}\hfill
  \includegraphics[width=0.5\textwidth]{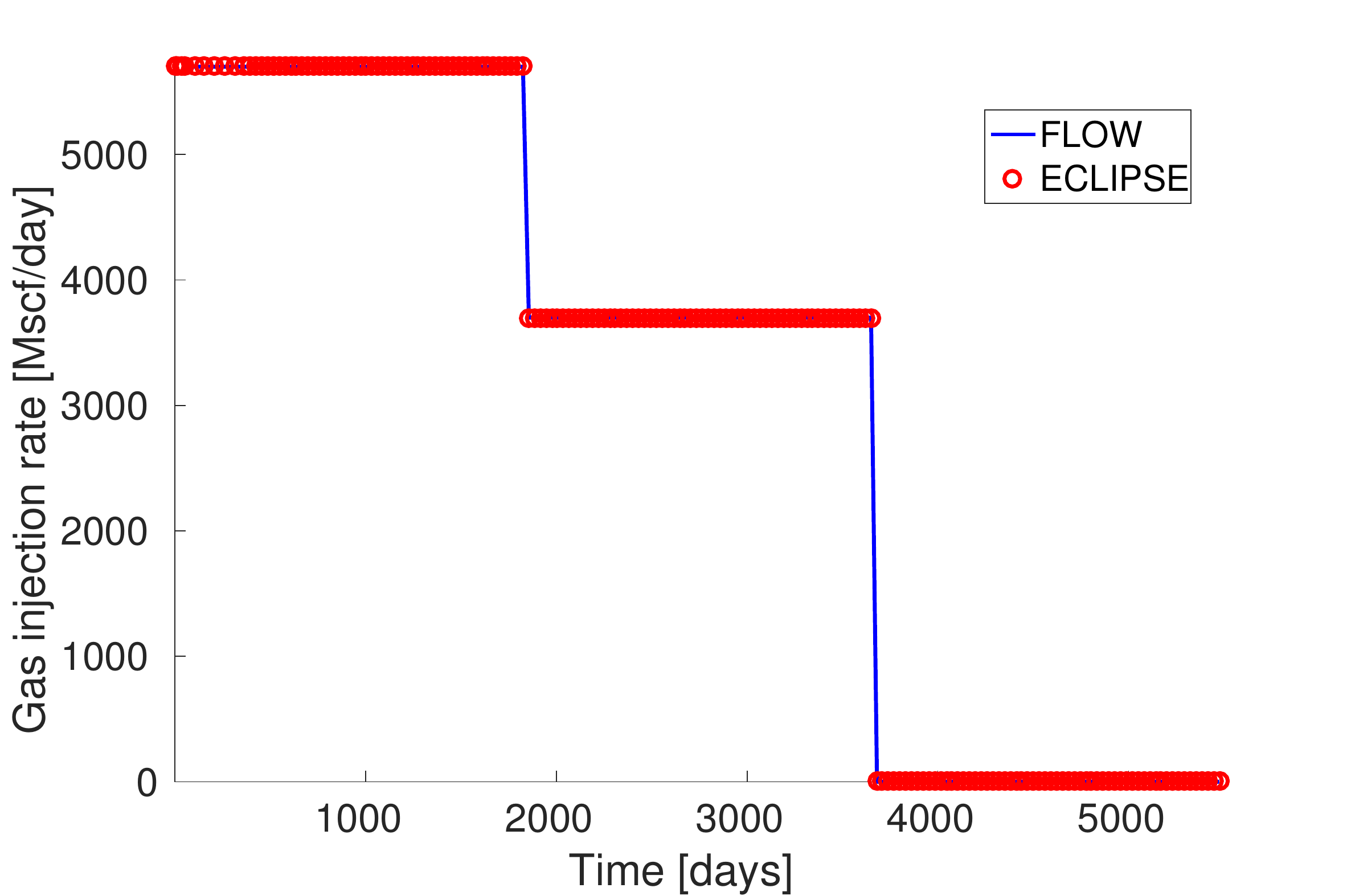}\\
  \includegraphics[width=0.5\textwidth]{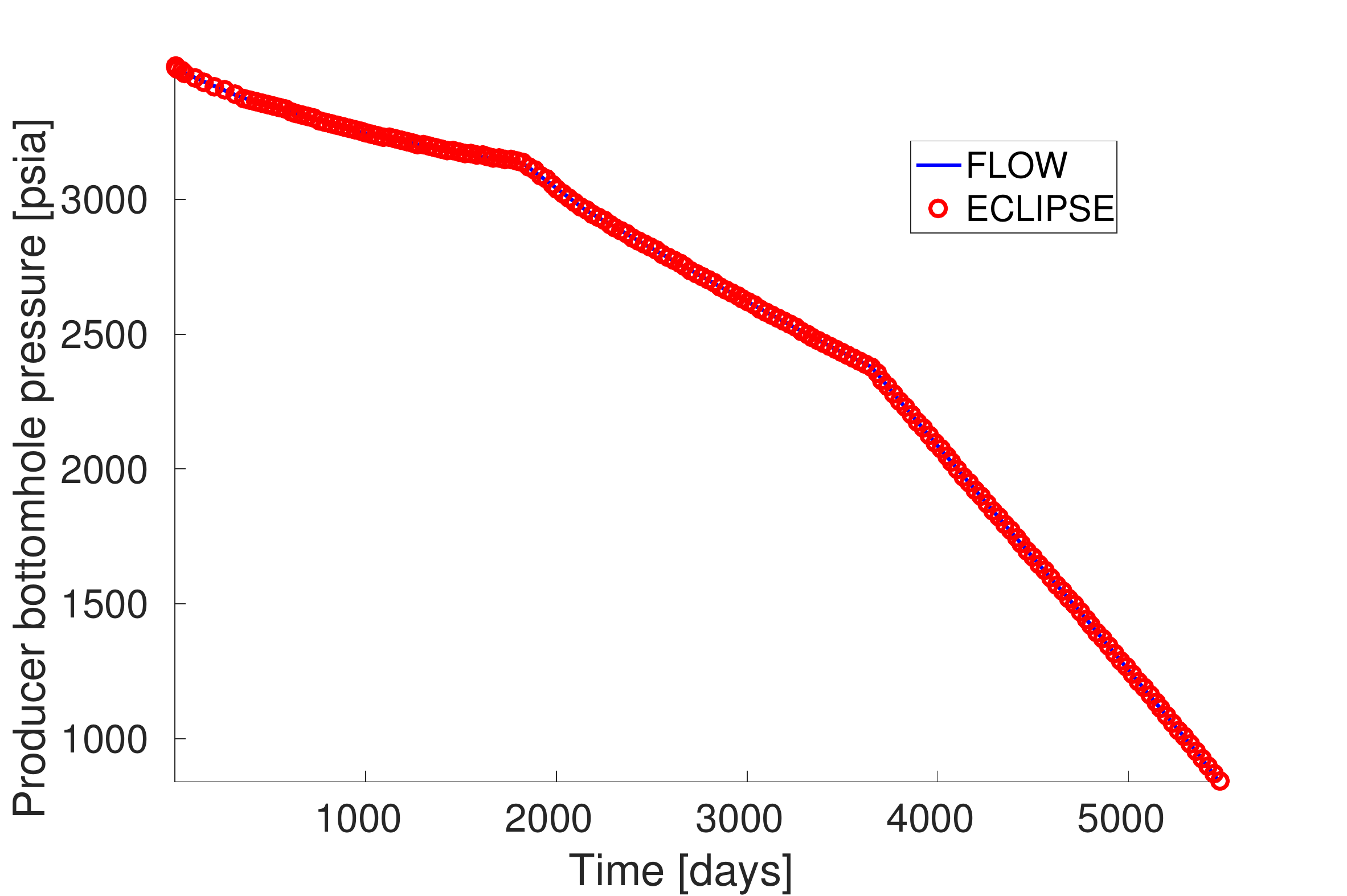}\hfill
  \includegraphics[width=0.5\textwidth]{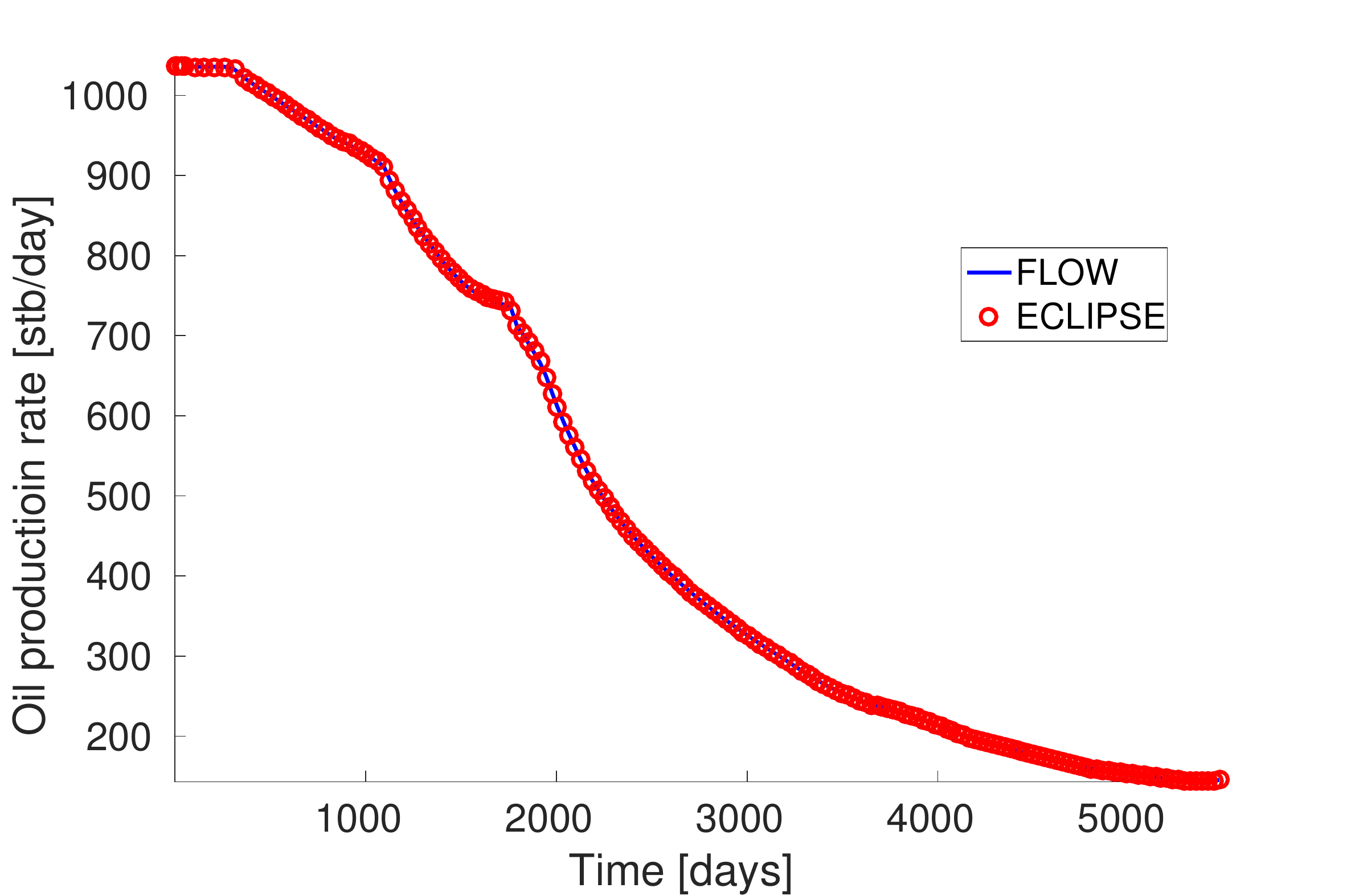}
\caption{Bottom-hole pressures and well rates of the SPE 3 benchmark, Case 2.}
\label{fig:spe3_results}
\end{figure}

An injector is located in the corner (cell column (1,1)) and perforates the top two
layers, whereas a producer perforates the bottom two layers of cell column (7,7). The
simulation time is 14 years in total.  The gas injection rate is 5700 Mscf/day for the
first four years, and 3700 Mscf/day for the next five years. No gas is injected the last
five years.  The BHP limit is a constant 4000 psia.  The production well operates with a
constant gas production limit 5200 Mscf/day and a constant BHP limit of 500 psia.
Figure~\ref{fig:spe3_results} confirms good agreement for bottom-hole pressures and rates
obtained by \opmflow and ECLIPSE.

\subsection{SPE 5 benchmark}
\label{sec:spe5}


The fifth comparative solution project \citep{killough16000fifth} is designed to compare
four-component miscible simulators with fully compositional simulators
for three different miscible gas injection scenarios. Herein,
we compare the solvent simulator in \opm is compared with the solvent simulator in
ECLIPSE for the three cases introduced in the SPE 5 paper. The same
grid with dimension 7$\times$7$\times$3
and the same fluid properties are used for all the cases. A
water-alternating-gas (WAG) injector is located in Cell (1,1,1). By
changing the alternation sequence and rates of water and gas, different
miscibility conditions are obtained in the three different scenarios. A
producer located in Cell (7,7,7) is controlled by the same oil
production rate in all three cases; see
\cite{killough16000fifth} for details, or confer the input deck given in
\opmdata.

\begin{figure}
  \includegraphics[width=.49\textwidth]{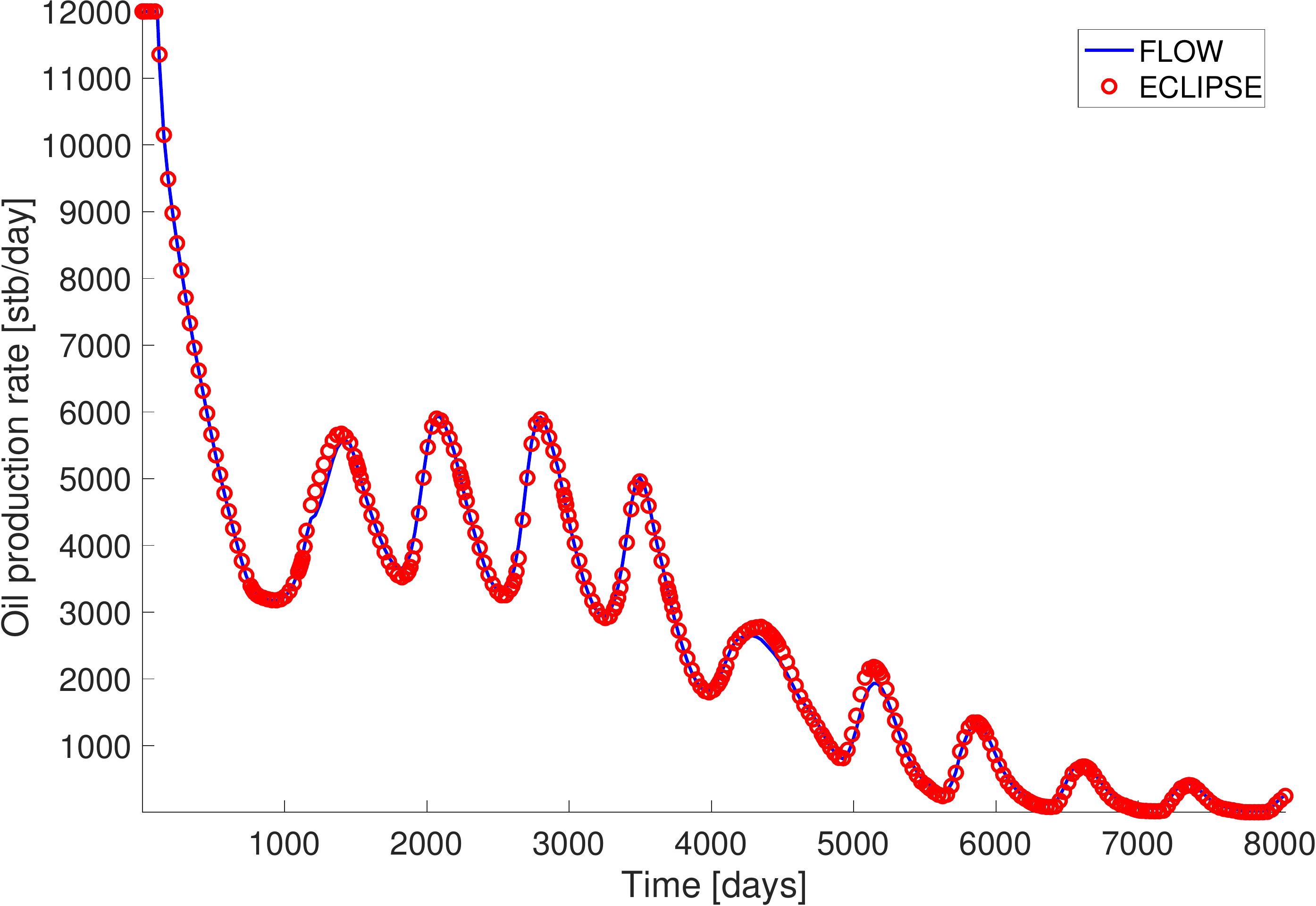}\hfill
  \includegraphics[width=.49\textwidth]{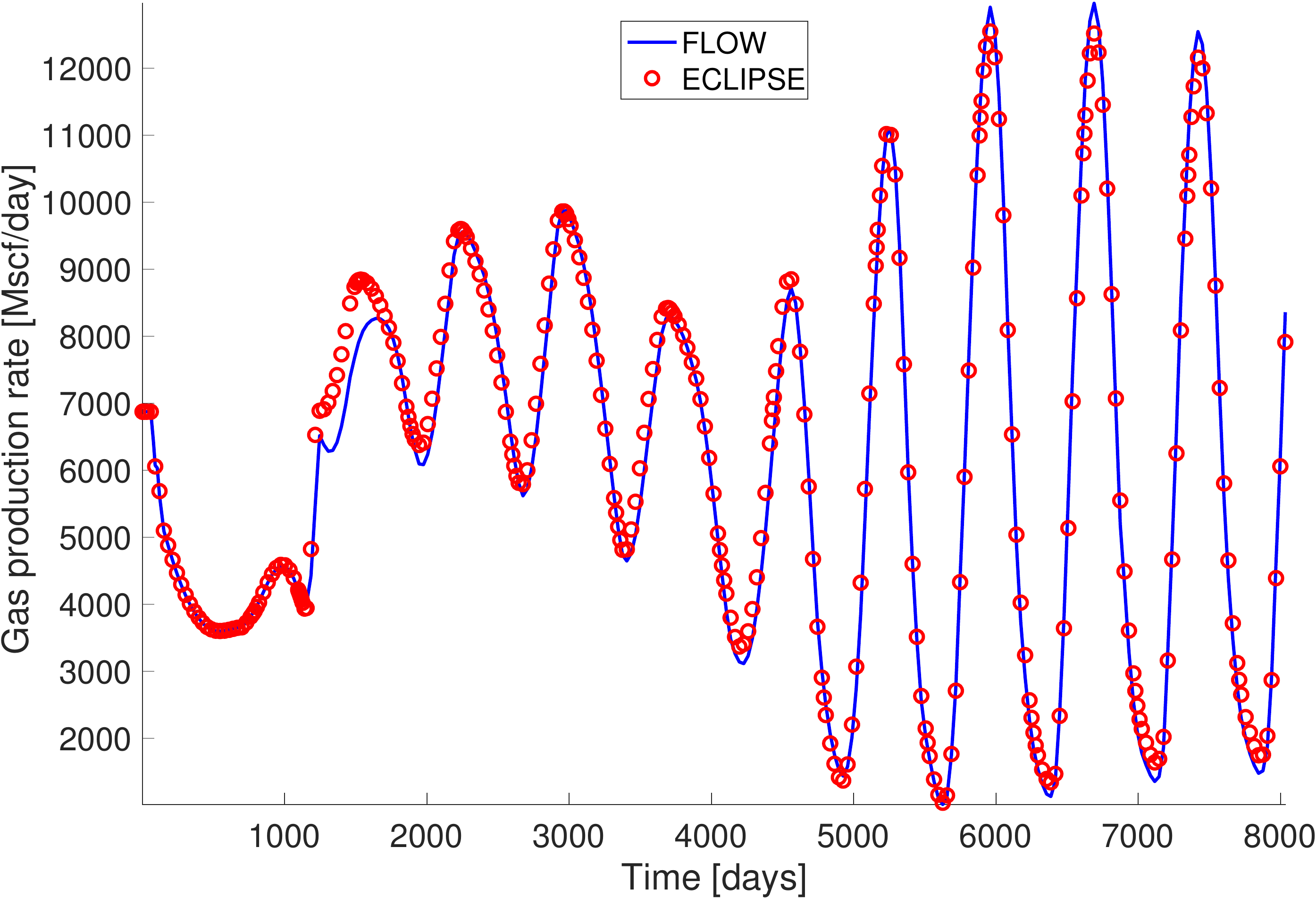}\\
  \includegraphics[width=.49\textwidth]{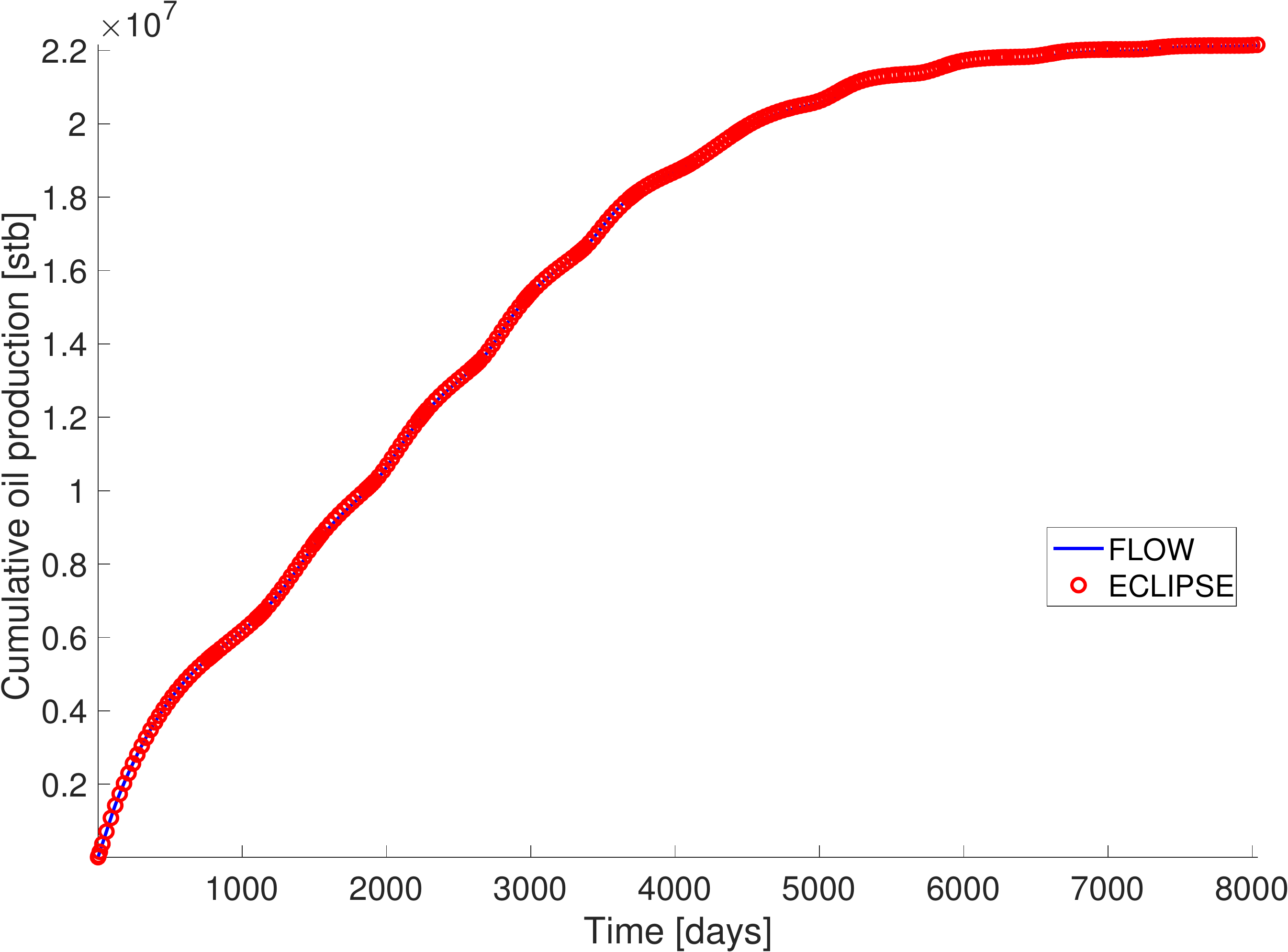}\hfill
  \includegraphics[width=.49\textwidth]{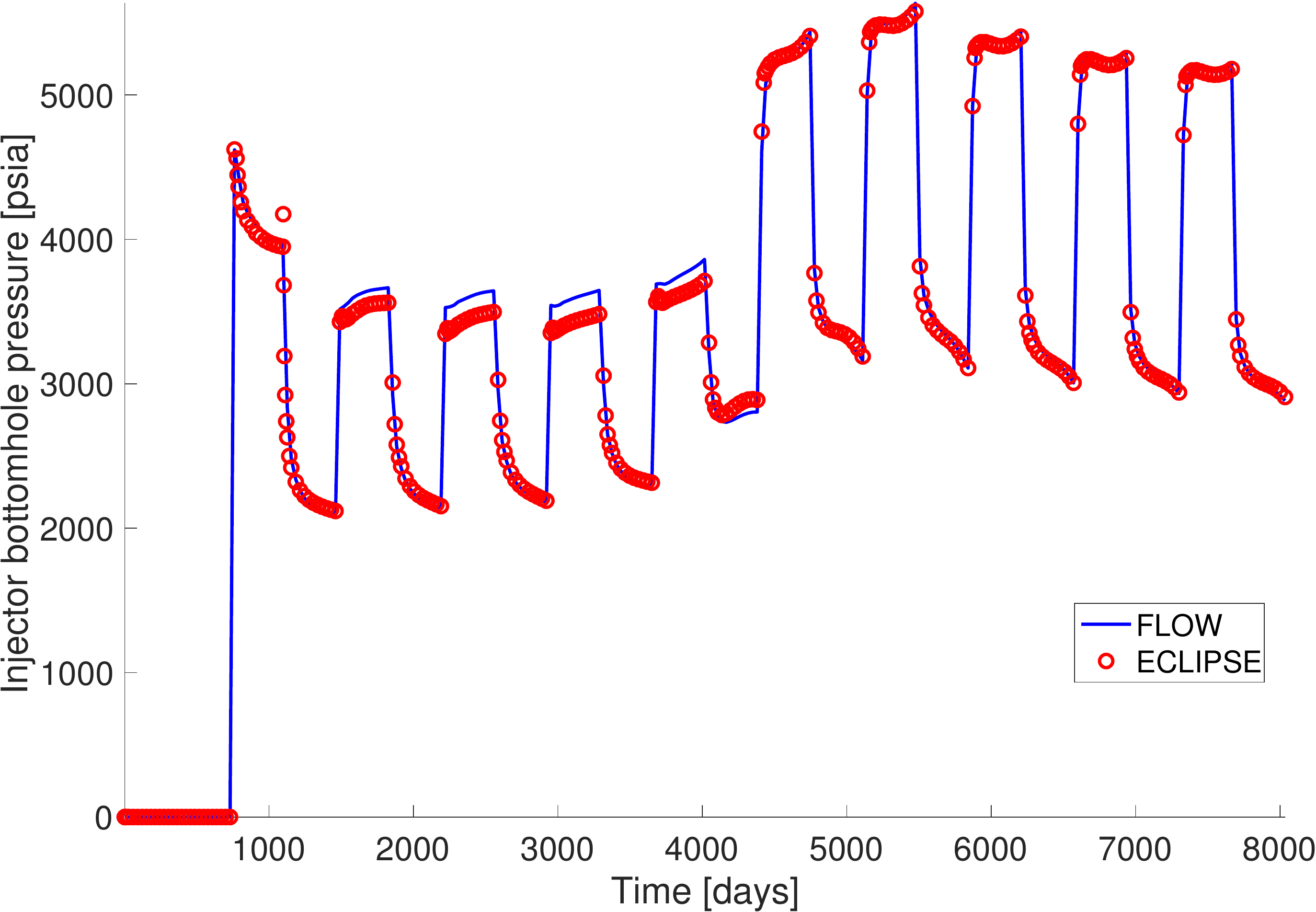}
  \caption{Oil production rate, gas production, and cumulative oil
    production of the producer, as well as bottom-hole pressure of the
    injector for the SPE 5 benchmark, Case 1.}
  \label{fig:spe5_results}
\end{figure}

Figure~\ref{fig:spe5_results} shows that, except for some minor discrepancies in the gas
production rate and the injector bottom-hole pressure, there is general agreement between
the well responses computed by \opmflow and those computed by the solvent module in
ECLIPSE. A closer investigation reveals that the discrepancies are mainly caused by
differences in how the density mixture of the Todd--Longstaff model is implemented. To the
best of our knowledge, both simulators uses the same model, but the complexity of the
model may lead to differences in the actual implementation.

\subsection{SPE 9 benchmark}
\label{sec:spe9}

\begin{figure}
  \centering
  \includegraphics[width=0.57\textwidth]{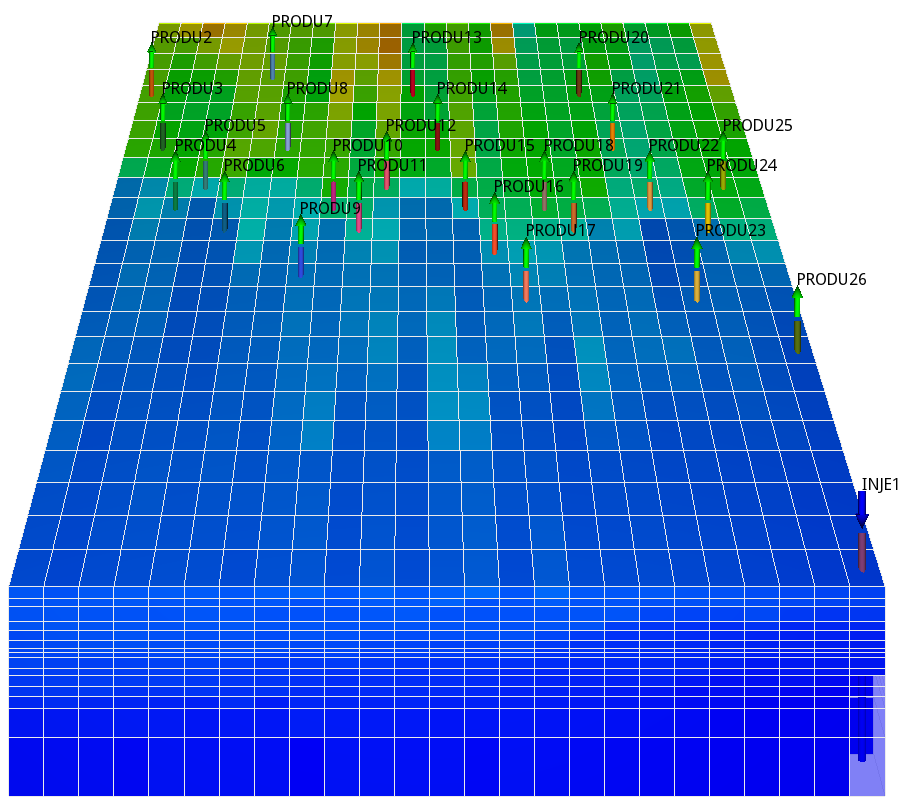}
  \includegraphics[width=0.4\textwidth]{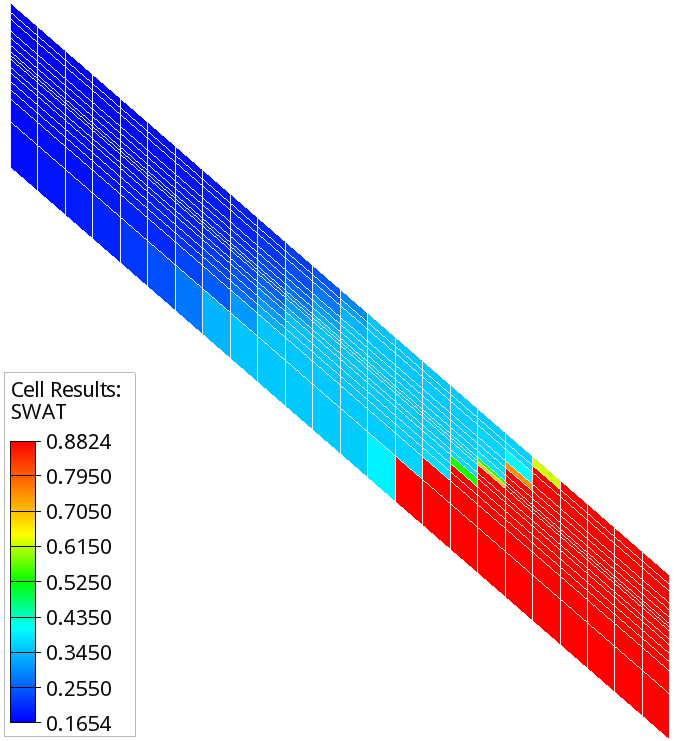}
  \caption{Left: frontal view of the SPE 9 benchmark case showing free
    gas saturation at the end of the simulation, and all wells. Right:
    side view showing initial water
    saturation. The z-axis is exaggerated 5 times in both plots.}
  \label{fig:spe9_sgas_swat}
\end{figure}

The ninth comparative solution project \citep{killough1995ninth} is designed to re-examine
black-oil simulation. The model includes a rectilinear grid with 9\,000 grid cells
and the dimension is 24$\times$25$\times$15 (Figure~\ref{fig:spe9_sgas_swat}).
Originally, the grid was provided in conventional rectangular
coordinates. In \opmdata, we provide data sets also in the corner-point grid format.
Cell (1, 1, 1) is at a depth of 9\,000 feet, and the remaining part of the grid is dipping in the
$x$-direction with an angle of 10 degrees. The model has a layered porosity
structure and highly heterogeneous permeability field. The well pattern
consists of 25 rate-controlled producers and one injector in the corner of the grid.

The total simulation time is 900 days. The injector is controlled by water injection
rate with target rate 500 stb/day and BHP limit 4\,000 psia. The producers are
under oil rate control with BHP limit 1\,000 psia. The production rate target is
1\,500 stb/day for all the producers, except between 300 and
360 days, when the rate target is temporarily lowered to 100 stb/day.

\begin{figure}
  \includegraphics[width=.49\textwidth]{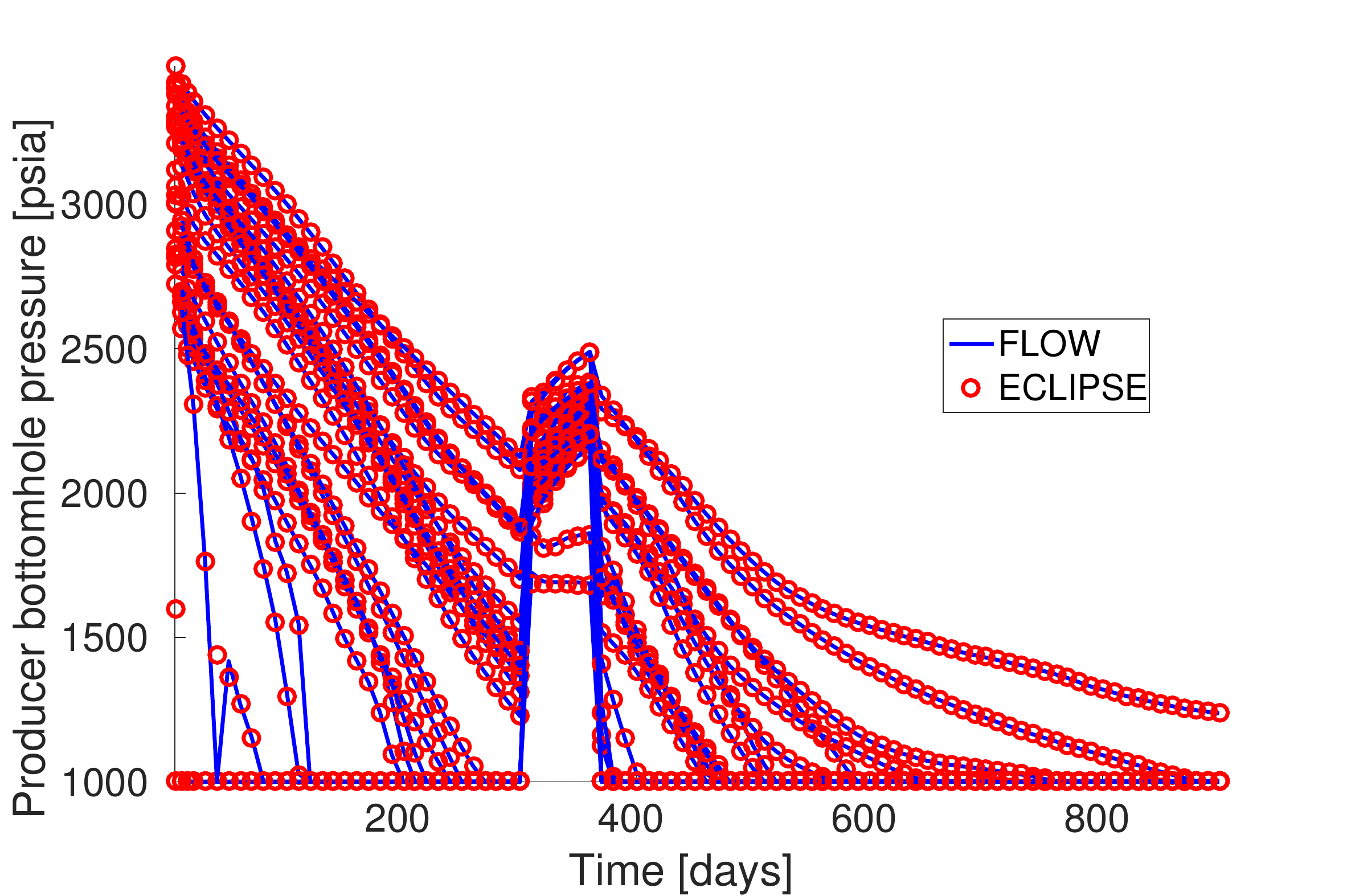} \hfill
  \includegraphics[width=.49\textwidth]{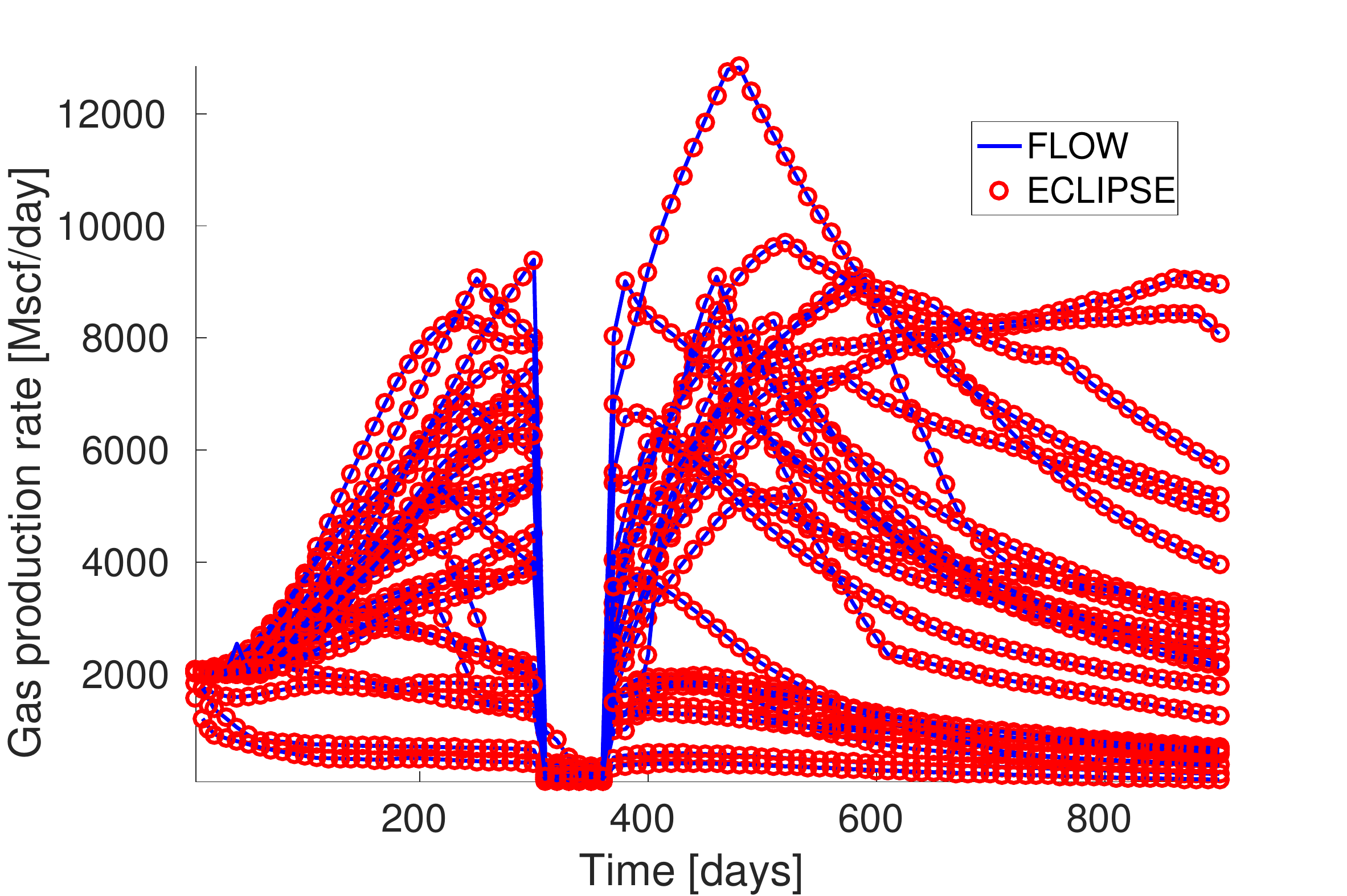}\\
  \includegraphics[width=.49\textwidth]{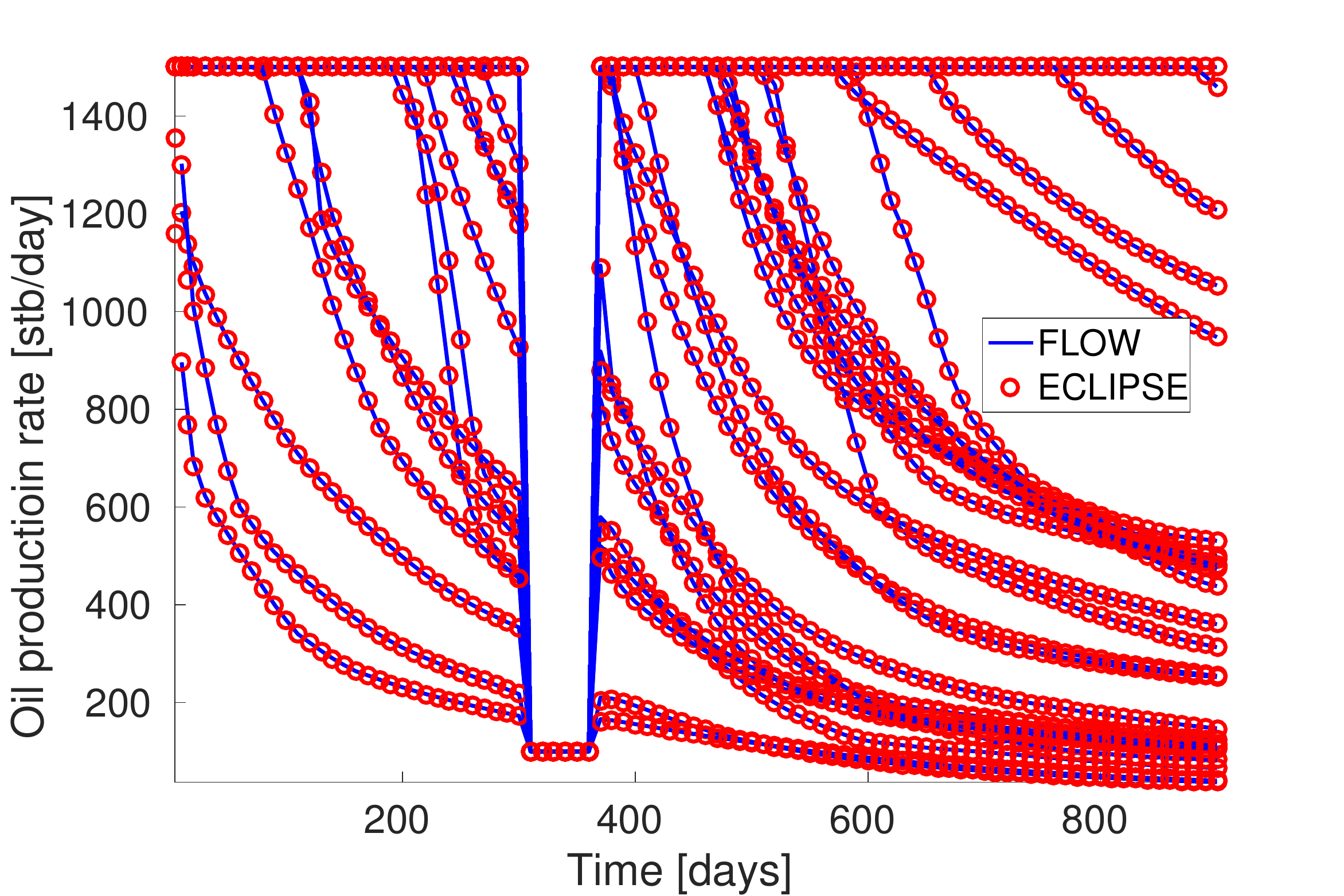}\hfill
  \includegraphics[width=.49\textwidth]{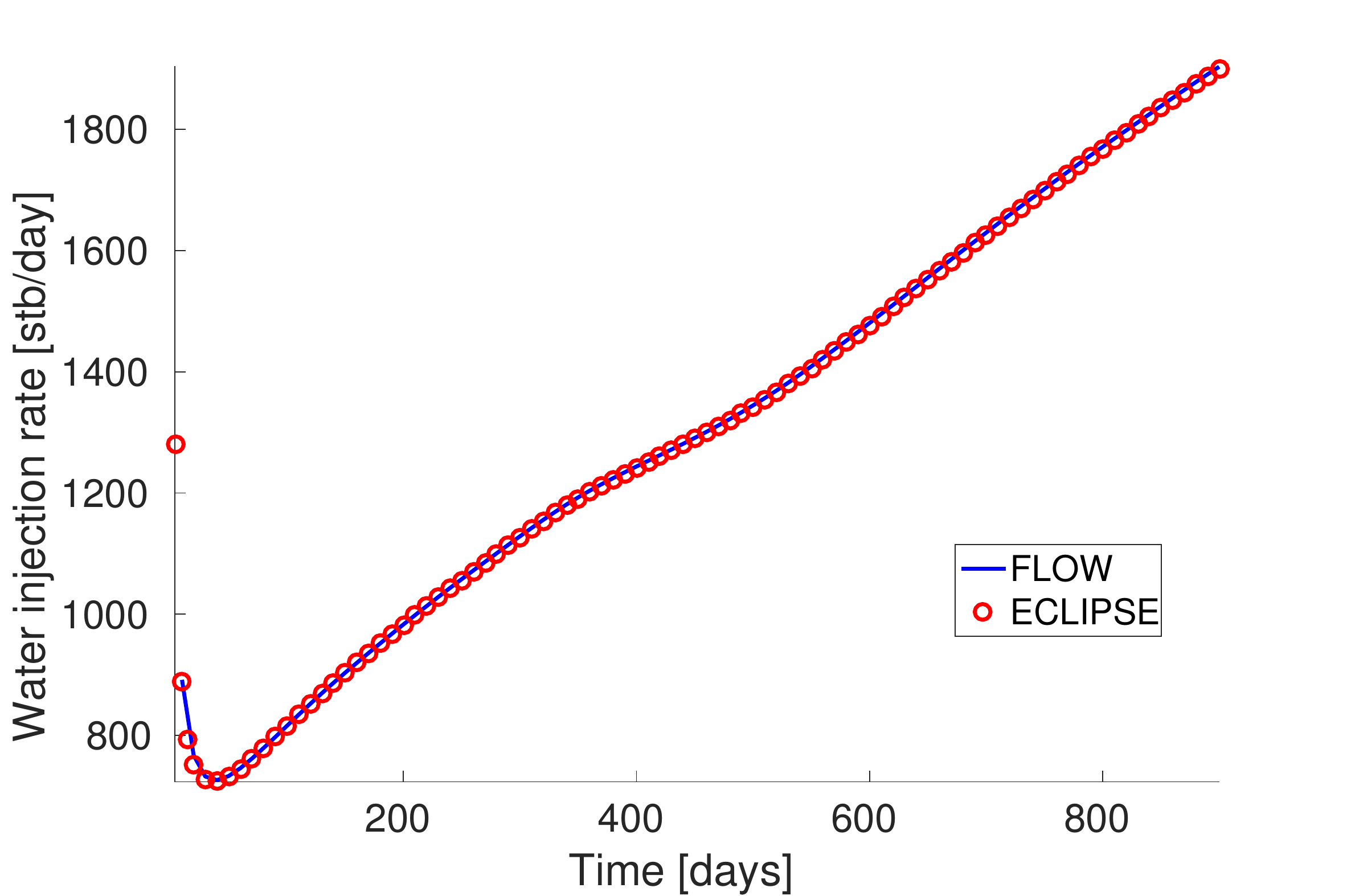}
  \caption{Bottom hole pressures, gas production rates, and oil production rates for all
    producers in the SPE 9 benchmark. Water rate are also reported for the injector.}
  \label{fig:spe9_results}
\end{figure}

This benchmark exercises in particular the well model of the simulator
and its interaction with the reservoir. It also has nontrivial phase
behaviour, as initially there is no free gas in the reservoir, but as
pressure is reduced below the bubble point, free gas appears near the
top, as can be seen in Figure~\ref{fig:spe9_sgas_swat}.
Figure~\ref{fig:spe9_results} shows that the well curves computed by
\opmflow and ECLIPSE are in excellent agreement.

\subsection{Norne}
\label{sec:norne}

\begin{figure}
  \centering
  \includegraphics[width=1.0\textwidth]{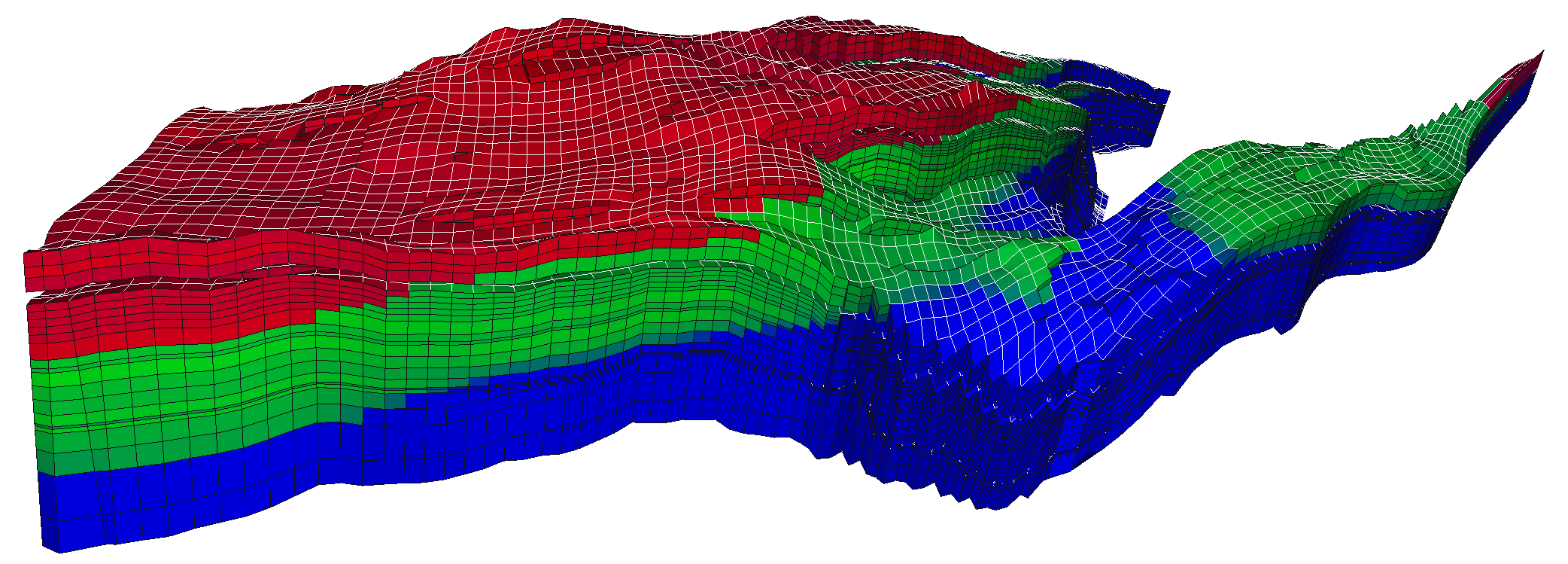}
  \caption{Grid for the Norne field case, with initial fluid
    distribution. Red is gas, green is oil, and blue is water.}
  \label{fig:norne_grid}
\end{figure}

Norne is a Norwegian Sea oilfield that has been in
production since 1997. The operators of the field have chosen to share
data, making it the only real field simulation case that was
openly available until 2018, when the Volve field data were also made open.
The Norne field was initially operated using alternating
water and gas injection, and in total 36 wells have been opened, not
all being active at the same time. The simulation model uses a
corner-point grid with sloping pillars and several faults, so the grid
has many non-logical-Cartesian connections and is therefore completely
unstructured and challenging to process.

\begin{figure}
  \includegraphics[width=.49\textwidth]{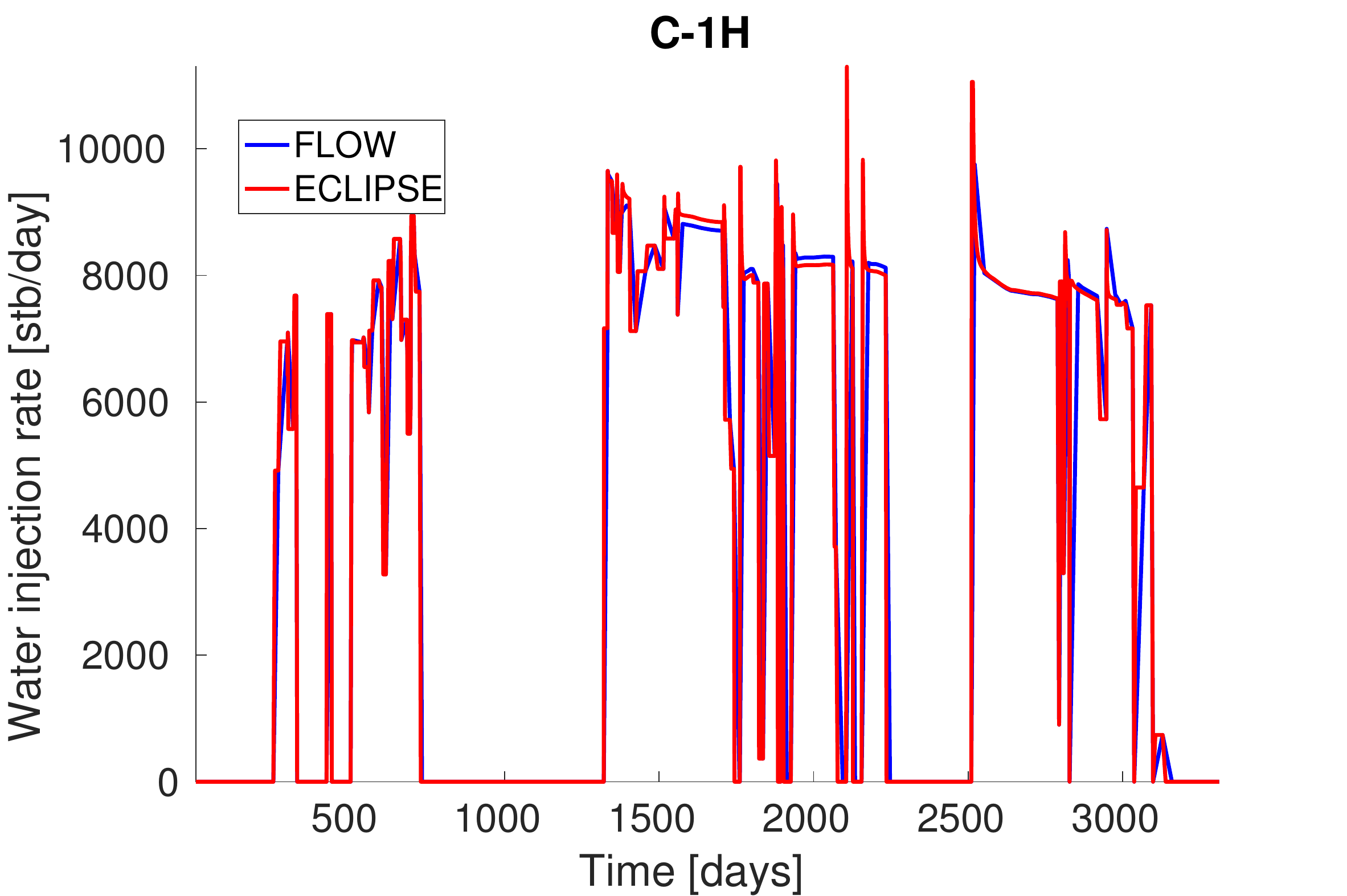}\hfill
  \includegraphics[width=.49\textwidth]{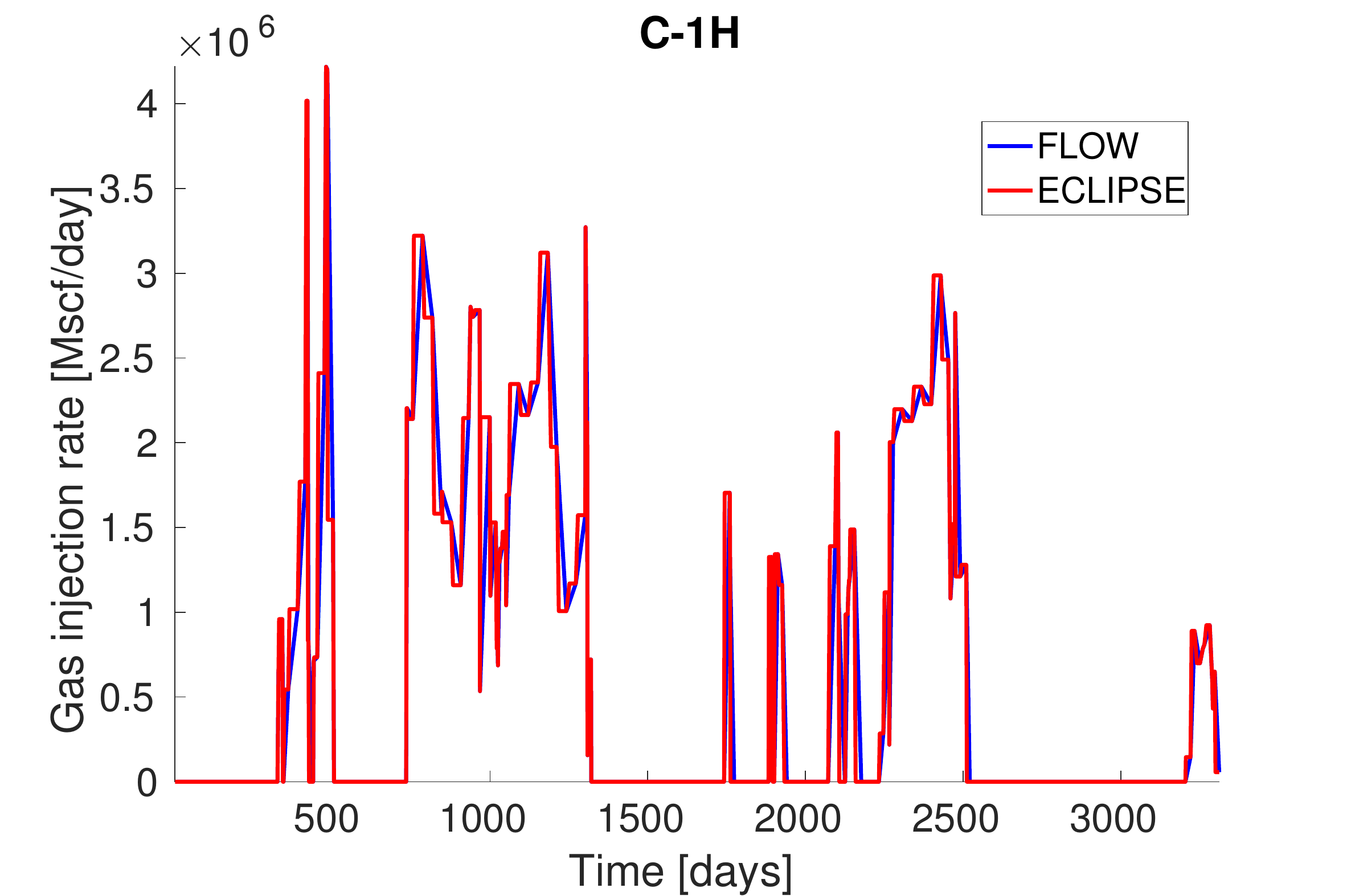}\\
  \includegraphics[width=.49\textwidth]{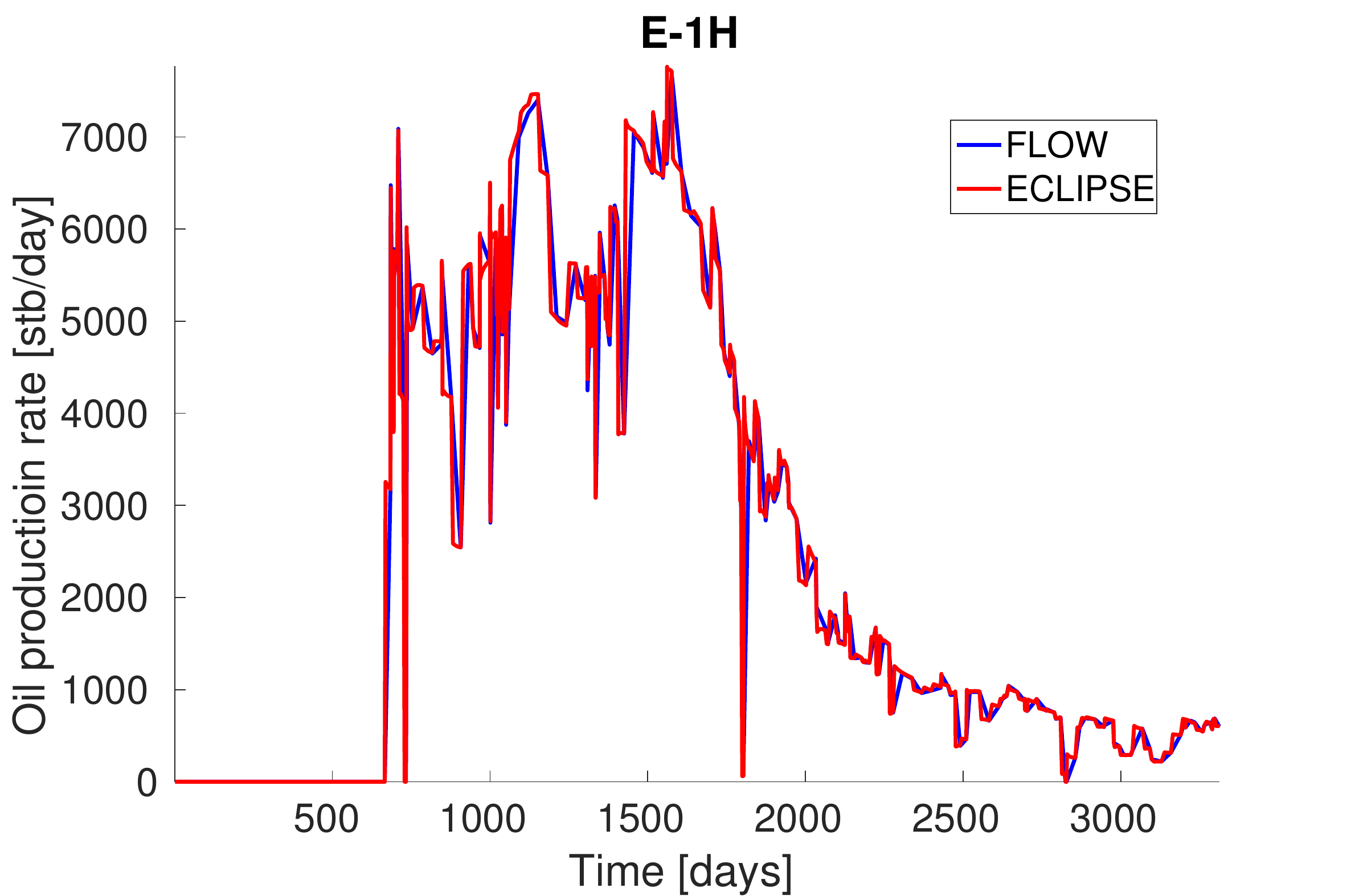}\hfill
  \includegraphics[width=.49\textwidth]{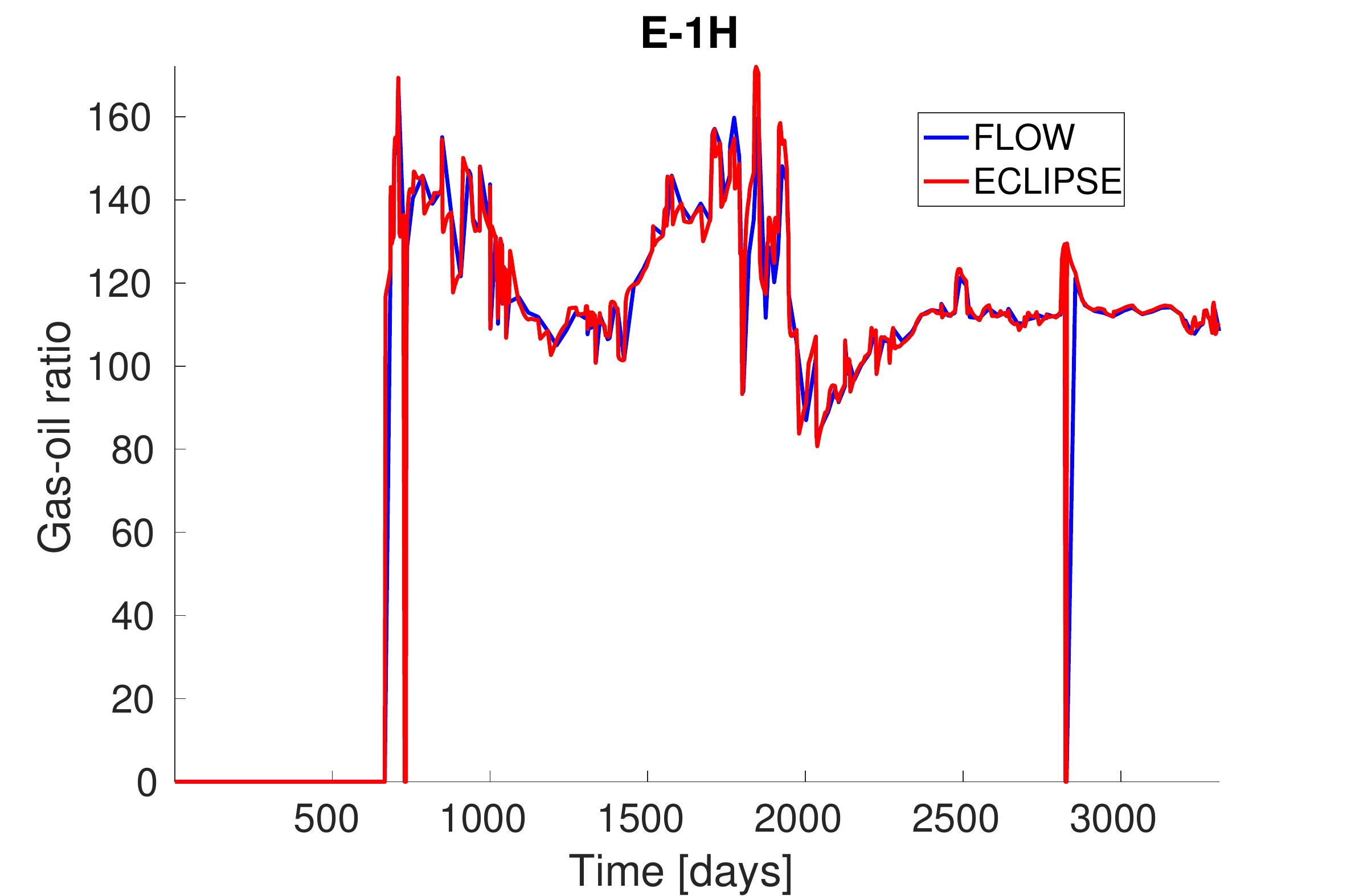}
  \caption{Well responses for injector C-1H and producer E-1H of the Norne field.}
  \label{fig:norne-match}
\end{figure}

The Norne field model includes features such as hysteresis, end-point scaling,
oil vaporization control, and multiple saturations regions.  For a full 
description or to test the model, confer to \href{http://github.com/OPM}{github.com/OPM}. 
Figure~\ref{fig:norne-match} shows that there also in this case is excellent match between
\opmflow and ECLIPSE, despite the complexity of the model.


\subsection{Polymer injection enhanced oil recovery (EOR) example}
\label{sec:polymer}

\begin{figure}
{\includegraphics[width=0.55\textwidth]{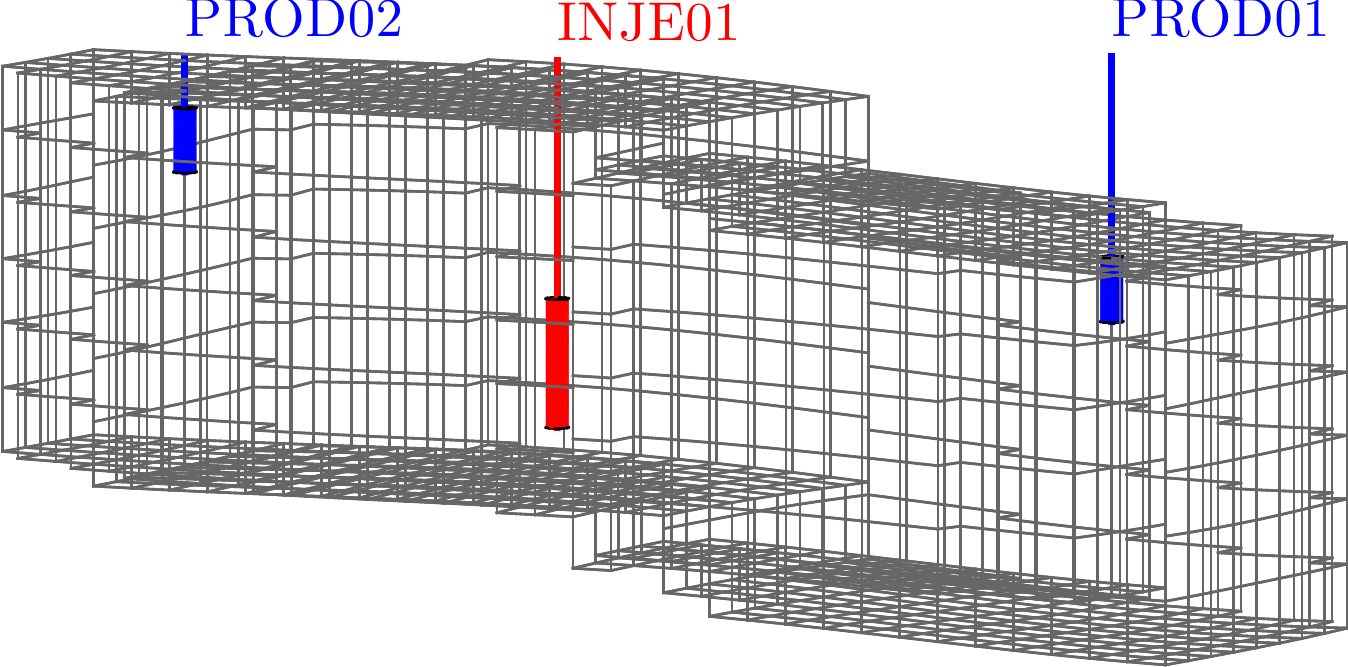}} \
{\includegraphics[width=0.45\textwidth]{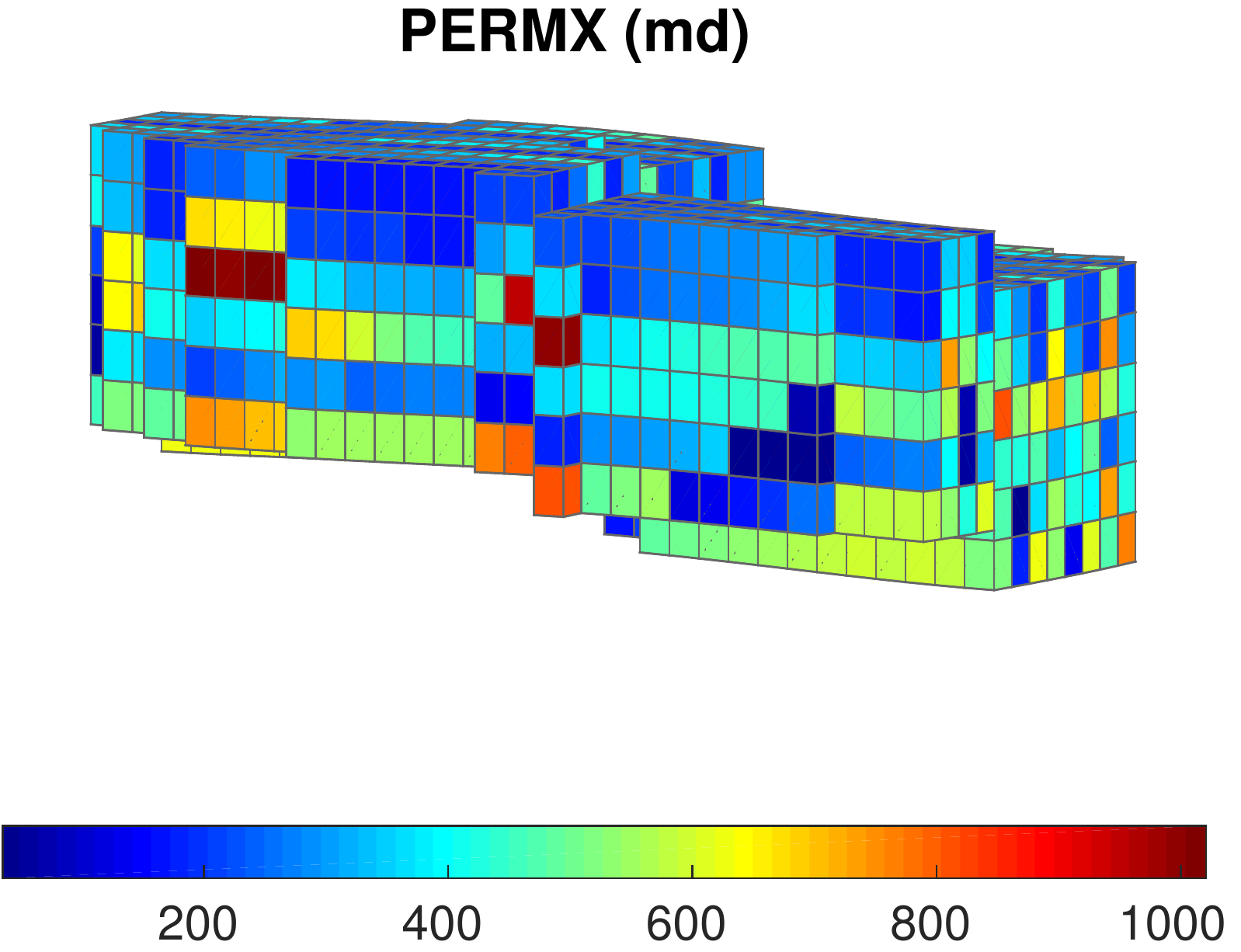}}
\caption{Computing grid and horizontal permeability of the polymer example.}
\label{fig:polymer_example_setup}
\end{figure}

To demonstrate the use of the polymer functionality of \opm, we
consider a simple 3D example taken from
\citep{Bao2017}. The grid consists of 2\,778 active cells and is generated with MRST
\citep{MRST}. The reservoir has physical extent of 1\,000m $\times$ 675m $\times$ 212m and
contains one injector and two producers (Figure~\ref{fig:polymer_example_setup}). The
injector is under rate control with target rate 2\,500\ m$^3$/day and bottom-hole pressure
limit 290 bar. The producers are under bottom-hole pressure control with a bottom-hole
pressure target of 230 bar.

The flooding process begins with a 560-day pure water injection, then a 400-day polymer
injection with concentration 1.0 kg/m$^3$, followed by a 1\,530-day pure water
injection. Four different kinds of configurations compared.  In the first, no polymer is
injected. The second includes polymer, but does not account for any shear effect. The
third models shear-thinning effects, whereas the fourth includes shear-thickening effects.

\begin{figure}
  \includegraphics[width=.49\textwidth]{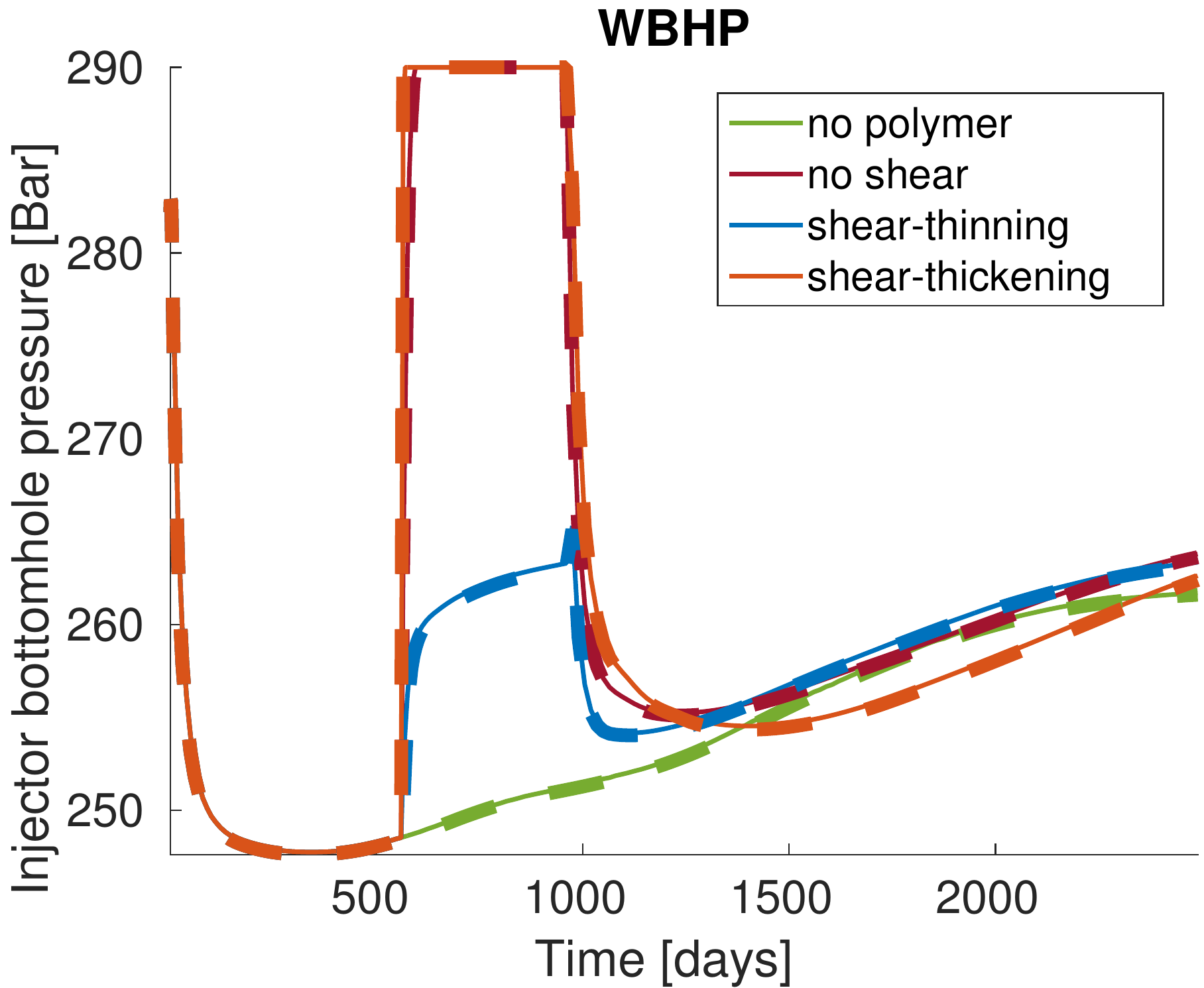} \hfill
  \includegraphics[width=.49\textwidth]{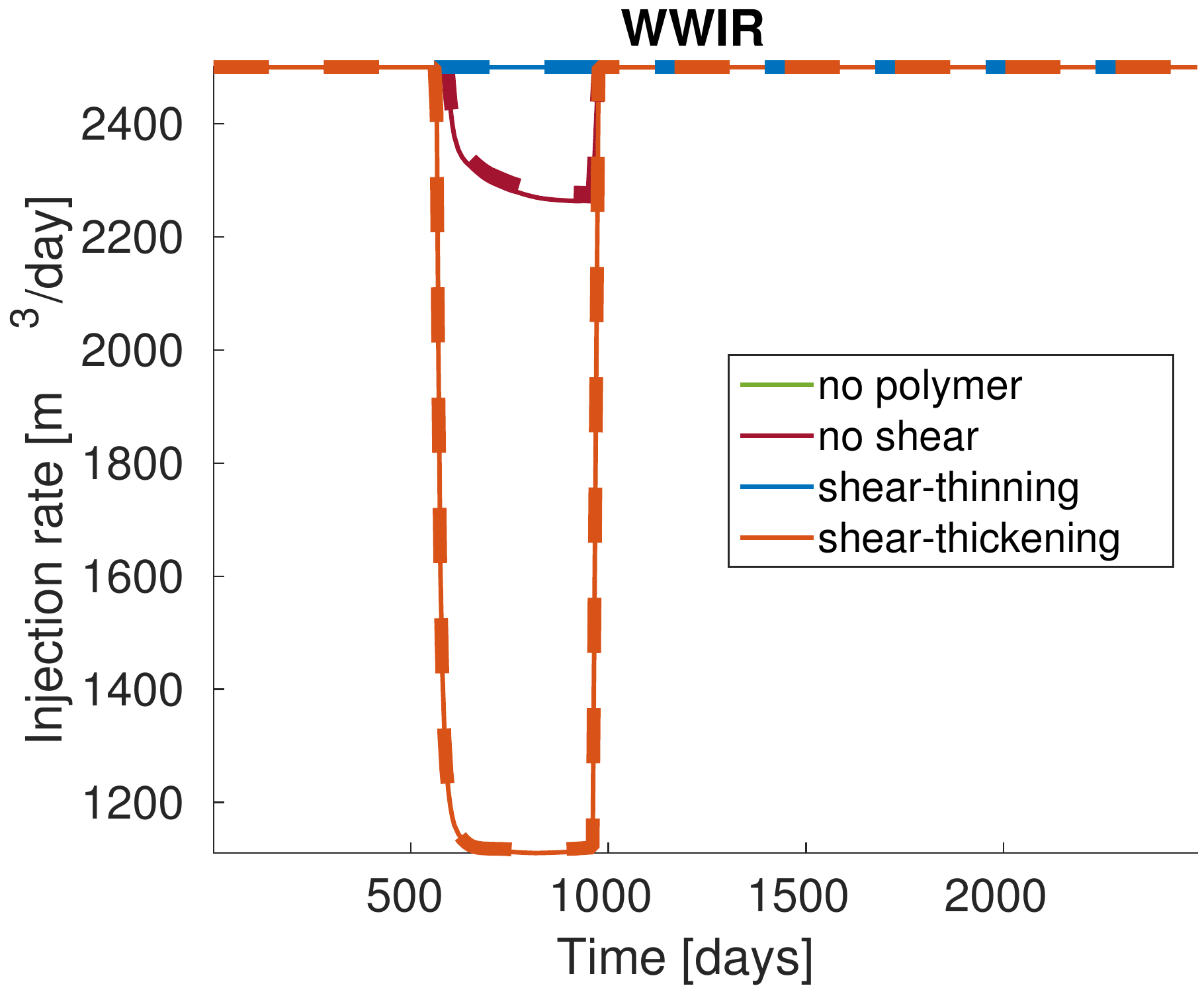} \\
  \includegraphics[width=.49\textwidth]{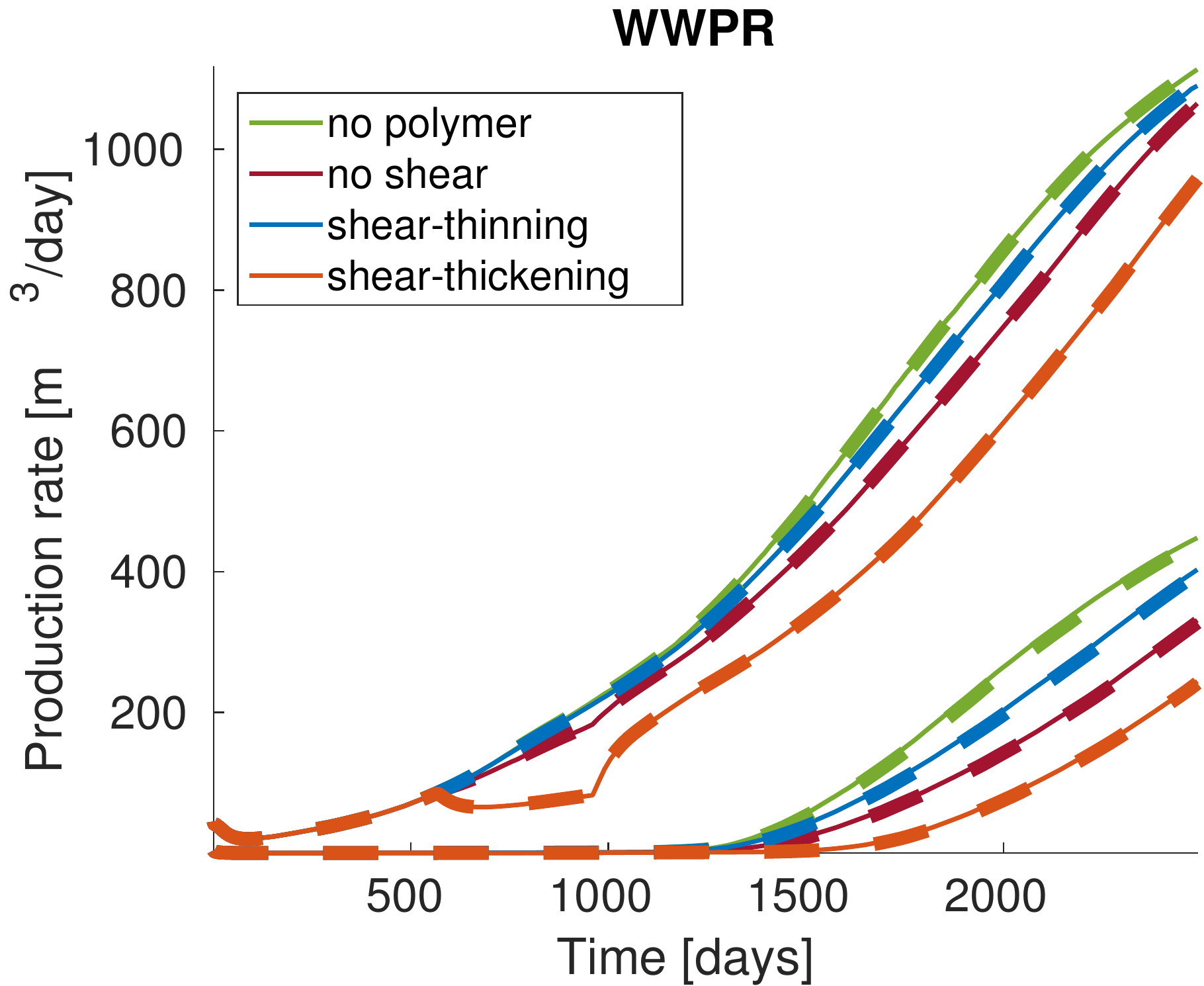} \hfill
  \includegraphics[width=.49\textwidth]{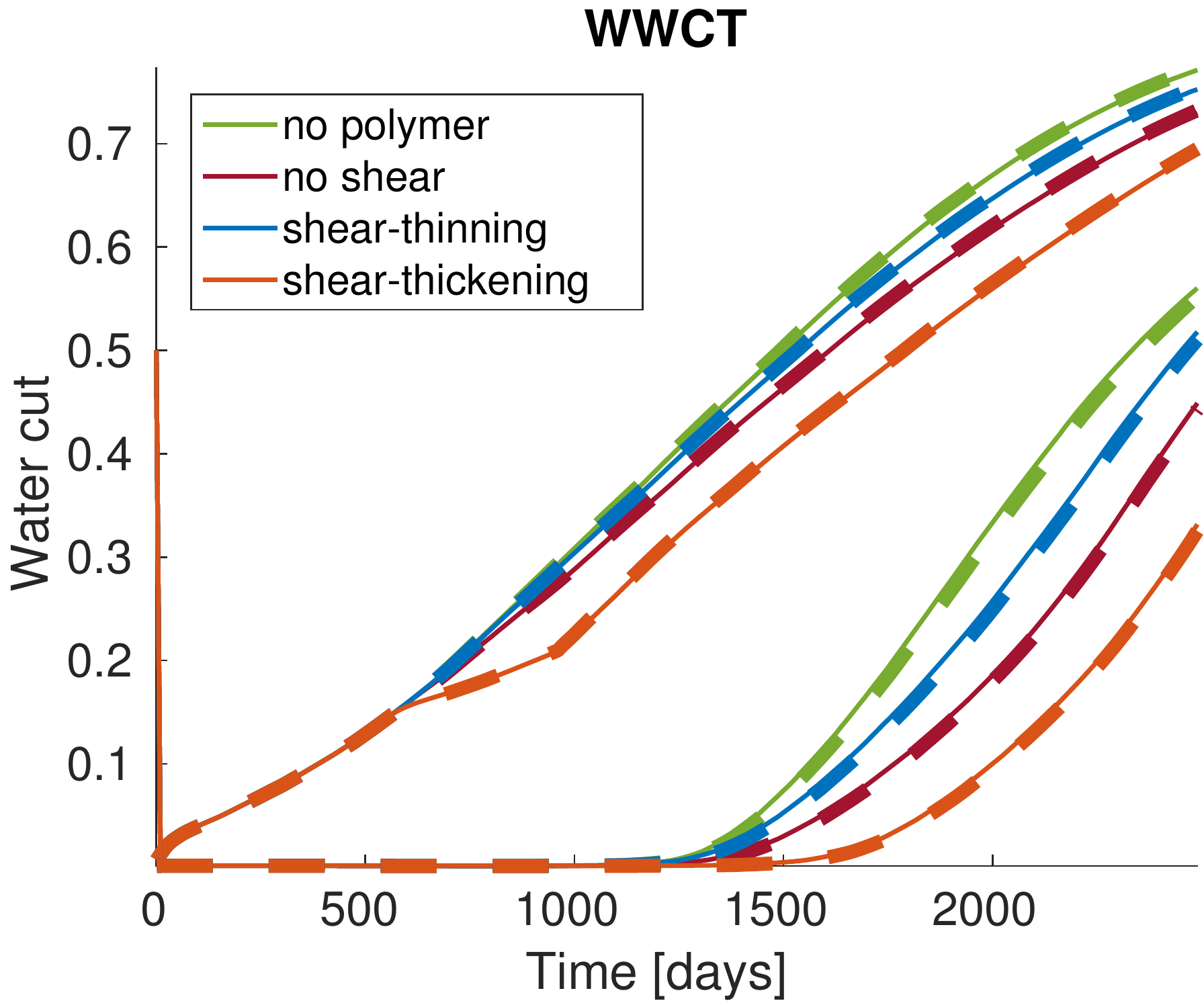}
  \caption{Well responses for the polymer flooding case. The top plots show 
    bottom-hole pressure (left) and water rate (right) for the injector.
    The bottom plots show  water rates and water cuts for the two producers.
    Results from \opmflow are shown with thin lines, ECLIPSE results with thick
    dashed lines.}
  \label{fig:polymer_example_plotting}
\end{figure}

Figure~\ref{fig:polymer_example_plotting} reports well responses for all four
configurations. The bottom-hole pressure (bhp) of the injector increases when the polymer
injection process starts. When no-shear effect is involved, the bhp reaches the limit, and
the injector switches to bhp control mode, which causes decreased injection rate. This is
a natural consequence of reduced injectivity from the polymer injection. With
shear-thinning effect, the bhp of the injector still increases but manages to stay below
the limit. As a result, the target injection rate is maintained. With shear-thickening
effect, the injection rate decreases further compared with the simulation that disregarded
the shear effect. We emphasize that this example is designed for verification of the
polymer functionality of \opm and not related to any real polymer flooding practice.

For comparison purposes, results from ECLIPSE are also
included in Figure~\ref{fig:polymer_example_plotting} with thick dashed lines.
Good agreement is observed between the simulators for all four configurations.

\subsection{Numerical performance}
Reservoir simulators are complex, and analysing numerical performance is involved.  Hence,
we can only present a brief overview here. In terms of performance, a well-implemented
reservoir simulator can be divided in two operations of almost equal importance. The first
is the assembly or linearization, i.e., calculating the Jacobian matrix.  The second is
the linear solve with preconditioning. For \opmflow, the assembly part may take
approximately half the simulation time with the linear solver accounting for the remaining
half. However, this is strongly case-dependent, and the balance changes with the
complexity of the fluids used, length of time steps, etc.  Before it makes sense to
compare \opmflow computationally to other implementations, we need to point our a few
prominent implementation choices.  Use of automatic differentiation makes the calculation
of the Jacobian less error prone, but requires more computations than a direct approach,
even though implementation of automatic differentiation using SIMD instructions reduces
parts of this overhead, as shown in \cite{lauser2018local}.

For the linear solves, it is the preconditioner that uses most of the computational
time. Multiple preconditioners are available in \opmflow. The default is incomplete
LU-factorization with zero fill-in.  This seems to be a good choice for small and medium
sized problems. For larger cases, CPR precoditioning with AMG is preferable, and \opmflow
provides this as an option.  For the linear solver itself, a biconjugate gradient method
is used by default.

\subsubsection{Serial performance} \label{sec:serial-performance}
For very small cases like SPE1 and SPE3, the serial performance of \opmflow is comparable
to that of ECLIPSE 100. The runs typically finish faster with \opmflow, since the
simulator does not need to check for licenses. Since both cases finish in a couple seconds
on a modern CPU, performance is typically not important, though. Moving over to SPE9,
which is more complex, the picture changes in favor of ECLIPSE 100. Actually, ECLIPSE 100
is approximately twice as fast on SPE9 compared to \opmflow. Still, the case finishes in
around ten seconds on a modern CPU, so it has not been considered important for
performance tuning in \opmflow. So far, the numerical efforts in \opmflow have been
centered around two full-field simulation cases. The first is Norne, which is openly
shared with the community. The Norne model is very different from the SPE cases, both in
terms of being more complex, and in terms of run times. A serial run of the Norne
case typically takes around ten minutes on a fast CPU. The run times of \opmflow and
ECLIPSE 100 are similar, with about ten percent run time advantage to ECLIPSE 100 when
using \opmflow from precompiled packages. Compiling \opmflow from source with more recent
compiler and more aggressive compiler switches, the tables turn in favour of \opmflow.

In addition, a proprietary full-field model has been used to guide implementation, a model
which is computationally more intensive than Norne. The model is unfortunately only
available through a non-disclosure agreement, and hence cannot be shared openly. The model
has more than 100\,000 cells, dozens of injectors and producers that penetrate the
reservoir vertically and horizontally, and complex geology with faults and thin layers.
The run time is approximately four times that of the Norne model. On single-case and ensemble
runs of this model, \opmflow performance is generally better than ECLIPSE 100.

A comparison of performance of the solvent extension of the \opmflow simulator is reported
in \cite{sandve2018open} and shows that \opmflow is significantly faster than ECLIPSE 100
for the solvent extension.

\subsubsection{Parallel scaling results}

Most commercial reservoir simulators use distributed memory with
message passing interface (MPI) for their parallel
implementation. Furthermore, domain decomposition of the reservoir
grid is done to split the load between the processes. How this is done
varies between simulators. The ECLIPSE 100 simulator decomposes the
domain along one axis, while newer simulators tend to use graph-based
partitioners such as Zoltan (used by \opmflow) or Metis (used by
INTERSECT). Similarly to ECLIPSE 100, the current version of \opmflow
does not support splitting a well between domains.

Tests of parallel scaling have been performed both on the Norne model
and on the proprietary model.  Scaling on both models are similar, and
as of the \opmversion release, \opmflow scales well on both models up
to sixteen cores. Scaling beyond sixteeen cores cannot be expected at
this point because of convergence issues. Comparing to ECLIPSE 100,
the the scaling of ECLIPSE 100 is similar to \opmflow up to sixteen
CPUs on the Norne model despite its simpler partitioning
approach. However, the limitations really show for the more
computationally intensive proprietary model, where ECLIPSE 100 is back
to the single CPU run time when using four CPUs. Attempting to use
eight or sixteen CPUs, ECLIPSE 100 will only increase run times to be
much slower than the single CPU time.

Benchmark timing results are available for both models, with up to
four and eight CPU cores, on
\href{http://linuxbenchmarking.com}{linuxbenchmarking.com}. These
results are used by the developers for quality assurance, making sure
that changes to the simulator do not have unintended adverse
consequences for serial or parallel performance.

\begin{table}
  \center
  \caption{\opmflow performance indicators for simulations of the Norne model. The columns
    display number of MPI-processes (procs), threads per process
    during assembly (thr(A)), compute nodes involved,  assembly execution
    time (AT), linear solver execution time (LST), total execution time (TT), total number
    of linear iterations (iter), speedup and parallel efficiency (eff) with respect to
    serial simulation. The computations were performed on four dual AMD EPYC 7601
    2.2GHz 32-core processors using 1, 2, \dots, 64 MPI-processes. The AMD nodes have 8
    memory lanes per socket and a 170.6 GB/s memory bandwidth, and the interconnect
    between the nodes is HDR InfiniBand (200Gbits/s)}\label{tab:NorneScale1}
  \small\medskip
\begin{tabular}{ lll | rrrlll}
  \hline
  procs & thr(A) & nodes & AT (s) &  LST (s) &  TT (s) &   iter $\quad$ &  speedup &       eff \\
  \hline
  1  & 1 & 1 & 351.49 &   300.59 &  691.26 &  25744 &       -- &        -- \\
2  & 1 & 1 & 180.73 &   191.14 &  401.89 &  25331 &  1.72 &  0.86 \\
4  & 1 & 1 &  95.71 &   112.49 &  229.94 &  25148 &  3.00 &   0.75 \\
8  & 1 & 1 &  63.51 &    86.43 &  168.35 &  25388 &  4.11 &  0.51 \\
16 & 4 & 1 &  21.73 &    50.93 &   88.72 &  25736 &  7.79&  0.49 \\
32 & 4 & 2 &  15.82 &    54.16 &   85.67 &  26123 &  8.07 &  0.25 \\
48 & 4 & 3 & 10.98 &    45.65 &   71.84 &  25920 &  9.62 &   0.20 \\
64 & 4 & 4 &   8.90 &    49.92 &   74.16 &  26461 &  9.32 &  0.15 \\
\hline
\end{tabular}
\end{table}

Table~\ref{tab:NorneScale1} reports parallel performance for Norne. Each simulation was
repeated several times, and the best time measurements are displayed. Because 16
MPI-processes per node is sufficient to fully utilize a node's memory bandwidth, adding
additional MPI-processes after 16 on a single node will not improve the performance of the
bandwidth-bound linear solver. Simulations using more than 16 MPI-processes were therefore
run on multiple nodes. Since the matrix assembly is also parallelized
using threads in addition to the MPI parallelization, idle cores can be partially exploited by using
additional threads to improve matrix assembly performance. We therefore use 4 threads per
MPI-processes when assembling for simulations with
16 or more MPI-processes. Hence all available cores of a compute node
are used at this stage. In addition to time measurements, we report speedup and efficiency
with respect to the sequential simulation, as well as the total number of ILU0-BiCG-stab
iterations needed to complete each simulation.

We observe improved total execution time for the parallel simulations, but also notice
that the parallel efficiency drops when the number of MPI-processes increases. The reason
for the drop in efficiency is not related to poor convergence in the parallel linear
solver, since the total number of linear iterations needed to complete each simulation
remains almost constant for all simulations.  We also observe that the matrix assembly
code scales fairly well. Unfortunately, the linear solver does not see much improvement in
performance when using more than 16 MPI-processes. The poor scaling of the linear solver
for more than 16 processors can mainly be attributed to communication overhead. The lack
of scaling in the I/O operations also impacts the overall parallel performance of
\opmflow. For example, when using 48 and 64 processors, the I/O operations account for a
larger proportion of total execution time than the matrix assembly.

\subsubsection{Ensemble simulation performance}

Running ensembles of model realizations is commonly done in history 
matching and optimization studies. The run time of each realization can vary 
significantly due to instability in the nonlinear solution process. As a measure of the 
performance of \opmflow, we therefore show results for a full 
ensemble and compare these with the ECLIPSE simulator. 
All ensemble members use the same grid, but permeability, relative permeability endpoints,
and fault multipliers are varied to represent the uncertainty in the geological properties. 
The ensemble can be acquired from \href{http://github.com/rolfjl/Norne-Initial-Ensemble}{github.com/rolfjl/Norne-Initial-Ensemble};
for more details we refer to  \cite{lorentzen2017history}. 

\begin{figure}
  \includegraphics[width=.49\textwidth]{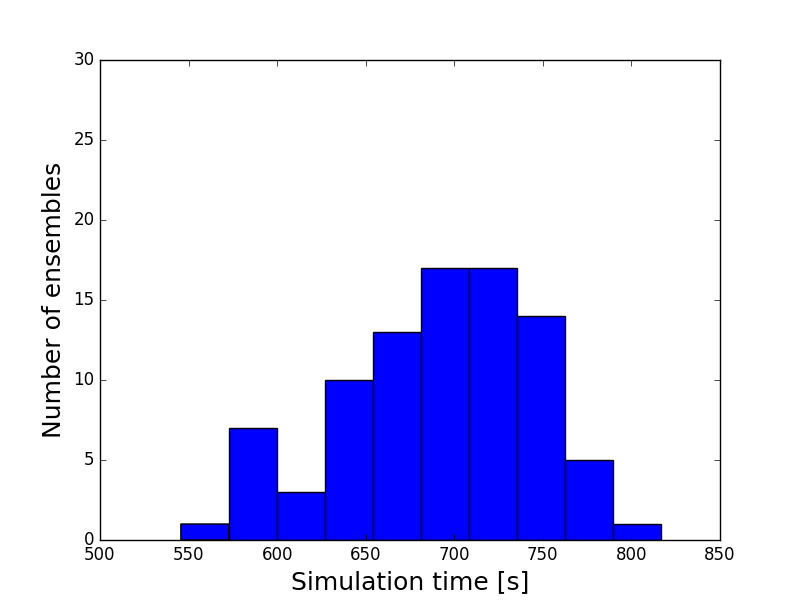} \hfill
  \includegraphics[width=.49\textwidth]{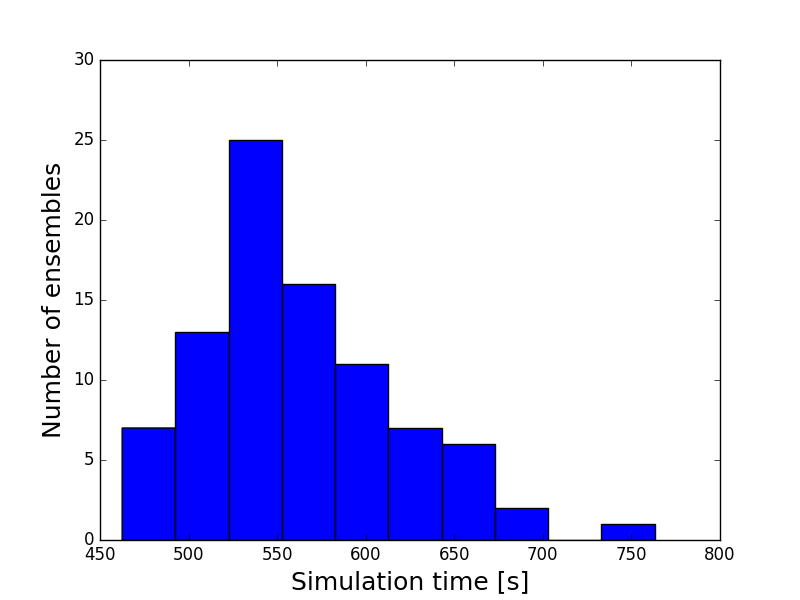}
  \caption{Histogram over simulation times for an ensemble of Norne realizations for
    \opmflow (left) and ECLIPSE (right).}
  \label{fig:norne_performance}
\end{figure}

\begin{figure}
  \centering
  \includegraphics[width=.6\textwidth]{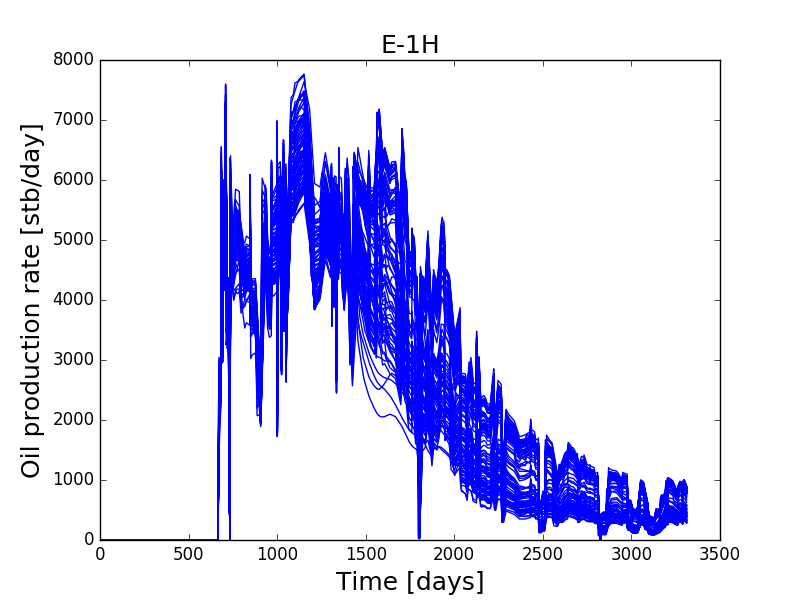}
  \caption{ Oil production rates for 100 realizations of the Norne model. }
  \label{fig:norne_ensemble}
\end{figure}

The individual simulations were run in serial mode on an Intel i7-6700 CPU @ 3.40 Ghz with 16 GiB memory. 
Figure~\ref{fig:norne_performance} shows the resulting run time distributions.
The mean run time for \opmflow on the ensembles is 690 seconds, whereas for ECLIPSE it
is 562 seconds, making \opmflow approximately 25\% slower than ECLIPSE on average.
The ensemble is modified compared to the base case referred to
in Section~\ref{sec:serial-performance}, so the results are not
completely equivalent. We still expect that the difference can be reduced with
aggressive compiler options and light tuning.
Both simulators run through all ensemble members without having severe convergence issues.
Tuning of numerical parameters such as tolerances can sometimes
improve simulator performance. For this comparison, both simulators
have been run with their respective default tuning parameters.
Figure~\ref{fig:norne_ensemble} report oil production results for a single well for the whole ensemble. 
Since unmatched realizations are used, there is a large spread in the results. This spread will typically be
reduced if the models are constrained on the historical data using a history matching procedure.

\section{Summary and Outlook}

This paper has described \opmflow at the time of writing. The software is under continuous
development, so changes and additions should be expected. New features in development
include adjoint calculations, higher-order discretizations in space and time
\cite{Kvashchuk:rsc19}, new fluid models for {\COto} behavior, and sequential implicit
methods.  We think that at this point \opm forms a robust base for both research and
industrial applications, and that going forward this combination will help shorten the
research cycle by giving researchers an industrial-strength platform on which to build and
test their methods, as well as satisfying engineering requirements.

\section{Acknowledgements}

The authors thank Schlumberger for providing us with academic software licenses to
ECLIPSE, used for comparing simulation results with \opmflow.  Work on \opmflow and this
article has been supported by Gassnova through the Climit Demo program and by Equinor.

Robert Kl{\"o}fkorn, Tor Harald Sandve, and Ove S{\ae}vareid acknowledge the Research
Council of Norway and the industry partners (ConocoPhillips Skandinavia AS, Aker BP ASA,
V{\aa}r Energi AS, Equinor ASA, Neptune Energy Norge AS, Lundin Norway AS, Halliburton AS,
Schlumberger Norge AS, and Wintershall DEA) of the National IOR Centre of Norway for
support.

\bibliography{paper}

\appendix

\section{Nomenclature}
\label{appendix:nomenclature}

\subsection{Symbol definitions for the continuous equations}
\label{appendix:nomenclature:cont}

\begin{description}
\item[$\phi$] {\em porosity}, can be a function of (oil) pressure:
  $\phi = m_\phi(p_o)\phi_{\rm ref}$.
\item[$m_\phi$] {\em pore volume multiplier} as function of pressure.
\item[$\phi_{\rm ref}$] {\em reference porosity}, a constant in time
  but varying in space.
\item[$p_\alpha$] {\em phase pressure} for phase $\alpha$.
\item[$b_\alpha$] {\em shrinkage/expansion factor} for phase $\alpha$
  defined as the ratio of surface volume at standard conditions
  to reservoir volume for a given amount of fluid:
  $b_\alpha = V_{\rm{surface}, \alpha}/V_{\rm{reservoir}, \alpha}$.
  For the oil phase, $b_o$ is called {\em shrinkage factor}, whereas
  $b_g$ is called {\em expansion factor}.
  The reciprocal quantity is called {\em formation volume
  factor}, and is usually denoted with a capital $B$:
  $B_\alpha = 1/b_\alpha$.  Usually a function of phase pressure and
  composition: $b_\alpha = b_\alpha(p_\alpha, r_{go}, r_{og})$. See
  Section~\ref{sec:pvt}.
\item[$r_{go}$] {\em ratio of dissolved gas to oil} in the oleic
  phase. Often called $r_S$ in other literature.
\item[$r_{og}$] {\em ratio of vaporized oil to gas} in the gaseous
  phase. Often called $r_V$ in other literature.
\item[$s_\alpha$] {\em saturation} of phase $\alpha$, the pore volume
  fraction occupied by the phase. The saturations of all phases sum to 1.
\item[$p_{c,\alpha \beta}$] {\em capillary pressure} between phases
  $\alpha$ and $\beta$. Typically a function of
  saturation: $p_{c,\alpha\beta} = p_{c,\alpha\beta}(s_\alpha)$.
\item[$\Kb$] {\em permeability} of the porous medium.
\item[$k_{r,\alpha}$] {\em relative permeability} for phase $\alpha$,
  modeling the reduction in effective permeability for a fluid phase
  in the presence of other phases. Typically a function of saturation,
  see Section~\ref{sec:relperm-cappress}.
\item[$\mu_\alpha$] {\em viscosity} of phase $\alpha$, typically a
  function of phase pressure and composition.
\item[$\lambda_\alpha$] {\em mobility} of phase $\alpha$, given by
  $\lambda_\alpha = k_{r,\alpha}/\mu_\alpha$.
\item[$\vb_\alpha$] {\em phase velocity} of phase $\alpha$.
\item[$\ub_\alpha$] {\em component velocity} of component $\alpha$.
\item[$\rho_{S, \alpha}$] {\em surface density} of phase $\alpha$ at
  one atmosphere, a given constant.
\item[$\rho_\alpha$] {\em density} of phase $\alpha$ in the
  reservoir. For water, $\rho_w = b_w\rho_{S,w}$. For the oleic phase
  the relationship is more complex since it must include dissolved
  gas: $\rho_o = b_o(\rho_{S,o} + r_{go}\rho_{S,g})$, similar for the
  gaseous phase: $\rho_g = b_g(\rho_{S,g} + r_{og}\rho_{S,o})$.
\item[$\gb$] {\em gravitational acceleration} vector.
\item[$q_{\alpha}$] {\em well outflux density} of pseudo component $\alpha$,
  (negative for well inflows). The form of this term depends on the
  well model used, see Section~\ref{sec:wellmodels}.
\end{description}

\subsection{Symbol definitions for the discrete equations}
\label{appendix:nomenclature:disc}

\begin{description}
\item[$V$] {\em cell volume}.
\item[$\Delta t$] {\em time step length} for the current Euler step.
\item[$v_\alpha$] {\em volume flux} of phase $\alpha$ (oriented quantity).
\item[$u_\alpha$] {\em surface volume flux} of pseudocomponent $\alpha$ (oriented quantity).
\item[$T_{ij}$] {\em transmissibility} factor for a connection,
  derived from permeability, see~\ref{appendix:trans}.
\item[$m_T$] {\em transmissibility multiplier} as function of
  pressure.
\item[$C(i)$] {\em connections} from cell $i$, i.e. the set of cells
  connected to it.
\item[$U(\alpha, ij)$] {\em upwind cell} for phase $\alpha$ for the
  connection between cells $i$ and $j$.
\item[$\Delta \Phi_{\alpha, ij}$] {\em potential difference} for phase $\alpha$ for the
  connection between cells $i$ and $j$ (oriented quantity).
\item[$g$] {\em gravitational acceleration} in the $z$-direction.
\item[$z_i$] {\em depth} of center of cell $i$.
\end{description}

\subsection{Symbol definitions for the well models}
\label{appendix:nomenclature:well}

\begin{description}
\item[$T_{w,j}$]  connection transmissibility factor.
\item[$M_{\alpha,j}$] the
  mobility for phase $\alpha$ at the connection $j$.
\item[$p_j$]  pressure of the grid block that contains the connection $j$.
\item[$p_{bhp, w}$] the bottom-hole pressure of the well $w$.
\item[$h_{w,j}$] pressure difference within the wellbore between connection $j$ and the
  well's bottom-hole datum depth.
\end{description}

\section{Transmissibility}
\label{appendix:trans}

The transmissibility $T_{ij}$ of a connection between two cells
is a discrete measure of the fluid flow capacity of that
connection. We can define $T_{ij}$ by requiring that the discrete flow
equations \eqref{eq:flux1}, \eqref{eq:flux2}, and \eqref{eq:head}
should be satisfied for a piecewise homogenous medium, by a velocity
field satisfying Darcy's law. For a single, incompressible fluid
$b = 1$ and the discrete equations reduce to
\begin{equation}
  v_{ij} = \frac{1}{\mu} T_{ij}\big( p_i - p_j - g\rho_{ij}(z_i - z_j) \big).
\end{equation}
We can ignore gravity since the formula must hold also
normal to the gravity direction, and assume without loss of generality
that the viscosity is 1, yielding the simplified discrete equation
\begin{equation}
  v_{ij} = T_{ij}\big( p_i - p_j\big)
\end{equation}
and the simplified Darcy law
\begin{equation}\label{eq:simpledarcy}
  \vb = -\Kb \nabla p.
\end{equation}

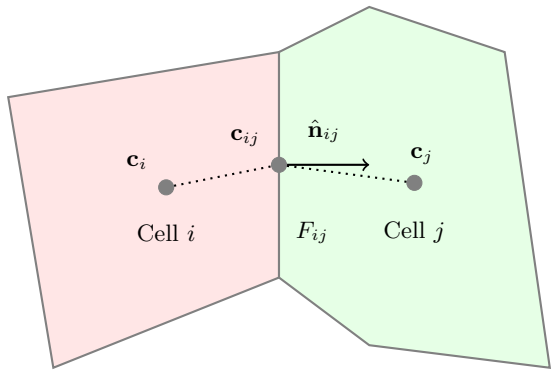
\begin{figure}
  \center
  \tikzstyle{edge}=[gray, thick, draw]
\tikzstyle{point}=[gray, circle, minimum size=2mm, inner sep=0mm, fill, draw]
\tikzstyle{dash}=[black, thick, dashed, draw]
\begin{tikzpicture}[scale=0.6,font=\small]
  \draw [edge,fill=red!10] (0,-2) -- (5,0) -- (5,5) -- (-1,4) -- cycle;
  \draw [edge,fill=green!10] (5,5) -- (5,0) -- (7,-1.5) -- (11,-2) -- (10,5) -- (7,6) -- cycle;

  \node [point] (c1) at (2.5,2) {};
  \node [point] (c12) at (5,2.5) {};
  \node [point] (c2) at (8,2.1) {};
  \node [above left=5pt of {(2.5,2)}] {$\cb_i$};
  \node [above left=5pt of {(5,2.5)}] {$\cb_{ij}$};
  \node [above right=5pt of {(7.5,2.1)}] {$\cb_j$};

  \node  [right=1pt of {(5,1)}, outer sep=2pt] {$F_{ij}$};

  \node (x1) at (2.5, 1) {Cell $i$};
  \node (x1) at (8, 1) {Cell $j$};

  \draw [black, thick, dotted] (c1) -- (c12);
  \draw [black, thick, dotted] (c12) -- (c2);

  \draw [black, thick, ->] (c12) -- (7,2.5);
  \node [above=5pt of {(6,2.5)}] {$\hat \nb_{ij}$};
\end{tikzpicture}
  \caption{Quantities used to compute the transmissibility between two cells.}
  \label{fig:transmissibility}
\end{figure}
 
Consider Figure~\ref{fig:transmissibility}, where $\cb_i$ and
$\cb_j$ are the centroids of cells $i$ and $j$ and $\cb_{ij}$ is the
centroid of their common face $F_{ij}$. Then the
flux $v_{ij}$ is given by
\begin{equation}
  v_{ij}  = \int_{F_{ij}} \vb \cdot \hat\nb_{ij} \, dA
\end{equation}
where $\hat\nb_{ij}$ is the unit normal of the face $F_{ij}$. This can
be approximated by evaluating at the face centroid and using the Darcy
formula \eqref{eq:simpledarcy} in each of the cells $i$ and $j$ by
\begin{align}
  v_{ij} & = -|F_{ij}| (p_{ij} - p_i) \frac{\Kb_i (\cb_{ij} -
           \cb_i)}{|\cb_{ij} - \cb_i|^2} \cdot \hat\nb_{ij},  &
  v_{ij} & = -|F_{ij}| (p_{ji} - p_{j}) \frac{\Kb_j (\cb_{ij} -
           \cb_{j})}{|\cb_{ij} - \cb_{j}|^2} \cdot \hat\nb_{ij},
\end{align}
where $|F_{ij}|$ is the area of the face $F_{ij}$ between cells $i$
and $j$, $p_{ij}$ is the pressure in cell $i$ at the point $\cb_{ij}$,
and $p_{ji}$ the pressure in cell $j$ at $\cb_{ij}$. We then define
the half-transmissibility $t_{ij}$ of cell $i$ with respect to
$F_{ij}$ by
\begin{align}
  t_{ij} & =  |F_{ij}| \frac{\Kb_i (\cb_{ij} -
           \cb_i)}{|\cb_{ij} - \cb_i|^2} \cdot \hat\nb_{ij}
\end{align}
with a similar definition for $t_{ji}$ notably using
$\hat\nb_{ji} = -\hat\nb_{ij}$. Then
\begin{align}
  v_{ij}  = -t_{ij} (p_{ij} - p_i), \qquad
  v_{ij}  = t_{ji} (p_{ji} - p_{j}).
\end{align}
Requiring that the interface pressures $p_{ij}$ and $p_{ji}$ are
equal, we get
\begin{align}
  v_{ij}  = T_{ij}(p_i - p_j) \qquad 
  T_{ij}  = \frac{t_{ij} t_{ji}}{t_{ij} + t_{ji}}
\end{align}
Finally, as a modeling tool one may multiply the transmissibilities by
a pressure-dependent multiplier $m_T$. This can be used for example to
model sub-scale features such as fractures opening and closing due to
pressure changes, thereby changing overall permeability, without
explicitly representing such fractures.
\end{document}